\documentclass[12pt]{article}
\usepackage[dvips]{graphicx}  

\begin{document}

.

\vspace{20mm}

\begin{center}

{\Large\bf  Dynamical 3-Space: A Review
\rule{0pt}{13pt}}\par

\bigskip

Reginald T. Cahill \\ 

{\small\it School of Chemistry, Physics and Earth Sciences, 

 Flinders University, Adelaide 5001, Australia\rule{0pt}{13pt}}\\

\vspace{10mm}

{\footnotesize\parbox{11cm}{%
For some 100 years physics has modelled space and time via the {\it spacetime} concept, with space being  merely an observer dependent perspective effect of that spacetime - space itself had no observer independent existence - it had no ontological status, and it certainly had no dynamical description. In recent years this has all changed.  In 2002 it was discovered that a dynamical 3-space had been detected many times, including the Michelson-Morley 1887 light-speed anisotropy experiment.  Here we review the dynamics of this 3-space, tracing its evolution from that of an emergent phenomena in the information-theoretic {\it Process Physics} to the phenomenological  description in terms of a velocity field describing the relative internal motion of the structured 3-space.
The new physics of the dynamical 3-space  is extensively tested against experimental and astronomical observations, including the necessary generalisation of the Maxwell, Schr\"{o}dinger and Dirac equations,  leading to a derivation and explanation of gravity as a refraction effect of the quantum matter waves. Phenomena now explainable include the bore hole anomaly, the systematics of black hole masses, the flat rotation curves of spiral galaxies, gravitational light bending and lensing, and the supernova and Gamma-Ray Bursts magnitude-redshift data, for the dynamical 3-space possesses a Hubble expanding 3-space solution. Most importantly none of these phenomena now require dark matter nor dark energy.  The flat and curved spacetime formalism is derived from the new physics, so explaining the apparent many successes of those formalisms, but which have now proven to be ontologically and experimentally flawed.

\rule[0pt]{0pt}{0pt}}}\bigskip

\end{center}

\newpage

\small{\tableofcontents}

\section{Introduction}
We review here some of the new physics emerging from the discovery that there exists a dynamical 3-space. This discovery changes all of physics. While at a deeper level this emerges from the information-theoretic Process Physics \cite{Book,PP2003,boot,srn,pp,R1,inertia} here we focus on the phenomenological description of this 3-space in terms of the velocity field that describes the internal dynamics of this structured 3-space.  It is straightforward to construct the minimal dynamics for this 3-space, and it involves two constants: $G$ - Newton's gravitational constant, and $\alpha$ - the fine structure constant.  $G$ quantifies the effect of matter upon the flowing 3-space, while $\alpha$ describes the self-interaction of the 3-space. Bore hole experiments and black hole astronomical observations give the value of $\alpha$ as the fine structure constant to within observational errors.  A major development is that the Newtonian theory of gravity \cite{Newton} is fundamentally flawed - that even in the non-relativistic limit it fails to correctly model numerous gravitational phenomena. So Newton's theory of gravity is far from being `universal'.  The Hilbert-Einstein theory of gravity (General Relativity  - GR), with gravity being  a curved spacetime effect, was based on the assumption that Newtonian gravity was valid in the non-relativistic limit. The ongoing effort to save GR against numerous disagreements with experiment and observation lead to the invention first of `dark matter' and then `dark energy'.  These effects are no longer required in the new physics. The 3-space velocity field has been directly detected in at least eight experiments including the Michelson-Morley experiment \cite{MM} of 1887, but most impressively by the superb experiment by Miller in 1925/1926 \cite{Miller}. The Miller experiment was one of the great physics experiments of the 20th century, but has been totally neglected by mainstream physics.  All of these experiments detected the dynamical 3-space by means of the light speed anisotropy - that the speed of light is different in different directions, and the anisotropy is very large, namely some 1 part in a 1000.  The existence of this 3-space as a detectable phenomenon implies that a generalisation of all the fundamental theories of physics be carried out.  The generalisation of the Maxwell equations leads to a simple explanation for gravitational light bending and lensing effects, the generalisation of the Schr\"{o}dinger equation leads to the first derivation of gravity - as a refraction effect of the quantum matter waves by the time dependence and inhomogeneities of the 3-space, leading as well to a derivation of the equivalence principle.  This generalised Schr\"{o}dinger equation also explains the Lense-Thirring effect as being caused by  vorticity in the flowing 3-space.  This effect is being studied by the Gravity Probe B (GP-B) gyroscope precession experiment. The generalisation of the Dirac equation  to take account of the interaction of the spinor with the dynamical 3-space results in the derivation of the curved spacetime formalism for the quantum matter geodesics, but without reference to the GR equations for the induced spacetime metric. What emerges from this derivation is that the spacetime is purely a mathematical construct - it has no ontological status. That discovery completely overturns the paradigm of 20th century physics. The  dynamical equation for the 3-space has black hole solutions with properties very different from the putative black holes of GR, leading to the verified prediction for the masses of the minimal black holes in spherical star systems. That same dynamics has an expanding 3-space solution - the Hubble effect for the universe. That solution has the expansion mainly determined by space itself. This expansion gives a extremely good account of the supernovae/Gamma-Ray Burst redshift data without the notion of `dark energy' or an accelerating universe.  This review focuses on the phenomenological modelling of the 3-space dynamics and its experimental checking. Earlier reviews are available in \cite{Book}(2005) and \cite{PP2003}(2003). Page limitations  mean that  some developments have not been discussed herein.

\section{Dynamics of  3-Space}\label{sect:dynamics}

\begin{figure}[t]
\vspace{0mm}\parbox{60mm}{\hspace{10mm}\includegraphics[width=55mm]{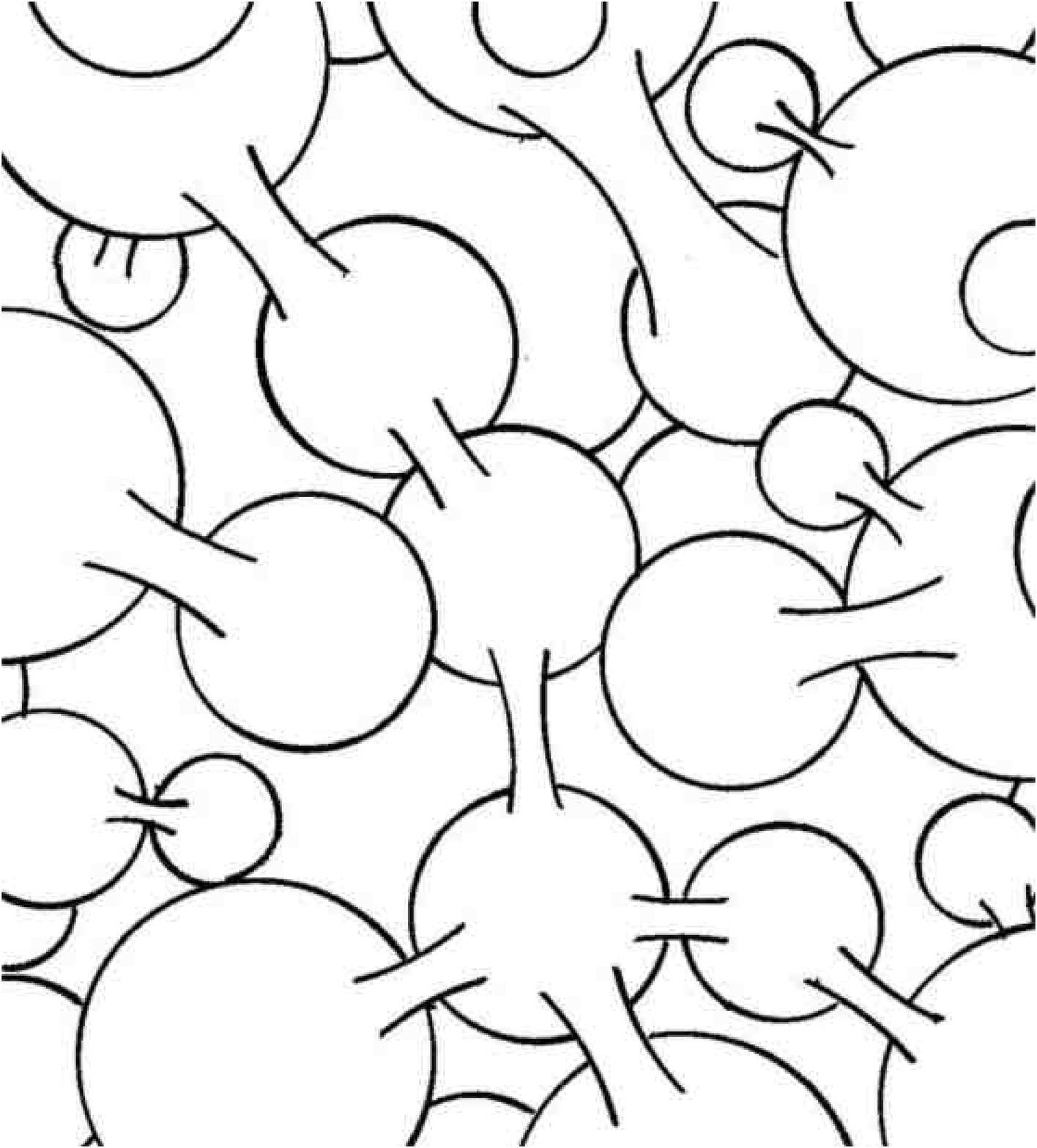}}\,
\parbox{70mm}{\vspace{0mm}\,\parbox{60mm}{\hspace{10mm}\includegraphics[width=60mm]{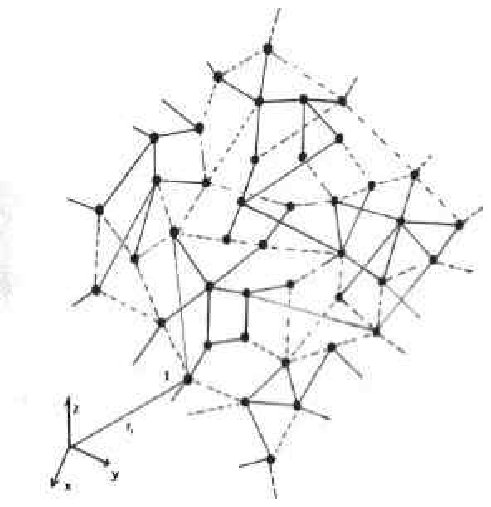}}\,}
\caption{\footnotesize{  This is an iconic representation of how a quantum foam dynamical  network (left), see \cite{Book} for details of the Quantum Homotopic Field Theory, has its inherent approximate 3-dimen\-sion\-al connectivity displayed by an embedding in a math\-em\-at\-ical   space, such as an $E^3$ or an $S^3$ as shown on the right.  The embedding space is not real; it is purely a mathematical artifact. Nevertheless this em\-bedd\-abi\-li\-ty helps de\-term\-ine the minimal dyn\-am\-ics for the network, as in (\ref{eqn:HubE1}).   The dyn\-am\-ical space is not an ether model, as the em\-bedd\-ing space does not exist.}}
\label{fig:Embedd}
\end{figure}

At a deeper level an information-theoretic approach to modelling reality, {\it Process Physics} \cite{Book},  leads to an emergent structured quantum foam `space'  which is 3-dimensional and dynamic, but where the 3-dimensionality is only approximate, in that if we ignore non-trivial topological aspects of the space, then it may be embedded in a 3-dimensional  geometrical manifold.  Here the space is a real existent discrete but fractal network of relationships or connectivities,  but the embedding space is purely a mathematical way of characterising the gross 3-dimensionality of the network.  This is illustrated in Fig.1. Embedding the network in the embedding space is very arbitrary; we could equally well rotate the embedding or use an embedding that has the network translated or translating.  These general requirements  then dictate the minimal dynamics for the actual network, at a phenomenological level.  To see this we assume  at a coarse grained level that the dynamical patterns within the network may be described by a velocity field ${\bf v}({\bf r},t)$, where ${\bf r}$ is the location of a small region in the network according to some arbitrary embedding.  The 3-space velocity field has been observed in at least 8 experiments 
\cite{MM,Miller,AMGE,MMCK,MMC,C5,C6,C7,Torr,DeWitte,Coax}. 
For simplicity we assume here that the global topology of the network   is not significant for the local dynamics, and so we embed in an $E^3$, although a generalisation to an embedding in $S^3$ is straightforward and might be relevant to cosmology.  The minimal dynamics is then obtained by  writing down the lowest-order zero-rank tensors, of dimension $1/t^2 $, that are invariant under translation and rotation, giving
\begin{equation}
\nabla.\left(\frac{\partial {\bf v} }{\partial t}+({\bf v}.{\bf \nabla}){\bf v}\right)
+\frac{\alpha}{8}(tr D)^2 +\frac{\beta}{8}tr(D^2)=
-4\pi G\rho; \mbox{\   \ }D_{ij}=\frac{1}{2}\left(\frac{\partial v_i}{\partial x_j}+
\frac{\partial v_j}{\partial x_i}\right)
\label{eqn:HubE1}\end{equation}
where $\rho({\bf r},t)$ is the matter and EM energy densities expressed as an effective matter density.  The embedding space coordinates provide a coordinate system or frame of reference that is convenient to describing the velocity field, but which is not real.  
In Process Physics  quantum matter  are topological defects in the network, but here it is sufficient to give a simple description in terms of an  effective density. 

We see that there are only four possible terms, and so we need at most three possible constants to parametrise the dynamics of space: $G, \alpha$ and $\beta$. $G$ turns out  to be Newton's gravitational constant, and describes the rate of non-conservative flow of space into matter.  To determine the values of $\alpha$ and $\beta$ we must, at this stage, turn to experimental data.  
However most experimental data involving the dynamics of space is observed by detecting the so-called gravitational  acceleration of matter, although increasingly light bending is giving new information.  Now the acceleration ${\bf a}$ of the dynamical patterns in space is given by the Euler  convective expression
\begin{equation}
{\bf a}({\bf r},t)= \lim_{\Delta t \rightarrow 0}\frac{{\bf v}({\bf r}+{\bf v}({\bf r},t)\Delta t,t+\Delta
t)-{\bf v}({\bf r},t)}{\Delta t}
=\frac{\partial {\bf v}}{\partial t}+({\bf v}.\nabla ){\bf v}
\label{eqn:HubE3}\end{equation} 
and this appears in one of the terms in (\ref{eqn:HubE1}). As shown in \cite{Schrod} and discussed later herein the acceleration  ${\bf g}$ of quantum matter is identical to this acceleration, apart from vorticity and relativistic effects, and so the gravitational acceleration of matter is also given by (\ref{eqn:HubE3}).

Outside of a spherically symmetric distribution of matter,  of total mass $M$, we find that one solution of (\ref{eqn:HubE1}) is the velocity in-flow field  given by
\begin{equation}
{\bf v}({\bf r})=-\hat{{\bf r}}\sqrt{\frac{2GM(1+\frac{\alpha}{2}+..)}{r}}
\label{eqn:HubE4}\end{equation}
but only when $\beta=-\alpha$,  for only then is the acceleration of matter, from (\ref{eqn:HubE3}), induced by this in-flow of the form
\begin{equation}
{\bf g}({\bf r})=-\hat{{\bf r}}\frac{GM(1+\frac{\alpha}{2}+..)}{r^2}
\label{eqn:HubE5}\end{equation}
 which  is Newton's Inverse Square Law of 1687 \cite{Newton}, but with an effective  mass $M(1+\frac{\alpha}{2}+..)$ that is different from the actual mass $M$.  So the success of Newton's law in the solar system informs us that  $\beta=-\alpha$ in (\ref{eqn:HubE1}). But we also see modifications coming from the 
$\alpha$-dependent terms.

In general because (\ref{eqn:HubE1}) is a scalar equation it is only applicable for vorticity-free flows $\nabla\times{\bf v}={\bf 0}$, for then we can write ${\bf v}=\nabla u$, and then (\ref{eqn:HubE1}) can always be solved to determine the time evolution of  $u({\bf r},t)$ given an initial form at some time  $t_0$.
The $\alpha$-dependent term in (\ref{eqn:HubE1})  (with now $\beta=-\alpha$) and the matter acceleration effect, now also given by (\ref{eqn:HubE3}),   permits   (\ref{eqn:HubE1})   to be written in the form
\begin{equation}
\nabla.{\bf g}=-4\pi G\rho-4\pi G \rho_{DM},
\label{eqn:HubE7}\end{equation}
where
\begin{equation}
\rho_{DM}({\bf r},t)\equiv\frac{\alpha}{32\pi G}( (tr D)^2-tr(D^2)),  
\label{eqn:HubE7b}\end{equation}
which  is an effective matter density that would be required to mimic the
 $\alpha$-dependent spatial self-interaction dynamics.
 Then (\ref{eqn:HubE7}) is the differential form for Newton's law of gravity but with an additional non-matter effective matter density.  So we label this as $\rho_{DM}$ even though no matter is involved \cite{alpha,DM}. This effect has been shown to explain the so-called `dark matter' effect in spiral galaxies, bore hole $g$ anomalies, and the systematics of galactic black hole masses.  
 
 The spatial dynamics  is non-local.  Historically this was first noticed by Newton who called it action-at-a-distance. To see this we can write  (\ref{eqn:HubE1}) as an integro-differential equation
 \begin{equation}
 \frac{\partial {\bf v}}{\partial t}=-\nabla\left(\frac{{\bf v}^2}{2}\right)+G\!\!\int d^3r^\prime
 \frac{\rho_{DM}({\bf r}^\prime, t)+\rho({\bf r}^\prime, t)}{|{\bf r}-{\bf r^\prime}|^3}({\bf r}-{\bf r^\prime})
 \label{eqn:HubE8}\end{equation}
This shows a high degree of non-locality and non-linearity, and in particular that the behaviour of both $\rho_{DM}$ and $\rho$ manifest at a distance irrespective of the dynamics of the intervening space. This non-local behaviour is analogous to that in quantum systems and may offer a resolution to the horizon problem.


 However  (\ref{eqn:HubE1}) needs to be further generalised \cite{Book} to include
vorticity, and also the effect of the motion of matter through this substratum via  
\vspace{-2mm}
\begin{equation}
{\bf v}_R({\bf r}_0(t),t) ={\bf v}_0(t) - {\bf v}({\bf r}_0(t),t),
\label{eqn:VortCG$}
\end{equation}
where ${\bf v}_0(t)$ is the velocity of an object, at ${\bf r}_0(t)$, relative to the same frame of reference that defines
the flow field; then ${\bf v}_R$ is the velocity of that matter relative to the substratum. One possible generalisation of the flow equation
(\ref{eqn:HubE1}) is, with $d/dt=\partial/\partial t +{\bf v}.\nabla$  the Euler fluid or total
derivative,
\begin{eqnarray}
&&\frac{d D_{ij}}{dt}+ \frac{\delta_{ij}}{3}tr(D^2) + \frac{tr D}{2}
(D_{ij}-\frac{\delta_{ij}}{3}tr D)+\frac{\delta_{ij}}{3}\frac{\alpha}{8}((tr
D)^2 -tr(D^2))\nonumber \\ && +(\Omega D-D\Omega)_{ij}=-4\pi
G\rho(\frac{\delta_{ij}}{3}+\frac{v^i_{R}v^j_{R}}{2c^2}+..),\mbox{ } i,j=1,2,3. 
\label{eqn:VortCG4a}\end{eqnarray}
\begin{equation}\nabla \times(\nabla\times {\bf v}) =\frac{8\pi G\rho}{c^2}{\bf v}_R,
\label{eqn:VortCG4b}\end{equation}
\begin{equation}
\Omega_{ij}=\frac{1}{2}(\frac{\partial v_i}{\partial x_j}-\frac{\partial v_j}{\partial
x_i})=-\frac{1}{2}\epsilon_{ijk}\omega_k=-\frac{1}{2}\epsilon_{ijk}(\nabla\times {\bf v})_k,
\label{eqn:VortBS}\end{equation}
and the vorticity vector field is $\vec{\omega}=\nabla\times {\bf v}$. For zero vorticity and
$v_R\ll c$ (\ref{eqn:VortCG4a}) reduces to (\ref{eqn:HubE1}).   We obtain from
(\ref{eqn:VortCG4b}) the Biot-Savart form for the vorticity 
\begin{equation}
\vec{\omega}({\bf r},t)
=\frac{2G}{c^2}\int d^3 r^\prime \frac{\rho({\bf r}^\prime,t)}
{|{\bf r}-{\bf r}^\prime|^3}{\bf v}_R({\bf r}^\prime,t)\times({\bf r}-{\bf r}^\prime).
\label{eqn:Vortomega}\end{equation} 
Eqn.(\ref{eqn:Vortomega}) has been applied to the precession of gyroscopes in the GP-B satellite experiment, see Sect.\ref{subsect:lense}.

\section[Generalised Schr\"{o}dinger Equation and Emergent Gravity]{Generalised Schr\"{o}dinger Equation and Emergent Gravity}\label{sect:schrodinger}

Let us consider what might be regarded as the conventional `Newtonian' approach to including gravity in the
Schr\"{o}dinger equation \cite{Schrod}.  There gravity is described by the Newtonian potential energy field
$\Phi({\bf r},t)$, such that ${\bf g}=-\nabla \Phi$, and we have for a `free-falling' quantum system, with mass
$m$,  
\begin{equation}
i\hbar\frac{\partial \psi({\bf r},t)}{\partial t}=-\frac{\hbar^2}{2m}\nabla^2\psi({\bf r},t)+
m\Phi({\bf r},t)\psi({\bf r},t)\equiv H(t)\Psi, 
\label{eqn:Schrodequiv1}\end{equation}
where the hamiltonian is in general now time dependent. The classical-limit trajectory  is obtained via the usual Ehrenfest method \cite{Ehrenfest}:
we first compute the time rate of change of the  so-called position `expectation value'
\begin{equation}
 \frac{d\!\!<\!\!{\bf r}\!\!> }{dt} \equiv \frac{d }{dt}(\psi,{\bf
r}\psi)=\frac{i}{\hbar}(H\psi,{\bf r}\psi)-\frac{i}{\hbar}(\psi,{\bf r}H\psi)
=\frac{i}{\hbar}(\psi,[H,{\bf r}]\psi),
\label{eqn:Schrodequiv2}\end{equation}
which is valid for a normalised state $\psi$.  The norm is time invariant when $H$ is hermitian
($H^\dagger=H$) even if $H$  itself is time dependent,
\begin{equation}
\frac{d}{dt}(\psi,\psi)=\frac{i}{\hbar}(H\psi,\psi)-\frac{i}{\hbar}(\psi,H\psi)=
\frac{i}{\hbar}(\psi,H^\dagger\psi)-\frac{i}{\hbar}(\psi,H\psi)=0.
\label{eqn:Schrodequiv3}\end{equation}
Next we compute the matter `acceleration' from (\ref{eqn:Schrodequiv2}).
\begin{eqnarray}
\frac{d^2\!\!<\!\!{\bf r}\!\!> }{dt^2} &=& \frac{i}{\hbar}\frac{d }{dt}(\psi,[H,{\bf r}]\psi),\nonumber\\ 
&=&\left(\frac{i}{\hbar}\right)^2(\psi,[H,[H,{\bf r}]]\psi)+\frac{i}{\hbar}(\psi,[\frac{\partial H(t)
}{\partial t},{\bf r}]\psi),\nonumber\\
&=&-(\psi,\nabla \Phi\psi)=(\psi,{\bf g}({\bf r},t)\psi)=<\!{\bf g}({\bf r},t)\!>.
\label{eqn:Schrodequiv4}\end{eqnarray}
 In the classical limit
$\psi$ has the form of a wavepacket where the spatial extent of $\psi$ is much smaller than the spatial region over
which  ${\bf g}({\bf r},t)$ varies appreciably.  Then we have the approximation $<\!{\bf g}({\bf r},t)\!>\approx {\bf
g}(<\!{\bf r}\!>,t)$, and finally we arrive at the Newtonian 2nd-law equation of motion for the wavepacket,
 \begin{equation}
\frac{d^2\!\!<\!\!{\bf r}\!\!> }{dt^2}\approx {\bf g}(<\!{\bf r}\!>, t).
\label{eqn:Schrodequiv5}\end{equation}
 In this classical
limit we obtain the equivalence principle, namely that the acceleration is independent of the mass $m$ and of
the velocity of that mass. But of course that followed by construction, as the equivalence principle is built
into (\ref{eqn:Schrodequiv1}) by having
$m$ as the coefficient of $\Phi$. In Newtonian gravity there is no explanation for the origin of $\Phi$ or
${\bf g}$. In the new theory gravity is explained in terms of a velocity field, which in turn has a deeper
explanation within {\it Process Physics}.

The key insight is that conventional physics has neglected the interaction of various systems with the dynamical
3-space.  Here we  generalise the Schr\"{o}dinger equation to take account of this new physics. Now gravity is a
dynamical effect arising from the time-dependence and spatial inhomogeneities of the 3-space  velocity field ${\bf
v}({\bf r},t)$, and  for a `free-falling' quantum system with mass
$m$ the Schr\"{o}dinger equation now has the generalised form  
\begin{equation}
i\hbar\left(\frac{\partial}{\partial t} +{\bf
v}.\nabla+\frac{1}{2}\nabla.{\bf v}\right) \psi({\bf r},t)=-\frac{\hbar^2}{2m}\nabla^2\psi({\bf r},t), 
\label{eqn:Schrodequiv6}\end{equation}
which we write as 
\begin{equation}
i\hbar\frac{\partial  \psi({\bf r},t)}{\partial t}=H(t)\psi({\bf r},t),
\label{eqn:Schrodequiv7}  \mbox{\ \ where \ \ }
H(t)=-i\hbar\left({\bf
v}.\nabla+\frac{1}{2}\nabla.{\bf v}\right)-\frac{\hbar^2}{2m}\nabla^2
\label{eqn:Schrodequiv8}\end{equation}
This form for $H$   specifies how the quantum system must couple to the velocity field, and it  uniquely 
follows from two considerations: (i) the generalised Schr\"{o}dinger equation must remain form invariant under a
change of observer, i.e. with $t
\rightarrow t$, and ${\bf r}
\rightarrow {\bf r}+{\bf V}t$, where ${\bf V}$ is the relative velocity of the two observers. Then we compute
that
$\displaystyle{\frac{\partial}{\partial t} +{\bf v}.\nabla +\frac{1}{2}\nabla.{\bf v} \rightarrow} $ $
\displaystyle{\frac{\partial}{\partial t} +{\bf v}.\nabla}+\frac{1}{2}\nabla.{\bf v}$, i.e. that it is an
invariant operator, and   (ii) require that
$H(t)$ be hermitian, so that the wavefunction norm is an invariant of the time evolution. This implies that the
$\frac{1}{2}\nabla.{\bf v}$ term must be included, as ${\bf v}.\nabla$ by itself is not hermitian for an
inhomogeneous ${\bf v}({\bf r},t)$. Then the consequences for the motion of wavepackets are uniquely determined;
they are fixed by these two quantum-theoretic requirements.

Then  again the classical-limit trajectory  is obtained via the position `expectation value', first with
\begin{eqnarray}
{\bf v}_O\equiv\frac{d\!\!<\!\!{\bf r}\!\!> }{dt} &=& \frac{d }{dt}(\psi,{\bf
r}\psi)=\frac{i}{\hbar}(\psi,[H,{\bf r}]\psi) =(\psi,({\bf v}({\bf r},
t)-\frac{i\hbar}{m}\nabla)\psi)\nonumber \\ &=&<\!\!{\bf v}({\bf r}, t)\!\!>-\frac{i\hbar}{m}<\!\!\nabla\!\!>,
\label{eqn:Schrodequiv9}\end{eqnarray}
on evaluating the commutator using  $H(t)$ in (\ref{eqn:Schrodequiv8}), and which is again valid for a normalised state
$\psi$. Then for the `acceleration' we obtain from (\ref{eqn:Schrodequiv9})  that\footnote{Care is needed to identify the range of the various $\nabla$'s.}
\begin{eqnarray}
\lefteqn{\frac{d^2\!\!<\!\!{\bf r}\!\!> }{dt^2} = \frac{d }{dt}(\psi,({\bf v}
-\frac{i\hbar}{m}\nabla)\psi)}\nonumber\\ 
& & =(\psi,\left(\frac{\partial {\bf v}({\bf r},t) }{\partial
t}+\frac{i}{\hbar}[H,({\bf v} -\frac{i\hbar}{m}\nabla)]\right)\psi),\nonumber\\
& &=(\psi,\frac{\partial {\bf v}({\bf r},t) }{\partial t}\psi)+
(\psi,
\left({\bf
v}.\nabla+\frac{1}{2}\nabla.{\bf v}-\frac{i\hbar}{2m}\nabla^2\right)\left({\bf v}
-\frac{i\hbar}{m}\nabla\right)\psi)-\nonumber\\
& &\mbox{\ \ \ }(\psi,\left.\left({\bf v}
-\frac{i\hbar}{m}\nabla\right)\left({\bf v}.\nabla+\frac{1}{2}\nabla.{\bf
v}-\frac{i\hbar}{2m}\nabla^2\right)\right)\psi), \nonumber\\ 
& & =(\psi,\left(\frac{\partial {\bf
v}({\bf r},t) }{\partial t}+(({\bf v}.\nabla){\bf v}) -\frac{i\hbar}{m}(\nabla\times{\bf
v})\times {\bf \nabla}\right)\psi)+\nonumber \\
&&+(\psi,\frac{i\hbar}{2m}(\nabla\times(\nabla\times {\bf v}))\psi),\nonumber \\
& &\approx\frac{\partial{\bf v}}{\partial t}+({\bf v}.\nabla){\bf v}+(\nabla\times{\bf
v})\times\left(\frac{d\!\!<\!\!{\bf r}\!\!> }{dt}-{\bf v}\right)+\frac{i\hbar}{2m}(\nabla\times(\nabla\times{\bf
v})),\nonumber \\ & &=\frac{\partial{\bf v}}{\partial t}+({\bf v}.\nabla){\bf v}+(\nabla\times{\bf
v})\times\left(\frac{d\!\!<\!\!{\bf r}\!\!> }{dt}-{\bf v}\right)\nonumber \\
& &=\frac{\partial{\bf v}}{\partial t}+({\bf v}.\nabla){\bf v}+
(\nabla\times{\bf v})\times{\bf v}_R
\label{eqn:Schrodequiv10}\end{eqnarray}
where in arriving at the 3rd last line we have invoked the small-wavepacket approximation, and 
also used (\ref{eqn:Schrodequiv9}) to identify 
\begin{equation} {\bf v}_R \equiv -\frac{i\hbar}{m}<\!\!\nabla\!\!>={\bf v}_O-{\bf v},
\label{eqn:Schrodequiv11}\end{equation}
where ${\bf v}_O$  is the velocity of the wavepacket or object `O' relative to the observer, so then ${\bf v}_R$
is the velocity of the wavepacket relative to the local 3-space. 
Then  all
velocity field terms are now evaluated at the location of the wavepacket. 
Note that the operator
\begin{equation}
-\frac{i\hbar}{m}(\nabla\times{\bf v})\times\nabla+\frac{i\hbar}{2m}(\nabla\times(\nabla\times{\bf v}))
\end{equation} is hermitian, but that separately neither of these two operators is hermitian. Then in general
the scalar product in  (\ref{eqn:Schrodequiv10}) is real. But then in arriving at the last line in (\ref{eqn:Schrodequiv10}) 
by means of the small-wavepacket approximation, we must then self-consistently use that 
$\nabla\times(\nabla\times{\bf v})={\bf 0}$, otherwise the acceleration acquires a spurious imaginary part.  
This is consistent with (\ref{eqn:VortCG4b}) outside of any  matter which contributes to the generation of the
velocity field, for  there
$\rho=0$. These observations point to a deep connection between quantum theory and the velocity field dynamics,
as already  argued in \cite{Book}.

We see that the
test `particle' acquires the acceleration of the velocity field, as in (\ref{eqn:HubE3}), and as well  an
additional vorticity induced acceleration which is the analogue of the Helmholtz acceleration in fluid mechanics. 
Then $\vec{\omega}/2$ is the instantaneous angular velocity of the local 3-space, relative to a distant
observer. Hence we find that the equivalence principle arises from the unique generalised Schr\"{o}dinger
equation and with the additional vorticity effect.  This vorticity effect depends on the absolute velocity ${\bf
v}_R$ of the object relative to the local space, and so requires a change in the Galilean or Newtonian form of
the equivalence principle.

The vorticity acceleration effect is the origin of the
Lense-Thirring so-called `frame-dragging' \footnote{In the spacetime formalism it is mistakenly argued that it
is `spacetime' that is `dragged'.} effect
\cite{LT} discussed later. While the generation of the vorticity is a relativistic effect,
as in (\ref{eqn:Vortomega}), the response of the  test particle to that vorticity is a non-relativistic effect, and
follows from the generalised Schr\"{o}dinger equation, and which is not present in the standard Schr\"{o}dinger
equation with coupling to the Newtonian gravitational potential, as in (\ref{eqn:Schrodequiv1}). Hence the generalised 
Schr\"{o}dinger equation with the new coupling to the velocity field is  more fundamental. The Helmholtz term in
(\ref{eqn:Schrodequiv10}) is being explored by the Gravity Probe B gyroscope precession experiment, however the
vorticity caused by the motion of the earth is extremely small, as discussed later in Sect.\ref{subsect:lense}. 

An important insight emerges from the above: the
generalised Schr\"{o}dinger equation involves two fields  ${\bf v}({\bf r},t)$ and  
$\psi({\bf r},t)$, where the coordinate ${\bf r}$ is merely a label to relate the two fields, and is
not itself the 3-space.  In particular while ${\bf r}$ relates to the embedding space,
the 3-space itself has time-dependence and inhomogeneities, and as well in the more general case will
exhibit vorticity $\vec{\omega}=\nabla\times{\bf v}$.  Only in the unphysical case does the description
of the 3-space become identified with the coordinate system ${\bf r}$, and that is when the velocity
field  ${\bf v}({\bf r},t)$ becomes uniform and time independent. Then by a suitable choice of
observer we may put  ${\bf v}({\bf r},t)={\bf 0}$, and the generalised Schr\"{o}dinger equation 
reduces to the usual `free' Schr\"{o}dinger equation.    As we discuss
later the experimental evidence is that  ${\bf v}({\bf r},t)$ is fractal and so cannot be removed by
a change to a preferred observer.  Hence the generalised Schr\"{o}dinger equation in
(\ref{eqn:Schrodequiv7}) is a major development for fundamental physics. Of course in general  other
non-3-space potential energy terms may be added to the RHS of (\ref{eqn:Schrodequiv8}). A prediction of this
new quantum theory, which also extends to a generalised Dirac equation, is that the fractal structure of space
implies that even at the scale of atoms etc there will be time-dependencies and inhomogeneities, and that these
will affect  transition rates  of quantum systems.  These effects are probably 
 those known as the Shnoll effects \cite{Shnoll1}.

\section{Generalised Dirac Equation and  Relativistic Gravity}\label{sect:dirac}

An analogous generalisation of the Dirac equation is also necessary giving the coupling of the spinor to the actual dynamical 
3-space, and again not to the embedding space as has been the case up until now, 
\begin{equation}
i\hbar\frac{\partial \psi}{\partial t}=-i\hbar\left(  c{\vec{ \alpha.}}\nabla + {\bf
v}.\nabla+\frac{1}{2}\nabla.{\bf v}  \right)\psi+\beta m c^2\psi
\label{eqn:Hub12}\end{equation}
where $\vec{\alpha}$ and $\beta$ are the usual Dirac matrices. Repeating the  Schr\"{o}dinger equation analysis for the space-induced acceleration we obtain
\begin{equation}\label{eqn:HubE12}
{\bf g}=\displaystyle{\frac{\partial {\bf v}}{\partial t}}+({\bf v}.{\bf \nabla}){\bf
v}+({\bf \nabla}\times{\bf v})\times{\bf v}_R-\frac{{\bf
v}_R}{1-\displaystyle{\frac{{\bf v}_R^2}{c^2}}}
\frac{1}{2}\frac{d}{dt}\left(\frac{{\bf v}_R^2}{c^2}\right)
\end{equation}
which generalises  (\ref{eqn:Schrodequiv10}) by having a term which limits the speed of the wave packet relative to space to be $<\!c$. This equation specifies the trajectory of a spinor wave packet in the dynamical 3-space.  

\section{Generalised Maxwell Equations and Light Lensing}\label{sect:maxwell}

\begin{figure}[t]
\vspace{0mm}\parbox{70mm}{  
\setlength{\unitlength}{2.0mm}
\vspace{0mm}
\hspace{-3mm}\begin{picture}(10,30)
\thicklines

\put(5,25){\line(1,0){12}}   
\put(10,25){\vector(1,0){1}}
\put(23,24){\line(4,-1){12}}
\put(23,24){\vector(4,-1){7}}
\put(23,25){\line(1,0){12}}
\put(27,23.0){ $\delta$}
\qbezier(17,25)(20,24.8)(23,24)  

\put(20,20){\line(1,0){10}}  
\put(20,20){\line(-1,0){10}}  
\put(20,20){\line(0,1){10}}  
\put(20,20){\line(0,-1){10}}  
\put(20,20){\line(1,1){7}}  
\put(20,20){\line(1,-1){7}}  
\put(20,20){\line(-1,1){7}}  
\put(20,20){\line(-1,-1){7}}  

\put(13,20){\vector(1,0){1}}
\put(27,20){\vector(-1,0){1}}
\put(20,27){\vector(0,-1){1}}
\put(20,13){\vector(0,1){1}}
\put(25,25){\vector(-1,-1){1}}
\put(25,15){\vector(-1,1){1}}
\put(15,15){\vector(1,1){1}}
\put(15.5,24.5){\vector(1,-1){1}}

\put(20,20){\circle{9}}  

\end{picture}
}\,
\parbox{60mm}{\vspace{-10mm}\caption{\footnotesize{ Shows bending of light through angle $\delta$ by the inhomogeneous spatial in-flow, according to the minimisation of the travel time in (\ref{eqn:lighttime}). This effect permits the in-flow speed at the surface of the sun to be determined to be 615km/s. The in-flow speed into the sun at the distance of the earth from the sun has been extracted from the Miller data, giving $50\pm10$km/s \cite{Book}.  Both speeds are in agreement with (\ref{eqn:HubE4}). }\label{fig:light}}}
\vspace{-14mm}\end{figure}
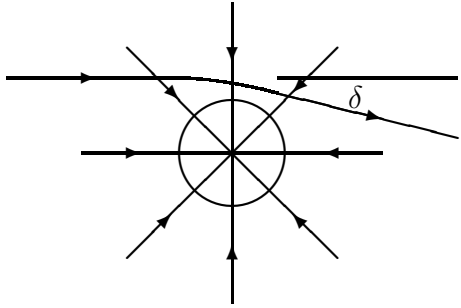

One of the putative key tests of the GR formalism was the gravitational bending of light. This also immediately follows from the new space dynamics once we also generalise the Maxwell equations so that the electric and magnetic  fields are excitations of the dynamical space. The dynamics of the electric and magnetic fields  must then have the form, in empty space,  
\begin{equation}
\displaystyle{ \nabla \times {\bf E}=-\mu\left(\frac{\partial {\bf H}}{\partial t}+{\bf v.\nabla H}\right)},
\displaystyle{ \nabla \times {\bf H}=\epsilon\left(\frac{\partial {\bf E}}{\partial t}+{\bf v.\nabla E}\right)},
\displaystyle{\nabla.{\bf H}={\bf 0}} , 
\displaystyle{\nabla.{\bf E}={\bf 0}}
\label{eqn:MaxE18}\end{equation}
which was first suggested by Hertz in 1890  \cite{Hertz}, but with $\bf v$ being a constant vector field. 
Suppose we have a uniform flow of space with velocity ${\bf v}$ wrt the embedding space or wrt an observer's frame of reference. Then we can find plane wave solutions for (\ref{eqn:MaxE18}):
\begin{equation}
{\bf E}({\bf r},t)={\bf E}_0e^{i({\bf k}.{\bf r}-\omega t)} \mbox{\ \ \ \  } {\bf H}({\bf r},t)={\bf H}_0e^{i({\bf k}.{\bf r}-\omega t)}
\label{eqn:pw}\end{equation}
with
\begin{equation}
\omega({\bf k},{\bf v})=c|\vec{{\bf k}}| +{\bf v}.{\bf k} \mbox{ \ \ \  where \ \ \  } c=1/\sqrt{\mu\epsilon}
\label{eqn:omega}\end{equation}
Then the EM group velocity is
\begin{equation}
{\bf v}_{EM}=\vec{\nabla}_k\omega({\bf k},{\bf v})=c\hat{\bf k}+{\bf v}
\label{eqn:groupv}\end{equation}
So the velocity of EM radiation ${\bf v}_{EM}$ has magnitude  $c$ only with respect to the space, and in general not with respect to the observer if the observer is moving through space, as experiment has indicated again and again, as discussed in Sect.\ref{sect:experimentalI}. These experiments show that the speed of light is in general anisotropic, as predicted by (\ref{eqn:groupv}).
The time-dependent and inhomogeneous  velocity field causes the refraction of EM radiation. This can be computed by using the Fermat least-time approximation. Then the EM ray paths  ${\bf r}(t)$ are determined by minimising  the elapsed travel time:
\begin{equation}
\tau=\int_{s_i}^{s_f}\frac{ds\displaystyle{|\frac{d{\bf r}}{ds}|}}{|c\hat{{\bf v}}_R(s)+{\bf v}(\bf{r}(s),t(s)|}
\mbox{ \ \ with \ \ }
{\bf v}_R=\left(  \frac{d{\bf r}}{dt}-{\bf v}(\bf{r}(t),t)\right)
\label{eqn:lighttime}\end{equation}
by varying both ${\bf r}(s)$ and $t(s)$, finally giving ${\bf r}(t)$. Here $s$ is a path parameter, and ${\bf v}_R$ is a 3-space tangent vector for the path.
As an example, the  in-flow in (\ref{eqn:HubE4}), which is applicable to light bending by the sun, gives  the angle of deflection 
\begin{equation}
\delta=2\frac{v^2}{c^2}=\frac{4GM(1+\frac{\alpha}{2}+..)}{c^2d}+...
\label{eqn:E19}\end{equation}
where $v$ is the in-flow speed at distance $d$  and $d$ is the impact parameter. This agrees with the GR result except for the $\alpha$ correction.  Hence the  observed deflection of $8.4\times10^{-6}$ radians is actually a measure of the in-flow speed at the sun's surface, and that gives $v=615$km/s.   These generalised Maxwell equations also predict gravitational lensing produced by the large in-flows associated with the new `black holes' in galaxies, see \cite{Ring}.  So again this effect permits the direct observation of the these  black hole effects with their non inverse-square-law accelerations.

\section{Free-Fall Minimum Proper-Time Trajectories\label{sect:free}}

The acceleration in (\ref{eqn:HubE12}) also arises from the following  argument, which is the
analogue of the Fermat least-time formalism for the quantum matter waves.  Consider the elapsed time for a comoving clock. Then taking account of the Lamour time-dilation effect that
time is given by 
\begin{equation}
\tau[{\bf r}_0]=\int dt \left(1-\frac{{\bf v}_R^2}{c^2}\right)^{1/2}
\label{eqn:SchrodG1}
\end{equation}  
with ${\bf v}_R$ given by (\ref{eqn:Schrodequiv11}) in terms of ${\bf v}_O$ and ${\bf v}$. Then this time effect
relates to the speed of the clock relative to the local 3-space, and that $c$ is the speed of light relative to
that local 3-space.  Under a deformation of the trajectory 
\begin{equation}{\bf r}_0(t)
\rightarrow  {\bf r}_0(t) +\delta{\bf r}_0(t),
\mbox{\ \ }{\bf v}_0(t) \rightarrow  {\bf v}_0(t) +\displaystyle\frac{d\delta{\bf r}_0(t)}{dt},\end{equation}
\begin{equation}\label{eqn:SchrodG2}
{\bf v}({\bf r}_0(t)+\delta{\bf r}_0(t),t) ={\bf v}({\bf r}_0(t),t)+(\delta{\bf
r}_0(t).{\bf \nabla}) {\bf v}({\bf r}_0(t),t)+... 
\end{equation}
Evaluating the change in proper travel time to lowest order
\begin{eqnarray*}\label{eqn:SchrodG3}
\delta\tau&=&\tau[{\bf r}_0+\delta{\bf r}_0]-\tau[{\bf r}_0] +... \nonumber\\
&=&-\int dt \:\frac{1}{c^2}{\bf v}_R. \delta{\bf v}_R\left(1-\displaystyle{\frac{{\bf
v}_R^2}{c^2}}\right)^{-1/2}+...\nonumber\\
&=&\int dt\frac{1}{c^2}\displaystyle{\frac{{\bf
v}_R.(\delta{\bf r}_0.{\bf \nabla}){\bf v}-{\bf v}_R.\displaystyle{\frac{d(\delta{\bf
r}_0)}{dt}}}{\sqrt{1-\displaystyle{\frac{{\bf v}_R^2}{c^2}}}}}+...\nonumber\\ 
&=&\int dt \frac{1}{c^2}\left(\frac{{\bf v}_R.(\delta{\bf r}_0.{\bf \nabla}){\bf v}}{ 
\sqrt{1-\displaystyle{\frac{{\bf
v}_R^2}{c^2}}}}  +\delta{\bf r}_0.\frac{d}{dt} 
\frac{{\bf v}_R}{\sqrt{1-\displaystyle{\frac{{\bf
v}_R^2}{c^2}}}}\right)+...\nonumber\\
&=&\int dt\: \frac{1}{c^2}\delta{\bf r}_0\:.\left(\frac{({\bf v}_R.{\bf \nabla}){\bf v}+{\bf v}_R\times({\bf
\nabla}\times{\bf v})}{ 
\sqrt{1-\displaystyle{\frac{{\bf v}_R^2}{c^2}}}}+\frac{d}{dt} 
\frac{{\bf v}_R}{\sqrt{1-\displaystyle{\frac{{\bf
v}_R^2}{c^2}}}}\right)+...\nonumber \\
\end{eqnarray*}
  Hence a 
trajectory ${\bf r}_0(t)$ determined by $\delta \tau=0$ to $O(\delta{\bf r}_0(t)^2)$
satisfies 
\begin{equation}\label{eqn:SchrodG4}
\frac{d}{dt} 
\frac{{\bf v}_R}{\sqrt{1-\displaystyle{\frac{{\bf v}_R^2}{c^2}}}}=-\frac{({\bf
v}_R.{\bf \nabla}){\bf v}+{\bf v}_R\times({\bf
\nabla}\times{\bf v})}{ 
\sqrt{1-\displaystyle{\frac{{\bf v}_R^2}{c^2}}}}.
\end{equation}
   Substituting ${\bf
v}_R(t)={\bf v}_0(t)-{\bf v}({\bf r}_0(t),t)$ and using 
\begin{equation}\label{eqn:SchrodG5}
\frac{d{\bf v}({\bf r}_0(t),t)}{dt}=\frac{\partial {\bf v}}{\partial t}+({\bf v}_0.{\bf \nabla}){\bf
v},
\end{equation}
we obtain
\begin{equation}\label{eqn:SchrodCG6}
 \frac{d {\bf v}_0}{dt}=\displaystyle{\frac{\partial {\bf v}}{\partial t}}+({\bf v}.{\bf \nabla}){\bf
v}+({\bf \nabla}\times{\bf v})\times{\bf v}_R-\frac{{\bf
v}_R}{1-\displaystyle{\frac{{\bf v}_R^2}{c^2}}}
\frac{1}{2}\frac{d}{dt}\left(\frac{{\bf v}_R^2}{c^2}\right).
\end{equation}
which is (\ref{eqn:HubE12}).
Then in the low speed limit  $v_R \ll c $   we  may neglect the last term, and we obtain
(\ref{eqn:Schrodequiv10}). Hence we see a close relationship between the geodesic equation, known first from General
Relativity, and the 3-space generalisation of the Schr\"{o}dinger equation, at least in the non-relativistic limit.
So in the classical limit, i.e when the   wavepacket approximation is valid, the wavepacket trajectory
 is specified by the least proper-time geodesic.    

The relativistic term in (\ref{eqn:SchrodCG6}) is responsible for the precession of
elliptical orbits and also for the event horizon effect. Hence the trajectory  in (\ref{eqn:Schrodequiv10})
is a non-relativistic minimum travel-time trajectory, which is Fermat's Principle.

\section{Deriving the Special Relativity Formalism}\label{sect:special}

The detection of absolute motion is not incompatible with  Lorentz symmetry; the contrary belief was postulated by Einstein, and has persisted for over 100 years, since 1905. So far the experimental evidence is that absolute motion and Lorentz symmetry are real and valid phenomena; absolute motion is motion relative to some substructure to space, whereas Lorentz symmetry parametrises dynamical effects caused by the motion of systems through that sub\-struc\-ture. 
Motion through the structured space, it is argued, induces actual dynamical time dilations and length
contractions in agree\-ment with the Lorentz interpretation of special relati\-vist\-ic
effects.  Then observers in  uniform motion `through' the space will, on measurement  of the speed of light  using the special but misleading Einstein measurement protocol, obtain
always the same numerical value  $c$.   To see this expli\-citly consider how various observers $P, P^\prime,\dots$
moving with  different  speeds through space, measure the speed of light.  They  each acquire a standard rod 
and an accompanying stan\-d\-ardised clock. That means that these standard  rods  would agree if they were brought
together, and at rest with respect to space they would all have length $\Delta l_0$, and similarly for
the clocks.    Observer $P$ and accompanying rod are both moving at  speed $v_R$ relative to space, with
the rod longitudinal to that motion. P  then  measures the time
$\Delta t_R$, with the clock at end $A$ of the rod,  for a light pulse to travel from  end $A$ to the other end
$B$  and back again to $A$. The  light  travels at speed $c$ relative to space. Let the time taken for
the light pulse to travel from
$A\,{\rightarrow}\, B\,$ be $t_{AB}$ and  from $B\,{\rightarrow}\, A$ be $t_{BA}$, as measured by a clock at rest with respect
to space\footnote{Not all clocks will behave in this same ``ideal'' manner.}$\!$. The  length of the rod 
moving at speed
$v_R$ is contracted to 
\vspace*{-3pt}
\begin{equation}
\Delta l_R=\Delta l_0\,\sqrt{1-\frac{v_R^2}{c^2}}\,.
\label{eqnum:c0}\end{equation}
In moving from  $A$ to $B$ the light must travel an extra  distance 
because the  end  $B$ travels a distance $v_Rt_{AB}$ in this time, thus the total distance that must be
traversed  is
\vspace*{-2pt}
\begin{equation}\label{eqnum:MM}
ct_{AB}=\Delta l_R+v_R\,t_{AB}\,,
\end{equation}
similarly on returning from $B$ to $A$ the light must travel the distance
\vspace*{-2pt}
\begin{equation}\label{eqnum:MMCK}
ct_{BA}=\Delta l_R-v_R\,t_{BA}\,.
\end{equation}
Hence the total travel time $\Delta t_0$ is
\begin{equation}\label{eqnum:c3}
\Delta t_0=t_{AB}+t_{BA}=\frac{\Delta l_R}{c-v_R}+\frac{\Delta l_R}{c+v_R}=\\[+4pt]
=\frac{2\Delta l_0}{c\,\sqrt{1-\displaystyle\frac{v_R^2}{c^2}}}\,.
\end{equation}
\vspace*{-4pt}
Because  of  the time dilation effect for the moving clock
\vspace*{-3pt}
\begin{equation}
\Delta t_R=\Delta t_0\,\sqrt{1-\displaystyle\frac{v_R^2}{c^2}}\,.
\label{eqnum:c4}\end{equation}
Then for the moving observer the speed of light is de\-fin\-ed as the distance the observer believes the light
travelled ($2\Delta l_0$) divided by the travel time according to the accom\-pa\-nying clock ($\Delta t_R$), namely 
$2\Delta l_0/\Delta t_R = c$, from above, which is thus the same speed as seen by an observer at rest in the space, namely $c$.  So the speed $v_R$ of the ob\-server through space is not revealed by this
pro\-cedure, and the observer is erroneously led to the conclusion that the speed of light is always $c$. 
This follows from two or more observers in manifest relative motion all obtaining the same speed c by this
procedure. Despite this failure  this special effect is actually the basis of the spacetime\index{spacetime}
Einstein mea\-su\-re\-ment protocol. That this protocol is blind to the ab\-sol\-ute motion has led to enormous confusion within physics.

To be explicit the Einstein measurement protocol\index{measurement protocol} actual\-ly inadvertently uses this
special effect by using the radar method for assigning historical spacetime coordinates to an event: the observer
records the time of emission and recep\-tion of radar pulses ($t_r \,{>}\, t_e$) travelling through  space, and then retrospectively assigns the time and distance of a distant event
$B$ according to (ignoring directional information for simplicity) 
\vspace*{-6pt}
\begin{equation}T_B=\frac{1}{2}\,\bigl(t_r+t_e\bigr)\,, \qquad
D_B=\frac{c}{2}\,\bigl(t_r-t_e\bigr)\,,\label{eqnum:25}\end{equation}
where each observer is now using the same numerical value of $c$.
 The event $B$ is then plotted as a point in 
an individual  geometrical construct by each  observer,  known as a space\-time record, with coordinates $(D_B,T_B)$. This
is the same as an historian recording events according to  some agreed protocol.  Unlike historians, who
don't confuse history books with reality, physicists do so. 
  We now show that because of this
protocol and the absolute motion dynamical effects, observers will discover on comparing their
historical records of the same events that the expression
\vspace*{-1pt}
\begin{equation}
 \tau_{AB}^2 =   T_{AB}^2- \frac{1}{c^2} \,D_{AB}^2\,,
\label{eqnum:26}\end{equation}
is an invariant, where $T_{AB}=T_A-T_B$ and $D_{AB}=D_A-D_B$ are the differences in times and distances
assigned to events $A$ and
$B$ using the Einstein measurement protocol (\ref{eqnum:25}), so long as both are sufficiently small
compared with the scale of inhomogeneities  in the velocity field. 

\begin{figure}
\vspace{0mm}\parbox{60mm}{{
\hspace{7mm}\begin{picture}(50,145)
\setlength{\unitlength}{1.4mm}
\thicklines
\put(-2,-2){{\bf $A$}}
\put(+1,30){{\bf $P(v_0=0)$}}
\put(16,14){\bf $B$ $(t^\prime_B)$}
\put(25,-3){\bf $D$}
\put(15,-3){\bf $D_B$}
\put(16,-0.5){\line(0,1){0.9}}
\put(-5,15){\bf $T$}
\put(26,24){$ P^\prime(v^\prime_0$)}
\put(0,0){\line(1,0){35}}
\put(0,0){\line(0,1){35}}
\put(0,35){\line(1,0){35}}
\put(35,0){\line(0,1){35}}
\put(0,0){\line(1,1){25}}
\put(0,8){\vector(2,1){16}}
\put(16,16){\vector(-2,1){16}}
\put(-3,8){\bf $t_e$}\put(-0.5,8){\line(1,0){0.9}}
\put(-3,16){\bf $T_B$}\put(-0.5,16){\line(1,0){0.9}}
\put(-3,24){\bf $t_r$}\put(-0.5,24){\line(1,0){0.9}}
\put(6,12){$\gamma$}
\put(6,22){$\gamma$}
\end{picture}}}
\parbox{70mm}{\caption{\footnotesize{  Here $T\,{-}\,D$ is the spacetime construct (from  the Einstein measurement protocol) of a special observer
$P$ {\it at rest} wrt space, so that $v_0\,{=}\,0$.  Observer $P^\prime$ is moving with speed
$v^\prime_0$ as determined by observer $P$, and therefore with speed $v^\prime_R=v^\prime_0$ wrt space. Two light
pulses are shown, each travelling at speed $c$ wrt both $P$ and space.   Event
$A$ is when the observers pass, and is also used to define zero time  for each for
convenience. \label{fig:spacetime1}}}}
\end{figure}
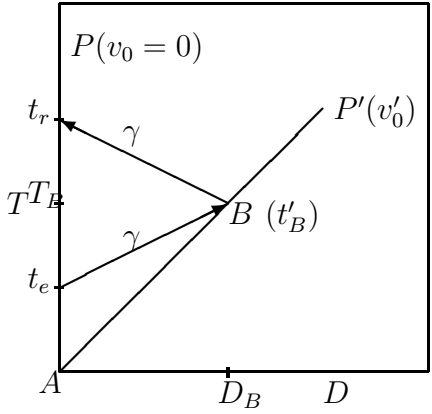

To confirm the invariant  nature of the construct in   (\ref{eqnum:26}) one must pay careful attention to
observational times as distinct from protocol times and distances, and this must be done separately for each
observer.  This can be tedious.  We now  demonstrate this for the situation illustrated in
Fig.~\ref{fig:spacetime1}. 

 By definition  the speed of
$P^\prime$ according to
$P$ is
$v_0^\prime =D_B/T_B$ and so
$v_R^\prime=v^\prime_0$,  where 
$T_B$ and $D_B$ are the protocol time and distance for event $B$ for observer $P$ according to
(\ref{eqnum:25}).  Then using (\ref{eqnum:26})  $P$ would find that
$(\tau^P_{AB})^2=T_{B}^2-\frac{1}{c^2}D_B^2$ since both
$T_A=0$ and $D_A$=0, and whence $(\tau^{P}_{AB})^2=(1-\frac{v_R^{\prime 2}}{c^2})T_B^2=(t^\prime_B)^2$ where
the last equality follows from the time dilation effect on the $P^\prime$ clock, since $t^\prime_B$ is the time
of event
$B$ according to that clock. Then $T_B$ is also the time that $P^\prime$  would compute for event $B$ when
correcting for the time-dilation effect, as the speed $v^\prime_R$ of $P^\prime$ through the quantum foam is
observable by $P^\prime$.  Then $T_B$ is the `common time' for event $B$ assigned by both
observers.  For
$P^\prime$ we obtain  directly, also from  (\ref{eqnum:25}) and (\ref{eqnum:26}), that
$(\tau^{P'}_{AB})^2=(T_B^\prime)^2-\frac{1}{c^2}(D^\prime_B)^2=(t^\prime_B)^2$, as $D^\prime_B=0$  and
$T_B^\prime=t^\prime_B$. Whence for this situation
\begin{equation}
(\tau^{P}_{AB})^2=(\tau^{P'}_{AB})^2,
\label{eqn:SRinvariant1}
\end{equation} and so the
 construction  (\ref{eqnum:26})  is an invariant.

While so far we have only established the invariance of the construct  (\ref{eqnum:26}) when one of the
observers is at rest in space, it follows that for two observers $P^\prime$ and
$P^{\prime\prime}$ both in absolute motion  it follows that they also agree on the invariance
of (\ref{eqnum:26}).  This is easily seen by using the intermediate step of  a stationary observer $P$:
\begin{equation}
(\tau^{P'}_{AB})^2=(\tau^{P}_{AB})^2=(\tau^{P''}_{AB})^2.
\label{eqn:SRinvariant2}
\end{equation}
Hence the protocol and Lorentzian absolute motion ef\-fects result in the construction in (\ref{eqnum:26})  being indeed an
invar\-iant\index{invariant interval} in general.  This  is  a remarkable and subtle result.  For Einstein this
invariance was a fundamental assumption, but here it is a derived result, but one which is nevertheless deeply
misleading. Explicitly indicating  small quantities  by $\Delta$ prefixes, and on comparing records
retrospectively, an ensemble of nearby observers  agree on the invariant
\vspace*{-2pt}
\begin{equation}
\Delta \tau^2=\Delta T^2-\frac{1}{c^2}\,\Delta D^2,
\label{eqnum:31}\end{equation} 
for any two nearby events.  This implies that their individual patches of spacetime records may be mapped one
into the other merely by a change of coordinates, and that collecti\-ve\-ly the spacetime patches  of all may
be represented by one pseudo-Riemannian manifold, where the choice of coord\-i\-na\-tes for this manifold is
arbitrary, and we finally arrive at the invariant 
\vspace*{-2pt}
\begin{equation}
\Delta\tau^2=g_{\mu\nu}(x)\,\Delta x^\mu \Delta x^\nu,
\label{eqnum:inv}\end{equation} 

\vspace*{3pt}\noindent
with $x^\mu=\{D_1,D_2,D_3,T\}$.  Eqn.~(\ref{eqnum:inv})  is invariant under the Lorentz transformations
\begin{equation}
x^{\prime\mu}={L^\mu}_{\!\nu}\, x^\nu,\rule[-8pt]{0pt}{0pt}
\label{eqnum:Lorentz}\end{equation}
where, for example for relative motion in the $x$ direction, ${L^\mu}_{\!\nu}$ is specified by
\vspace*{-4pt}
\begin{equation}
\displaystyle
x^\prime=\frac{x-vt}{\sqrt{1-v^2/c^2}}\,, \mbox{ \ \ }y^\prime=y, \mbox{ \ \ } z^\prime=z, \mbox{ \ \ }t^\prime=\frac{t-vx/c^2}{\sqrt{1-v^2/c^2}}\,
\label{eqnum:xLorentz}\
\end{equation}

So absolute motion and special relativity effects, and even Lorentz symmetry, are all compatible:  a possible pre\-ferr\-ed frame is  hidden by the Einstein measurement protocol. 

The experimental question is then whether or not a supposed preferred frame actually exists  or not --- can it be detected experimentally?  The answer is that there are now eight such consistent experiments.  

The notion that the special relativity formalism requires that the speed of light be isotropic, that it be $c$ in all frames, has persisted for most of the last century. The actual situation is that it only requires that the round trip speed be invariant. This means that the famous Einstein light speed postulate is actually incorrect.  This is discussed in \cite{AIP,light,ep,Levy,Guerra}.

\section{Deriving the General Relativity Formalism}\label{sect:general}

As discussed above the generalised Dirac equation gives rise to a trajectory determined by  (\ref{eqn:HubE12}),
which may be obtained by extremising the time-dilated elapsed time (\ref{eqn:SchrodG1}).
\begin{equation}
\tau[{\bf r}_0]=\int dt \left(1-\frac{{\bf v}_R^2}{c^2}\right)^{1/2}
\label{eqn:GRE13}\end{equation}  
 This happens because of the Fermat least-time effect for quantum matter waves: only along the minimal time trajectory do the quantum waves  remain in phase under small variations of the path. This again emphasises  that gravity is a quantum effect.   We now introduce a spacetime mathematical construct according to the metric
\begin{equation}
ds^2=dt^2 -(d{\bf r}-{\bf v}({\bf r},t)dt)^2/c^2
=g_{\mu\nu}dx^{\mu}dx^\nu
\label{eqn:GRE14}\end{equation}
Then according to this metric the elapsed time in (\ref{eqn:GRE13}) is
\begin{equation}
\tau=\int dt\sqrt{g_{\mu\nu}\frac{dx^{\mu}}{dt}\frac{dx^{\nu}}{dt}},
\label{eqn:GRE14b}\end{equation}
and the minimisation of  (\ref{eqn:GRE14b}) leads to the geodesics of the spacetime, which are thus equivalent to the trajectories from (\ref{eqn:GRE13}), namely  (\ref{eqn:HubE12}).
Hence by coupling the Dirac spinor dynamics to the 3-space dynamics we derive the geodesic formalism of General Relativity as a quantum effect, but without reference to the Hilbert-Einstein equations for the induced metric.  Indeed in general the metric of  this induced spacetime will not satisfy  these equations as the dynamical space involves the $\alpha$-dependent  dynamics, and $\alpha$ is missing from GR.   
So why did GR appear to succeed in a number of key tests where the Schwarzschild metric was used?  The answer is provided by identifying the induced spacetime metric corresponding to the in-flow in (\ref{eqn:HubE4}) outside of a spherical matter system, such as the earth.  Then (\ref{eqn:GRE14})  becomes
 \begin{equation}
ds^2=dt^{ 2}-\frac{1}{c^2}(dr+\sqrt{\frac{2GM(1+\frac{\alpha}{2}+..)}{r}}dt)^2
-\frac{1}{c^2}r^2(d\theta^{ 2}+\sin^2(\theta)d\phi^2),
\label{eqn:GRE15}\end{equation}
 Making the change of variables $t\rightarrow t^\prime$ and
$\bf{r}\rightarrow {\bf r}^\prime= {\bf r}$ with
\begin{equation}
t^\prime=t-
\frac{2}{c}\sqrt{\frac{2 GM(1{+}\frac{\alpha}{2}{+}\dots)r}{c^2}}+ 
\frac{4\ GM(1{+}\frac{\alpha}{2}{+}\dots)}{c^3}\,\mbox{tanh}^{-1}\sqrt{\frac{2 GM(1{+}\frac{\alpha}{2}{+}\dots)}{c^2r}}
\label{eqn:GRE16}\end{equation}
this becomes (and now dropping the prime notation)
\begin{eqnarray}
ds^2&=&\left(1-\frac{2GM(1+\frac{\alpha}{2}+..)}{c^2r}\right)dt^{ 2} 
-\frac{1}{c^2}r^{ 2}(d\theta^2+\sin^2(\theta)d\phi^2) \nonumber \\
&&-\frac{dr^{ 2}}{c^2\left(1-{\displaystyle\frac{
2GM(1+\frac{\alpha}{2}+..)}{ c^2r}}\right)}.
\label{eqn:GRE17}\end{eqnarray}
which is  one form of the the Schwarzschild metric but with the $\alpha$-dynamics induced effective mass shift. Of course this is only valid outside of the spherical matter distribution, as that is the proviso also on (\ref{eqn:HubE4}). As well the above particular change of coordinates also introduces spurious singularities at the event horizon\footnote{The event horizon of  (\ref{eqn:GRE17}) is at a different radius from the actual event horizon of the black hole solutions that arise from (\ref {eqn:HubE1}).},  but other choices do not do this. 
Hence in the case of the Schwarzschild metric the dynamics missing from both the Newtonian theory of gravity and General Relativity is merely hidden in a mass redefinition, and so didn't affect the various standard tests of GR, or even of Newtonian gravity.  Note that as well we see that the Schwarzschild metric is none other than Newtonian gravity in disguise, except for the mass shift.  While we have now explained why the GR formalism appeared to work, it is also clear that this formalism hides the manifest dynamics of the dynamical space, and which has also been directly detected in gas-mode interferometer and coaxial-cable experiments.

Nevertheless we now show \cite{Book} that in the limit $\alpha\rightarrow 0$  the induced metric in (\ref {eqn:GRE14}), with ${\bf v}$ from  (\ref {eqn:HubE1}), satisfies the Hilbert-Einstein equations so long as we use relativistic corrections for the matter density on the RHS of (\ref {eqn:HubE1}). This means that  (\ref {eqn:HubE1}) is consistent with for example the binary pulsar data - the relativistic aspects being associated with the matter effects upon space and the relativistic effects of the matter in motion through the dynamical 3-space.  The agreement of GR with the pulsar data is implying that the $\alpha$-dependent effects are small in this case, unlike in black holes and spiral galaxies.
The GR equations are
\begin{equation}
G_{\mu\nu}\equiv R_{\mu\nu}-\frac{1}{2}Rg_{\mu\nu}=\frac{8\pi G}{c^2} T_{\mu\nu},
\label{eqn:GR32}\end{equation}
where  $G_{\mu\nu}$ is  the Einstein tensor, $T_{\mu\nu}$ is the  energy-momentum tensor,
$R_{\mu\nu}=R^\alpha_{\mu\alpha\nu}$ and
$R=g^{\mu\nu}R_{\mu\nu}$ and
$g^{\mu\nu}$ is the matrix inverse of $g_{\mu\nu}$. The curvature tensor is
\begin{equation}
R^\rho_{\mu\sigma\nu}=\Gamma^\rho_{\mu\nu,\sigma}-\Gamma^\rho_{\mu\sigma,\nu}+
\Gamma^\rho_{\alpha\sigma}\Gamma^\alpha_{\mu\nu}-\Gamma^\rho_{\alpha\nu}\Gamma^\alpha_{\mu\sigma},
\label{eqn:GRcurvature}\end{equation}
where $\Gamma^\alpha_{\mu\sigma}$ is the affine connection
\begin{equation}
\Gamma^\alpha_{\mu\sigma}=\frac{1}{2} g^{\alpha\nu}\left(\frac{\partial g_{\nu\mu}}{\partial x^\sigma}+
\frac{\partial g_{\nu\sigma}}{\partial x^\mu}-\frac{\partial g_{\mu\sigma}}{\partial x^\nu} \right).
\label{eqn:GRaffine}\end{equation}
Let us  substitute the metric in (\ref{eqn:GRE14}) into  (\ref{eqn:GR32}) using (\ref{eqn:GRcurvature}) and (\ref{eqn:GRaffine}).   The various components of the
Einstein tensor are then found to be
\begin{eqnarray}\label{eqn:GRG}
G_{00}&=&\sum_{i,j=1,2,3}v_i\mathcal{G}_{ij}
v_j-c^2\sum_{j=1,2,3}\mathcal{G}_{0j}v_j-c^2\sum_{i=1,2,3}v_i\mathcal{G}_{i0}+c^2\mathcal{G}_{00}, 
\nonumber\\ G_{i0}&=&-\sum_{j=1,2,3}\mathcal{G}_{ij}v_j+c^2\mathcal{G}_{i0},   \mbox{ \ \ \ \ } 
G_{ij}=\mathcal{G}_{ij},   \mbox{ \ \ \ \ } i,j=1,2,3.
\end{eqnarray}
where the  $\mathcal{G}_{\mu\nu}$ are  given by
\begin{eqnarray}\label{eqn:GRGT}
\mathcal{G}_{00}&=&\frac{1}{2}((trD)^2-tr(D^2)), \mbox{\  \  }
\mathcal{G}_{i0}=\mathcal{G}_{0i}=-\frac{1}{2}(\nabla\times(\nabla\times{\bf v}))_i,   \mbox{ \  }
i=1,2,3.\nonumber\\ 
\mathcal{G}_{ij}&=&
\frac{d}{dt}(D_{ij}-\delta_{ij}trD)+(D_{ij}-\frac{1}{2}\delta_{ij}trD)trD
-\frac{1}{2}\delta_{ij}tr(D^2)+(\Omega D-D\Omega)_{ij}, \nonumber \\ 
 &&\mbox{ \ \ \ \ \ \ \ \ \ \ \ \ \ \ \ \ \ \ \ \ \ \ \ \ \ \ \ \  } i,j=1,2,3.
\end{eqnarray}
In vacuum, with $T_{\mu\nu}=0$, we find from (\ref{eqn:GR32}) and (\ref{eqn:GRG}) that $G_{\mu\nu}=0$ implies that  
$\mathcal{G}_{\mu\nu}=0$. 
We see that the Hilbert-Einstein equations demand that 
\begin{equation}
(trD)^2-tr(D^2)=0
\label{eqg:DMGR}\end{equation}  
but it is these terms in  (\ref {eqn:HubE1})  that explain the various gravitational anomalies. This simply corresponds to the fact that GR does not permit the `dark matter' effect, and this happens because GR was forced to agree with Newtonian
gravity, in the appropriate limits, and that theory also has no such effect. As well in GR the energy-momentum
tensor
$T_{\mu\nu}$ is not permitted to make any reference to absolute linear motion of the matter; only  the relative
motion of matter or absolute rotational motion is permitted, contrary to the experiments.

It is very significant to note that the above exposition of the GR formalism for  the  metric in (\ref{eqn:GRE14}) is exact. Then taking the trace of the $\mathcal{G}_{ij}$ equation in (\ref{eqn:GRGT}) we obtain,
also exactly, and in the case of zero vorticity, and outside of matter so that
$T_{\mu\nu}=0$,
\begin{equation}
\frac{\partial }{\partial t}(\nabla.{\bf v})+\nabla.(({\bf
v}.{\bf \nabla}){\bf v})=0
\label{eqn:GRf3vacuum}\end{equation}
which is the Newtonian `velocity field' formulation of Newtonian gravity outside of matter, as in (\ref {eqn:HubE1}) but with $\alpha=\beta=0$.   So GR turns out to be Newtonian gravity in a grossly overstructured mathematical formalism.

\section{Experimental and Observational Phenomena I}\label{sect:experimentalI}

We now briefly review the extensive range of light speed experiments that have detected that the speed of light is not isotropic - the speed is different in different directions when measured in a laboratory experiment on earth,  as predicted by the generalised Maxwell equations, Sect.\ref{sect:maxwell}. The most famous of these experiments was that of Michelson and Morley in 1887.
Contrary to often repeated claims, this experiment decisively detected the anisotropy. The cause of the misunderstanding surrounding this experiment is that the Newtonian based theory Michelson used for the calibration of the experiment is simply wrong, and of course not unexpectedly.  Clearly as the Michelson interferometer is a 2nd order  $v/c$ experiment its calibration requires a `relativistic' analysis, in particular one must take account of  arm contractions and also the Fresnel drag effect.  The Michelson-Morley fringe shift data then gives a speed in excess of 300km/s, as first discovered by Cahill and Kitto in 2002 \cite{MMCK}.

\subsection{Anisotropy of the Speed of Light}\label{subsect:anisotropy}

That the speed of light in vacuum is the same in all directions, i.e. isotropic,  for all observers has been taken as  a critical assumption  in  the standard formulation of fundamental physics, and was introduced by Einstein in 1905 as one of his key postulates when formulating his interpretation of Special Relativity.
The need to detect any anisotropy has challenged physicists from the 19th century to the present day, particularly  following the Michelson-Morley experiment of 1887. 
The problem arose when Maxwell in 1861 successfully computed the speed of light  $c$ from his unified theory of electric and magnetic fields: but what was the speed $c$ relative to?  There have been many attempts to detect any supposed light-speed anisotropy and  there have so far been 8 successful  and consistent such experiments, and as well numerous unsuccessful experiments, i.e. experiments in which no anisotropy was observed. The reasons for these different outcomes is now understood:  any light-speed anisotropy produces not only an expected `direct' effect, being that which is expected to produce a `signal', but also affects  the very physical structure of the apparatus, and with this effect usually overlooked in the design of some detectors.  In some designs these effects exactly cancel.  As already stated there is overwhelming evidence  from 8 experiments that the speed of light is anisotropic, and with a {\it large} anisotropy at the level of 1 part in $10^3$: so these experiments show that a dynamical 3-space exists, and that the spacetime concept was only a mathematical construct - it does not exist as an entity of reality, it has no ontological significance.   These developments have lead to a new physics in which the dynamics of the 3-space have been formulated, together with the required generalisations of the Maxwell  equations (as first suggested by Hertz in  1890 \cite{Hertz}), and of  the Schr\"{o}dinger and Dirac equations, which have lead to the new emergent theory and explanation of gravity, with numerous confirmations of that theory from the data from black hole systematics, light bending, spiral galaxy rotation anomalies, bore hole anomalies, etc. This data has revealed that the coupling constant for the self-interaction of the dynamical 3-space is none other than the fine structure constant  $\approx 1/137$ \cite{alpha,DM,galaxies,newBH}, which suggests an emerging unified theory of quantum matter and a quantum foam description of the dynamical 3-space. 

The most influential of the early attempts to detect any anisotropy in the speed of light was the Michelson-Morley experiment of 1887, \cite{MM}. Despite that, and its influence on physics, its operation was only finally understood in 2002 \cite{AMGE,MMCK,MMC}.     The problem has been that the Michelson interferometer has a major flaw in its design, when used to detect any light-speed anisotropy effect\footnote{Which also severely diminishes its use in long-baseline interferometers built to detect gravitational waves.}.  To see this requires use of Special Relativity effects.   The Michelson interferometer compares the round-trip light travel time in two orthogonal arms, by means of interference fringe shifts measuring time differences, as the device is rotated. However  if the device is operated in vacuum, any anticipated change in  the total travel times caused by the light travelling at different speeds in the outward and inward directions is exactly cancelled by the Fitzgerald-Lorentz mirror-supporting-arm contraction effect - a real physical effect. Of course this is precisely how Fitzgerald and Lorentz independently arrived at the idea of the length contraction effect.  In vacuum this means that the round-trip travel times in each arm {\it do not} change during rotation. This is the fatal  design flaw that has confounded physics for over 100 years.  However the cancellation  of a supposed change in the round-trip travel times and the Lorentz contraction effect is merely an incidental flaw of the Michelson interferometer. The critical observation  is that if we have a gas in the light path, the round-trip travel times are changed, but the Lorentz arm-length contraction effect is unchanged, and then these effects no longer exactly cancel. Not surprisingly the fringe shifts are now  proportional to $n-1$, where $n$ is the refractive index of the gas.   Of course with  a gas present one must also take account of the Fresnel drag effect, because the gas itself is in absolute motion. This is an important effect, so large  in fact that it reverses the sign of the time differences between the two arms, although in operation that is not a problem.  As well, since for example for air $n=1.00029$ at STP, the sensitivity of the interferometer is very low. Nevertheless the Michelson-Morley experiment as well as the Miller  interferometer experiment of 1925/1926 \cite{Miller} were done in air, which is why they indeed observed and reported  fringe shifts. As well  Illingworth \cite{C5} and Joos \cite{C6} used helium gas in the light paths in their Michelson interferometers; taking account of that brings their results into agreement with those of the air interferometer experiment, and so confirming the refractive index effect.   Jaseja {\it et al.} \cite{C7} used a He-Ne gas mixture of unknown refractive index, but again detected fringe shifts on rotation.  A re-analysis of the data from the above experiments, particularly from the enormous data set of Miller,  has revealed that a large light-speed  anisotropy had been detected from the very beginning of such experiments, where the speed is some $430\pm 20$km/s - this is in excess of 1 part in $10^3$, and the Right Ascension and Declination of the direction was determined by Miller \cite{Miller} long ago.  We also briefly review the RF coaxial cable speed experiments of Torr and Kolen \cite{Torr}, DeWitte \cite{DeWitte}  and Cahill \cite{Coax}, which agree with the gas-mode Michelson interferometer experiments.

\subsection{Michelson Gas-mode Interferometer}\label{subsect:michelsongas}

Let us first consider the new understanding of how the Mich\-elson interferometer works.  This brilliant but very subtle device was conceived by Michelson as  a means to detect the anisotropy of the speed of light, as was expected towards the end of the 19th century.   Michelson used Newtonian physics to develop the theory and hence the calibration \rule{-.2pt}{0pt}for \rule{-.2pt}{0pt}his \rule{-.2pt}{0pt}device.  However we now understand that this  device  detects 2nd order effects in $v/c$ to determine $v$, and  so we must take account of  re\-la\-tivistic effects.   However the application and analysis of data from various Michelson interferometer experiments  using a relativistic theory only occurred in 2002, some  97 years after the development of Special Relativity by Einstein, and some 115 years after the famous 1887 experiment.  As a consequence of the necessity of  using relativistic effects it was discovered in 2002 that the gas in the light paths plays a critical role, and that we finally understand how to calibrate the device, and we also discovered,  some  76 years after the 1925/26 Miller experiment, what determines the calibration constant  $k$ \rule{-.2pt}{0pt}that \rule{-.2pt}{0pt}Miller \rule{-.2pt}{0pt}had \rule{-.2pt}{0pt}determined \rule{-.2pt}{0pt}using \rule{-.2pt}{0pt}the \rule{-.2pt}{0pt}Earth's \rule{-.2pt}{0pt}rotation speed about the Sun to set the calibration.   This, as we discuss later, has enabled us to now appreciate that gas-mode Mich\-elson interferometer experiments have confirmed the reality of the Fitzgerald-Lorentz length contraction effect:  in the usual interpretation of Special Relativity this effect, and others, is usually regarded as an observer dependent effect, an illusion induced by the spacetime. But the experiments are to the contrary showing that the length contraction effect is an actual observer-independent dynamical effect, as Fitzgerald and Lorentz had proposed.

\begin{figure}
\setlength{\unitlength}{0.9mm}
\hspace{25mm}\begin{picture}(0,30)
\thicklines
\put(-10,0){\line(1,0){50}}
\put(-5,0){\vector(1,0){5}}
\put(40,-1){\line(-1,0){29.2}}
\put(15,0){\vector(1,0){5}}
\put(30,-1){\vector(-1,0){5}}
\put(10,0){\line(0,1){30}}
\put(10,5){\vector(0,1){5}}
\put(11,25){\vector(0,-1){5}}
\put(11,30){\line(0,-1){38}}
\put(11,-2){\vector(0,-1){5}}
\put(8.0,-2){\line(1,1){5}}
\put(9.0,-2.9){\line(1,1){5}}
\put(6.5,30){\line(1,0){8}}
\put(40,-4.5){\line(0,1){8}}
\put(5,12){ $L$}
\put(4,-5){ $A$}
\put(35,-5){ $B$}
\put(25,-5){ $L$}
\put(12,26){ $C$}
\put(9,-8){\line(1,0){5}}
\put(9,-9){\line(1,0){5}}
\put(14,-9){\line(0,1){1}}
\put(9,-9){\line(0,1){1}}
\put(15,-9){ $D$}
\put(50,0){\line(1,0){50}}
\put(55,0){\vector(1,0){5}}
\put(73,0){\vector(1,0){5}}
\put(85,0){\vector(1,0){5}}
\put(90,15){\vector(1,0){5}}
\put(100,-1){\vector(-1,0){5}}
\put(100,-4.5){\line(0,1){8}}
\put(68.5,-1.5){\line(1,1){4}}
\put(69.3,-2.0){\line(1,1){4}}
\put(70,0){\line(1,4){7.5}}
\put(70,0){\vector(1,4){3.5}}
\put(77.5,30){\line(1,-4){9.63}}
\put(77.5,30){\vector(1,-4){5}}
\put(73.5,30){\line(1,0){8}}
\put(83.3,-1.5){\line(1,1){4}}
\put(84.0,-2.0){\line(1,1){4}}
\put(100,-1){\line(-1,0){14.9}}
\put(73,3){$\alpha$}
\put(67,-5){ $A_1$}
\put(80,-5){ $A_2$}
\put(85,-8){\line(1,0){5}}
\put(85,-9){\line(1,0){5}}
\put(90,-9){\line(0,1){1}}
\put(85,-9){\line(0,1){1}}
\put(90,-9){ $D$}
\put(95,-5){ $B$}
\put(79,26){ $C$}
\put(90,16){ $v$}
\put(-8,8){(a)}
\put(55,8){(b)}

\end{picture}

\vspace{7mm}
\caption{\footnotesize{Schematic diagrams of the Michelson Interferometer, with beamsplitter/mirror at $A$ and
mirrors at $B$ and $C$ on arms  from $A$, with the arms of equal length $L$ when at rest.  $D$ is a 
screen or detector. In (a) the interferometer is
at rest in space. In (b) the interferometer is moving with speed $v$ relative to space in the direction
indicated. Interference fringes are observed at the  detector $D$.  If the interferometer is
rotated in the plane  through $90^o$, the roles of arms $AC$ and $AB$ are interchanged, and during the
rotation shifts of the fringes are seen in the case of absolute motion, but only if the apparatus operates
in a gas.  By counting fringe changes the speed $v$ may be determined.}  \label{fig:Minterferometer}}
\end{figure}
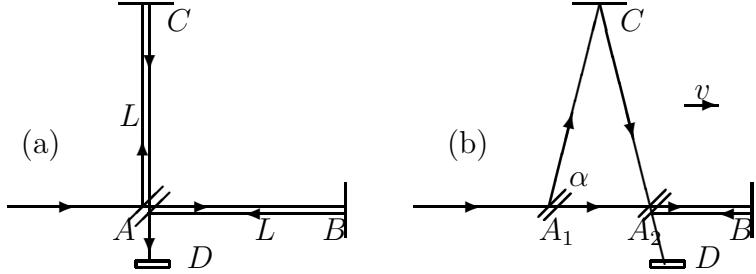

The Michelson interferometer compares the change in the difference between
travel times, when the device is ro\-tat\-ed, for two coherent beams of light that travel in
orthogonal directions between mirrors; the  changing time difference being indicated by the
shift of the interference fringes during the rotation.  This effect is caused by the
absolute motion of the device through 3-space with speed $v$, and that the speed of light is
relative to that 3-space, and not relative to the apparatus/observer. However to detect the
speed of the apparatus through that 3-space gas must be present in the light paths for
purely technical reasons. The post relativistic-effects theory for this device is remarkably
simple.    Consider here only the case where the arms are parallel/anti-parallel to the direction of absolute motion. The re\-la\-tivistic Fitzgerald-Lorentz contraction effect causes the arm $AB$ parallel to the
absolute velocity to be physically con\-tract\-ed to length (see Fig.\ref{fig:Minterferometer})
\vspace*{-2pt}
\begin{equation}
L_{||}=L\,\sqrt{1-\frac{v^2}{c^2}}\,.
\label{eqn:e1}\end{equation}
The time $t_{AB}$ to travel $AB$ is set by $Vt_{AB}=L_{||}+vt_{AB}$, while for $BA$ by 
$Vt_{BA}=L_{||}-vt_{BA}$, where $V=c/n$   is the speed of light,  with
$n$  the refractive index of the gas present. For simplicity we ignore here the Fresnel
drag effect, an effect caused by the gas also being in absolute motion, see \cite{Book}. The Fresnel drag effect is actually large, and results in a change of sign in (\ref{eqn:e4}) and (\ref{eqn:e5}).
For the total $ABA$ travel time we then obtain
\vspace*{-1pt}
\begin{equation}
t_{ABA}=t_{AB}+t_{BA}=\frac{2 LV}{V^2-v^2}\sqrt{1-\frac{v^2}{c^2}}\,.
\label{eqn:e2}\end{equation} 
For travel in the $AC$ direction we have, from the  Pyth\-a\-goras theorem for the
 right-angled triangle in Fig.\ref{fig:Minterferometer} that $(Vt_{AC})^2=L^2+(vt_{AC})^2$  and that
$t_{CA}=t_{AC}$. Then for the total $ACA$ travel time
\vspace*{-2pt}
\begin{equation}
t_{ACA}=t_{AC}+t_{CA}=\frac{2L}{\sqrt{V^2-v^2}}\,.
\label{eqn:e3}\end{equation} 
 Then the difference in travel time is
\begin{equation}
\Delta t=\frac{(n^2-1)\,L}{c}\frac{v^2}{c^2}+\mbox{O}\left(\frac{v^4}{c^4} \right).
\label{eqn:e4}\end{equation}     
after expanding in powers of $v/c$. This clearly shows that the interferometer
can only operate as a detector of absolute mo\-tion when not in vacuum ($n\,{=}\,1$), namely when
the light pass\-es through a gas, as in the early experiments (in transpa\-r\-ent solids a more
complex phenomenon occurs). A more general analysis \cite{Book} with the arms at angle $\theta$ to ${\bf v}$   gives
\begin{equation}
\Delta t=k^2\frac{Lv_P^2}{c^3}\cos
\bigl(2 (\theta-\psi)\bigr) ,
\label{eqn:e5}\end{equation}
where $\psi$ specifies the direction of   ${\bf v}$  projected onto the plane of the interferometer relative to the local meridian, and where $k^2\,{\approx}\, n(n^2\,{-}\,1)$. Neglect of the relativistic Fitz\-gerald-Lorentz contraction effect
gives   $k^2\,{\approx}\, n^3\,{\approx }\,1$
 for gases, which is essentially the Newtonian theory that Mich\-elson used.
 
 However the above analysis does not correspond to how the interferometer is actually operated. That analysis does not actually predict fringe shifts for the  field of view would be uniformly illuminated, and the observed effect would be a changing level of luminosity rather than fringe shifts. As Miller knew, the mirrors must be made slightly non-orthogonal with the degree of non-orthogonality determining how many fringe shifts were visible in the field of view. Miller exper\-i\-mented with this effect to determine a comfortable number of fringes: not too few and not too many.  Hicks \cite{Hicks} deve\-lop\-ed a theory for this effect --- however it is not necessary to be aware of this analysis in using the interferometer: the non-orthogonality   reduces the symmetry of the device, and instead of having period of 180$^\circ$ the symmetry now has a period of 360$^\circ$, so that to (\ref{eqn:e5})  we must add the extra term $a\cos(\theta-\beta)$ in
\vspace*{-4pt}
 \begin{equation}
\Delta t=k^2\frac{L(1+e\theta)v_P^2}{c^3}\cos\bigl(2(\theta-\psi)\bigr)+a(1+e\theta)\cos(\theta-\beta)+f
\label{eqn:e6}\end{equation}   
The term $1+e\theta$ models the temperature effects, namely that as the arms are uniformly rotated, one rotation taking several minutes, there will be a temperature induced change in the length of the arms. If the temperature effects are linear in time, as they would be for short time intervals, then they are linear in $\theta$. In the  Hick's term the parameter $a$ is proportional to the length of the arms, and so also has the temperature factor. The term $f$ simply models any offset  effect.
Michelson and Morley and Miller took these two effects  into account  when analysing his data. The Hick's effect is particularly apparent in the Miller  and Michelson-Morley data.

The interferometers are operated with the arms horizon\-tal. Then in (\ref{eqn:e6}) $\theta$ is the azimuth of one arm
relative to the local meridian, while $\psi$ is the azimuth of the absolute motion
velocity projected onto the plane of the interferometer, with projected component $v_P$.
Here the Fitzgerald-Lorentz con\-traction is a real dynamical effect of absolute motion,
unlike the Einstein spacetime view that it is merely a spacetime perspective artifact, and
whose magnitude depends on the choice of observer. The instrument is operated by rotating
at a rate of one rotation over several minutes, and observing the shift in the fringe
pattern through a telescope during the rotation.  Then fringe shifts from six (Michelson
and Morley) or twenty (Miller) successive rotations are averaged to improve the signal to noise ratio, and the average sidereal time noted.

\begin{figure}
\vspace{0mm}
\hspace{14mm}\includegraphics[scale=0.95]{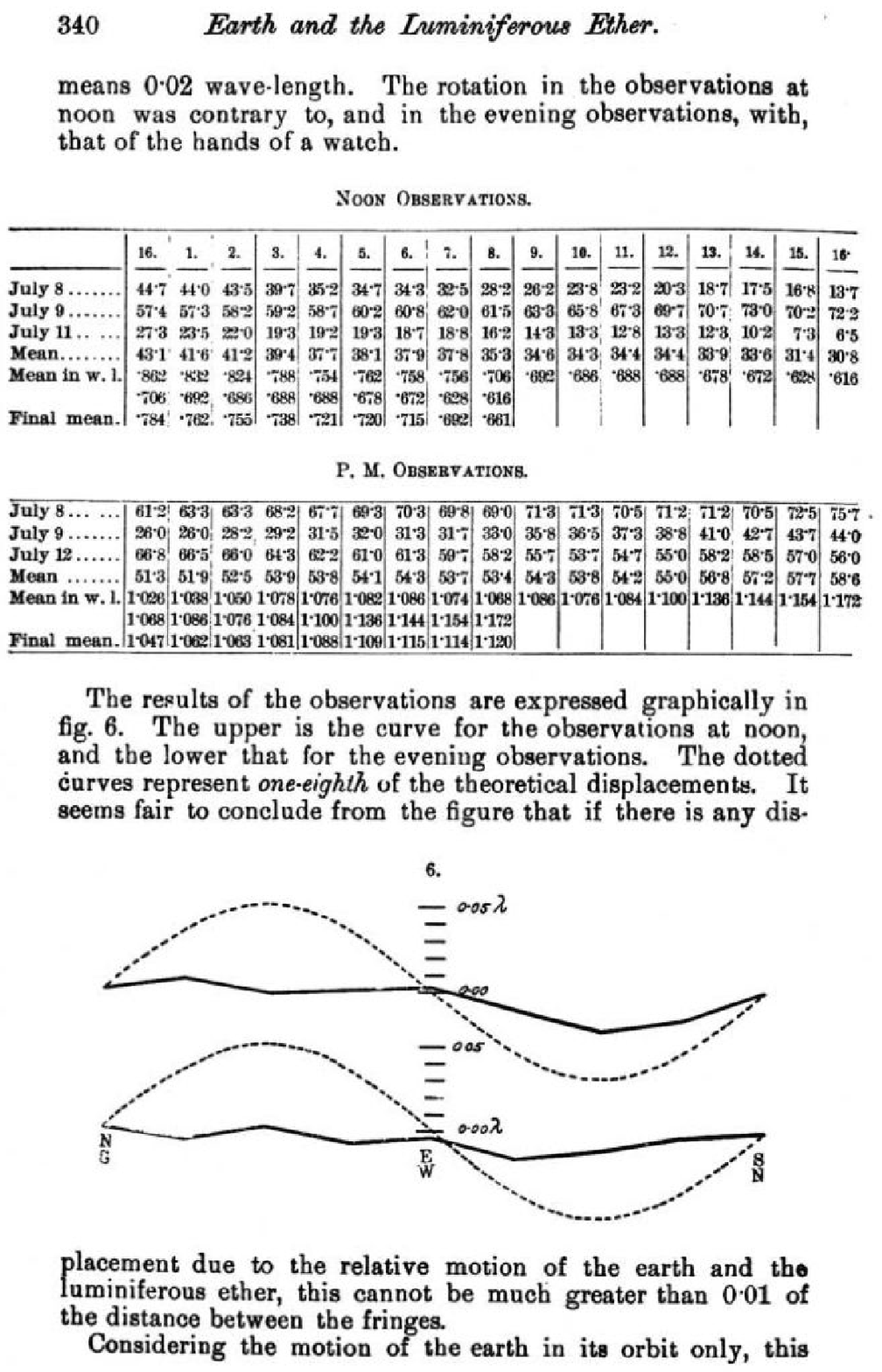}
\vspace{-10mm}\caption{\footnotesize{Page 340 from the 1887 Michelson-Morley paper \cite{MM} showing the table of observed fringe shifts, measured here in divisions of the telescope screw thread, and which is analysed using (\ref{eqn:e6}) with the results shown in Fig.\ref{fig:MM1887}.}}
\label{fig:MM1887page}\end{figure}

\subsection{Michelson-Morley Experiment 1887}\label{subsect:michelsonmorley}

\begin{figure}
\hspace{-10mm}\includegraphics[scale=0.7]{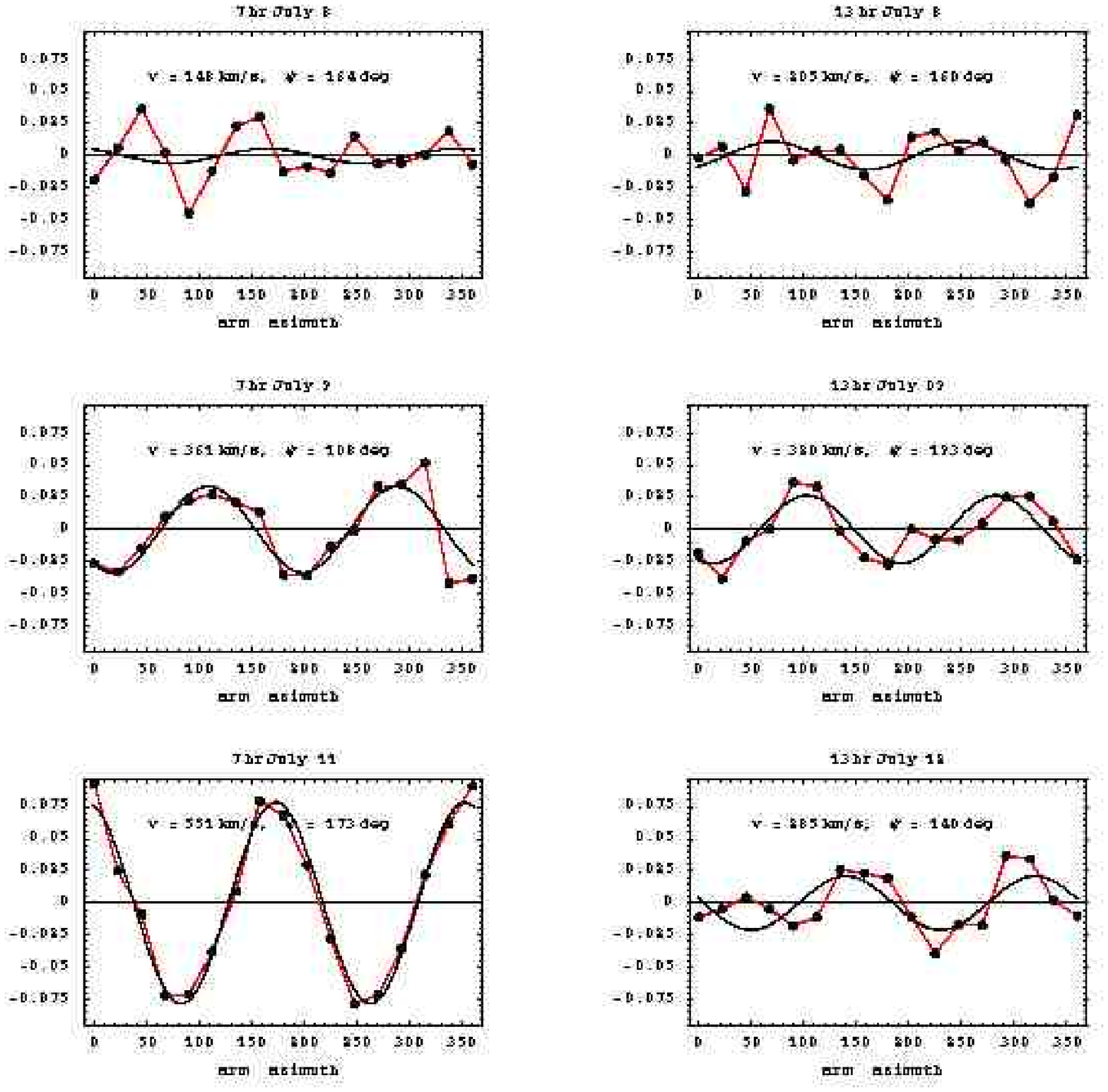}
\vspace{-12mm}\caption{\footnotesize{ Analysis of the Michelson-Morley fringe shift data from the table in Fig.\ref{fig:MM1887page}. The plots are for the sidereal times and days indicated, and each plot arises from averaging six successive rotations, i.e. only 36 rotations were performed in July 1887.   The data was fitted with (\ref{eqn:e6}) by a 6 parameter least-squares-fit by varying $v_P, \psi, a, \beta, e$ and $f$.  Only $v_P$ and $\psi$ are of physical interest, and are shown in each plot. $\psi$ is measured clockwise from North.  After these parameters have been determined the Hicks and temperature terms were subtracted from the data, and plotted above together with the $\cos\bigl(2(\theta-\psi)\bigr)$ expression.  This makes the fringe shifts more easily seen. We see that four of the plots show a good fit to the expected form, while the other two give a poor fit.  We also see that at the same time on successive days the speed and direction are significantly different. These are `gravitational wave' effects, and were seen in later experiments as well. }}
\label{fig:MM1887}\end{figure}

Page 340 of the Michelson-Morley 1887 paper reporting the observed fringe shifts is reproduced in Fig.\ref{fig:MM1887page}. Each row of the table is the average from six successive rotations. In the graphs Michelson and Morley are noting that the fringe shifts are much smaller than expected. But they were using Newtonian physics to calibrate the device. We now know that the detector is nearly 2000 times less sensitive than given by that calibration, and that these fringe shifts correspond to a speed in excess of 300km/s.   Michelson and Morley implicitly assumed the
Newtonian value  $k{=}1$, while Miller used an indirect method to estimate the value of $k$,
as he understood that the Newtonian theory was invalid, but had no other theory for the
interferometer.  Of course the Einstein postulates, as distinct from Special Relativity,  have that absolute motion has no meaning, and so effectively demands that $k=0$.  Using $k=1$  gives only a nominal value for $v_P$, being some 8--9\,km/s for the Michelson and Morley experiment, and some 10\,km/s from Miller; the  difference  arising  from  the 
 different  latitudes of Cleveland and Mt.\,Wilson, and from Michelson and Morley taking data at limited times.  So already Miller knew that his observations were consistent with those of Michelson and Morley, and so the  important need for reproducibility was being confirmed.

\subsection{Miller Experiment 1925/26}\label{subsect:miller}

\begin{figure}
\hspace{20mm}\includegraphics[scale=0.6]{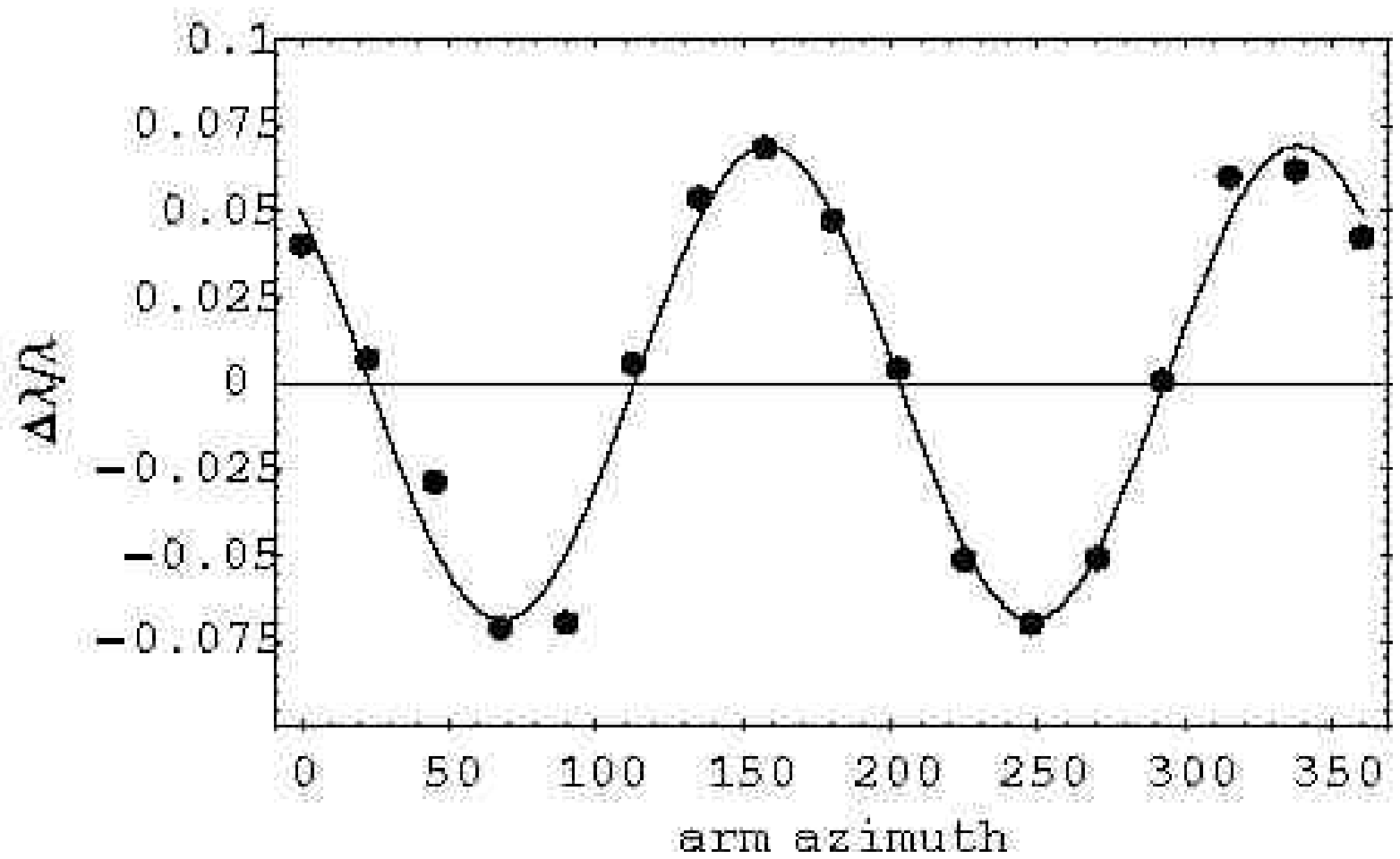}
\vspace{-6mm}\caption{\footnotesize{ Typical Miller rotation-induced fringe shifts from average of 20 rotations,  measured every 22.5$^\circ$, in fractions of a wavelength $\Delta \lambda/\lambda$, vs  arm azimuth $\theta$(deg), measured
clockwise from North, from Cleveland Sept. 29, 1929 16:24 UT; 11:29\,hrs average local sidereal time. The curve is the best fit using the form in (\ref{eqn:e6}), and then subtracting the  Hick's $\cos(\theta-\beta)$ and temperature terms from the data. Best fit gives $\psi=\mbox{158}^\circ$, or $22^\circ$  measured from South, and a projected speed of $v_P =\mbox{315}$\,km/s.  This plot shows the high quality of the Miller fringe shift observations.  In the 1925/26 run of observations the rotations were repeated some 8,000 times.  }
\label{fig:Millers}}\end{figure}

\begin{figure}[t]
\hspace{8mm}\includegraphics[scale=1.6]{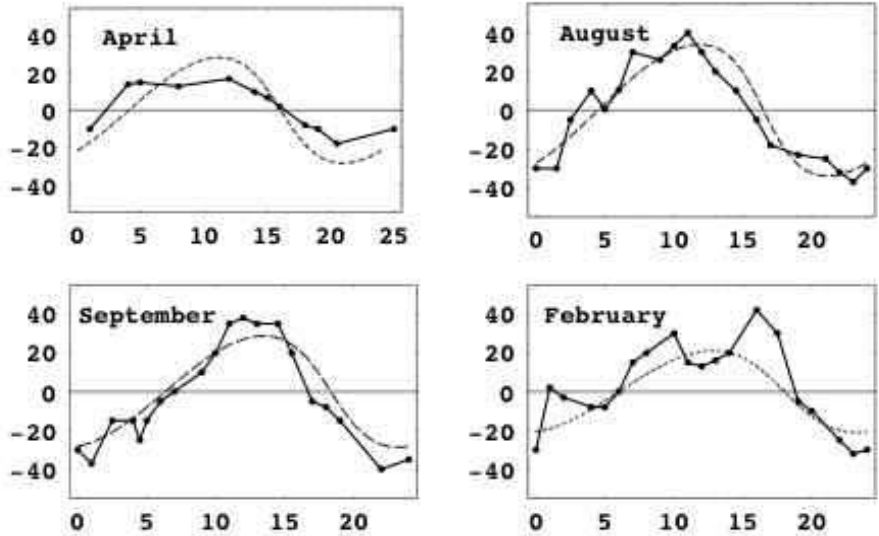}
\caption{\footnotesize Miller azimuths $\psi$, measured from south and plotted against sidereal time in hrs, showing both
data and best fit of theory giving
$v=433$ km/s in the direction ($\alpha=5.2^{hr}, \delta=-67^0$), using $n=1.000226$ appropriate for the
altitude of Mt. Wilson. The variation form month to month arises from the orbital motion of the earth about the sun: in different
months the vector sum of the galactic velocity of the solar system with the orbital velocity and sun in-flow velocity is different.  As shown
in Fig.\ref{fig:DeWitte} DeWitte using a completely different experiment detected the same direction and speed.}  
\label{fig:MillerAz}\end{figure}

The Michelson and Morley air-mode interferometer fringe shift data was 
based upon a total of only 36 rotations in July 1887, revealing the nominal speed of some
8--9\,km/s when analysed using the prevailing but incorrect Newtonian theory which has  $k\,{=}\,1$ in 
(\ref{eqn:e6}), and this value was known to Michelson and Morley. Including the Fitzgerald-Lorentz
dynamical contraction effect as well as the effect of the gas present as in (\ref{eqn:e6})  we find that
 $n_{air}\,{=}\,\mbox{1.00029}$ gives $k^2\,{=}\,\mbox{0.00058}$  for air, which explains why the observed fringe shifts were so
small.   They rejected their own data on the sole but spurious ground that the value of 8\,km/s was smaller than the speed of the Earth about the Sun of 30km/s.  What their result really showed was that (i) absolute motion had been detected because fringe shifts of
the correct form, as in  (\ref{eqn:e6}), had been detected, and (ii) that the theory giving $k^2\,{=}\,1$  was
wrong, that Newtonian physics had failed. Michelson and Morley in 1887 should have announced that
the speed of light did depend of the direction of travel, that the speed  was relative to an
actual physical 3-space. However contrary to their own data they concluded that absolute
motion had not been detected.  This  has had enormous implications for fundamental
theories of space and time over the last 100 years.

It was Miller \cite{Miller} who recognised that in  the 1887 paper   the theory for
the Michelson interferometer must be wrong.  To avoid using that theory Miller introduced
the scaling factor $k$, even though he had no theory for its value. He then used the effect of
the changing vector addition of the Earth's orbital velocity and the absolute galactic
vel\-oc\-ity of the solar system to determine the numerical value of  $k$, because the orbital
motion modulated the data, as shown in Fig.\ref{fig:MillerAz}. By making some 8,000 rotations of the
interferometer at Mt.\,Wilson in 1925/26 Miller determined the first estimate for $k$  and for
the absolute linear velocity of the solar system. Fig.\ref{fig:Millers} shows typical data from averaging
the fringe shifts from 20 rotations of the Miller interferometer, performed over a short
period of time, and clearly shows the expected form in (\ref{eqn:e6}). In Fig.\ref{fig:Millers} the fringe shifts during rotation are
given as fractions of a wavelength, $\Delta \lambda/\lambda\,{=}\,\Delta t/T$, where $\Delta t$ is given by 
(\ref{eqn:e6}) and $T$  is the period of the light. Such rotation-induced fringe shifts clearly show that
the speed of light is different in dif\-ferent directions. The claim that Michelson interferometers,
operating in gas-mode, do not produce fringe shifts under rotation is clearly incorrect. But it is that
claim that lead to the continuing belief, within physics, that absolute motion had never been detected,
and that the speed of light is invar\-iant. The value of $\psi$  from such rotations together lead
to plots like those in Fig.\ref{fig:MillerAz}, which show $\psi$ from the 1925/1926 Miller \cite{Miller}
interferometer data for four different months of the year, from which the RA\,$=$\,5.2\,hr is
readily apparent. While the orbital motion of the Earth about the Sun slightly affects the
RA in each month, and Miller used this effect to determine the value of  $k$, the new theory of
gravity required a reanalysis of the data , revealing that the solar system has
a large observed galactic velocity of some 420$\pm$30\,km/s in the direction (RA\,$=$\,5.2\,hr, Dec\,${=}{-}$67$^\circ$). This is different from the speed of 369\,km/s in the direction (RA\,${=}$\,11.20\,hr, Dec\,${=}{-}$7.22$^\circ$) extracted from the Cosmic Microwave Background (CMB) anisotropy, and which
de\-scribes a motion relative to the dis\-tant universe, but not relative to the local 3-space.
The Miller velocity is explained by galactic gravitational in-flows \cite{Book}.

An important implication of the new understanding of the Michelson interferometer is that vacuum-mode 
resonant cavity experiments should give a null effect, as is the case \cite{C3}.

\vspace*{-3pt}
\subsection{Other Gas-mode Michelson Interferometer  Experiments}\label{subsect:other}

\vspace*{-3pt}
Two old interferometer experiments, by Illingworth \cite{C5} and Joos \cite{C6}, used
helium, enabling the refractive index effect to be recently confirmed, because for helium,
with $n\,{=}$ ${=}\,\mbox{1.000036}$, we find that 
$k^2\,{=}\,\mbox{0.00007}$.  Until the refractive index effect was taken into account
the data from the helium-mode experiments appeared to be inconsistent with the data from the air-mode
experiments; now they are seen to be consistent. Ironically helium was introduced in place of air to
reduce any possible unwanted effects of a gas, but we now understand the essential role of
the gas. The data from an interferometer experiment by Jaseja {\it  et al.}
\cite{C7}, using two
orthogonal masers with a He-Ne gas mixture, also indicates that they detected absolute
motion, but were not aware of that as they used the incorrect Newtonian theory and so
considered the fringe shifts to be too small to be real, reminiscent of the same mistake by
Michelson and Morley. While the Michelson interferometer is a 2nd order device, as the effect of
absolute motion is proportional to  $(v/c)^2$, as in  (\ref{eqn:e6}), but 1st order devices are also possible
and the coaxial cable experiments described next are in this class. 

\vspace*{-3pt}
\subsection[Coaxial Cable Speed of EM Waves  Anisotropy  Experiments]{Coaxial Cable Speed of EM Waves  Anisotropy  \\ Experiments}\label{subsect:coaxial}

\vspace*{-3pt}
Rather than use light travel time experiments to demonstrate the anisotropy of the speed of light, another  technique is to measure the one-way speed of radio waves through a coaxial electrical cable.  While this not a direct `ideal' technique, as then the complexity of the propagation physics comes into play, it provides not only an independent confirmation of the light anisotropy effect, but also one which takes advantage of modern electronic timing technology.

\vspace*{-3pt}
\subsection{Torr-Kolen Coaxial Cable Anisotropy  Experiment}\label{subsect:torr}

\vspace*{-2pt}
The first one-way coaxial cable speed-of-propagation exper\-i\-ment was performed at the Utah University in 1981 by Torr and Kolen. This involved two rubidium  clocks placed approximately 500\,m apart with a 5\,MHz radio frequency (RF) signal propagating between the clocks via a buried EW nitrogen-filled coaxial cable maintained at a constant pres\-sure of 2\,psi. Torr and Kolen  found that, while the round-trip speed time remained constant within 0.0001\%\,$c$, as expected from Sect.\ref{sect:special}, variations in the one-way travel time  were observed. The maximum effect occurred, typically, at the times predicted using the Miller galactic velocity, although Torr and Kolen appear to have been unaware of the Miller experiment. As well Torr and Kolen reported fluctuations in both the magnitude, from 1--3\,ns, and the time of maximum variations in travel time.  These effects are interpreted as arising from the turbulence in the flow of space past the Earth.

\vspace*{2pt}
\subsection{De Witte Coaxial Cable Anisotropy Experiment}\label{subsect:dewitte}

\begin{figure}[t]
\vspace{3.2mm}\parbox{65mm}{\includegraphics[width=65mm,scale=0.2]{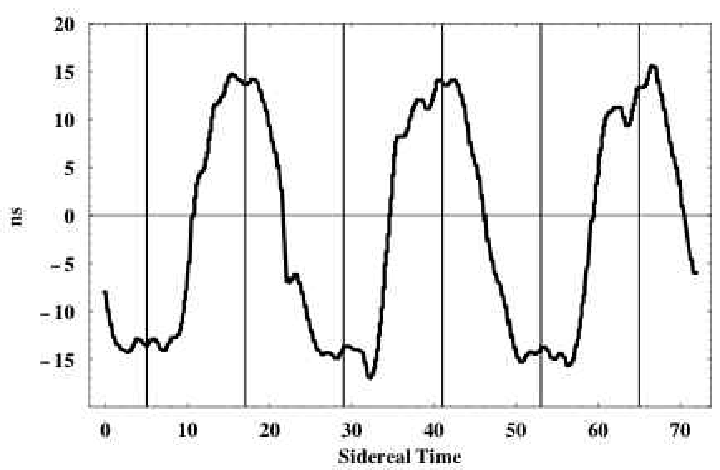}}
\parbox{70mm} {\includegraphics[width=70mm,scale=0.2]{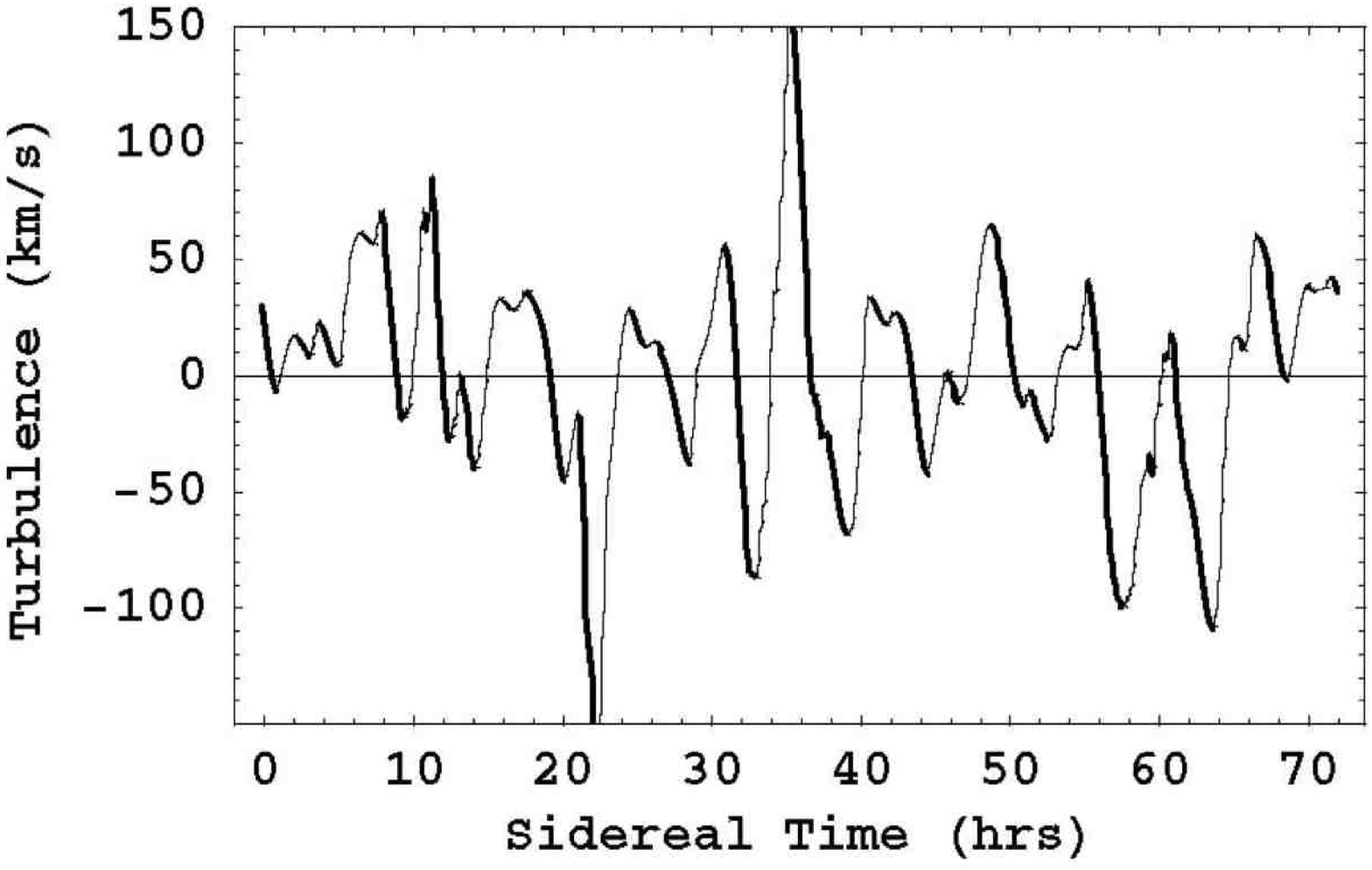}}
\caption{\footnotesize{  (a) Variations in twice the one-way travel time, in ns, for an RF signal to travel 1.5\,km through a coaxial cable between  Rue du Marais and Rue de la Paille, Brussels. 
An offset  has been used  such that the average is zero.   The cable has a
North-South  orientation, and the data is $\pm$ difference of the travel times  for NS and SN
propagation.  The sidereal time for maximum  effect of ${\sim}\,$5\,hr and   ${\sim}\,$17\,hr (indicated
by vertical lines) agrees with the direction found by Miller. Plot shows
data over 3 sidereal days  and is plotted against sidereal time. 
 The fluctuations are evidence of turbulence  of gravitational waves. (b) Shows the speed fluctuations, essentially `gravitational waves' observed by      De Witte in 1991 from the measurement of variations in the RF coaxial-cable travel times.  This data is obtained from that in (a) after removal of the dominant effect caused by the rotation of the Earth.  Ideally the velocity fluctuations are three-dimensional, but the De Witte experiment had only one  arm. This plot is suggestive of a fractal structure to the velocity field. This is confirmed by the power law analysis   in  \cite{Schrod, DeWitte}.}}
\label{fig:DeWitte}
\end{figure}
During 1991 Roland De Witte performed a most extensive RF coaxial cable travel-time anisotropy experiment, accu\-mu\-lating data over 178 days. His data is in complete agreement with the Michelson-Morley 1887 and  Miller 1925/26 inter\-ferometer experiments.   The Miller and De Witte experiments will eventually be recognised as two of the most significant experiments in physics, for independently and using different experimental techniques they detected essentially the same velocity of absolute motion.  But also they detected turbu\-lence in the flow of space past the Earth --- none other than gravitational waves.
The De Witte experiment was within Belgacom, the Belgium telecommunications company. This organisation had two sets of atomic clocks in two buildings in Brussels separated by 1.5\,km and the research project was an investigation of  the task of synchronising these two clusters of atomic clocks. To that end 5MHz RF signals were sent in both directions   through two buried coaxial cables linking the two clusters.   The atomic clocks were caesium beam atomic clocks, and there were three in each cluster: A1, A2 and A3 in one cluster, and B1, B2, and B3 at the other cluster. In that way the stability of the clocks could be established and monitored. One cluster was in a building on Rue du Marais and the second cluster was due south in a building on Rue de la Paille.  Digital phase comparators were used to measure changes in times between clocks within the same cluster and also in the one-way propagation times of the RF signals.  At both locations the comparison between local clocks, A1-A2 and A1-A3, and between B1-B2, B1-B3, yielded linear phase variations in agreement with the fact that the clocks have not exactly the same frequencies together with a short term and long term phase noise. But between distant clocks A1 toward B1 and B1 toward A1, in addition to the same linear phase variations, there is also an additional clear sinusoidal-like phase undulation with an ap\-proximate 24\,hr period of the order of 28\,ns peak to peak, as shown in Fig.~\ref{fig:DeWitte}. The experiment was performed over 178 days, making it possible to measure with an accuracy of  25\,s the period of the phase signal to be the sidereal day (23\,hr 56\,min).

Changes in propagation times were observed over 178 days from June 3 to November 27, 1991. A sample of the data, plotted against sidereal time for just three days, is shown in Fig.\ref{fig:DeWitte}.  De Witte recognised that the data was evi\-dence of absolute motion but he was unaware of the Miller experiment and did not realise that the Right Ascensions for minimum/maximum propagation time agreed almost exactly with that predicted using the Miller's direction (RA$\,{=}\,$5.2\,hr, Dec$\,{=}{-}$67$^\circ$). In fact De Witte expected that the direction of 
absolute motion should have been in the CMB direction, but that would have given the data a totally different sidereal time signature, namely the times for maximum/minimum would have been shifted by 6\,hrs.  The declination of the velocity observed in this De Witte experiment cannot be de\-t\-er\-mined from the data as only three days of data are avai\-lable.   The De Witte data is analysed in \cite{Coax} and as\-sum\-ing a declination of 60$^\circ$\,S a speed of 430\,km/s is obtained, in good agreement with the Miller speed and Michelson-Morley speed. So a different and non-relativistic technique is con\-firm\-ing the results of these older experiments. This is dramatic.

De Witte reported the sidereal time of the `zero' cross-over time, that is in Fig.\ref{fig:DeWitte} for all 178 days of data.  That showed  that the time variations are correlated with sidereal time and not local solar time. A least-squares best fit of a linear relation to that data gives that the cross-over time is retarded, on average, by 3.92 minutes per solar day. This is to be compared with the fact that a sidereal day is 3.93 minutes shorter than a solar day. So the effect is certainly galactic and not associated with any daily thermal effects, which in any case would be very small as the cable is buried.  Miller had also compared his data against sidereal time and established the same property, namely that the diurnal effects  actually tracked sidereal time and not solar time,  and that orbital effects were also apparent, with both effects apparent   in Fig.\ref{fig:MillerAz}. 

\begin{figure}
\hspace{0mm}\parbox{75mm}{\includegraphics[width=70mm,scale=0.21]{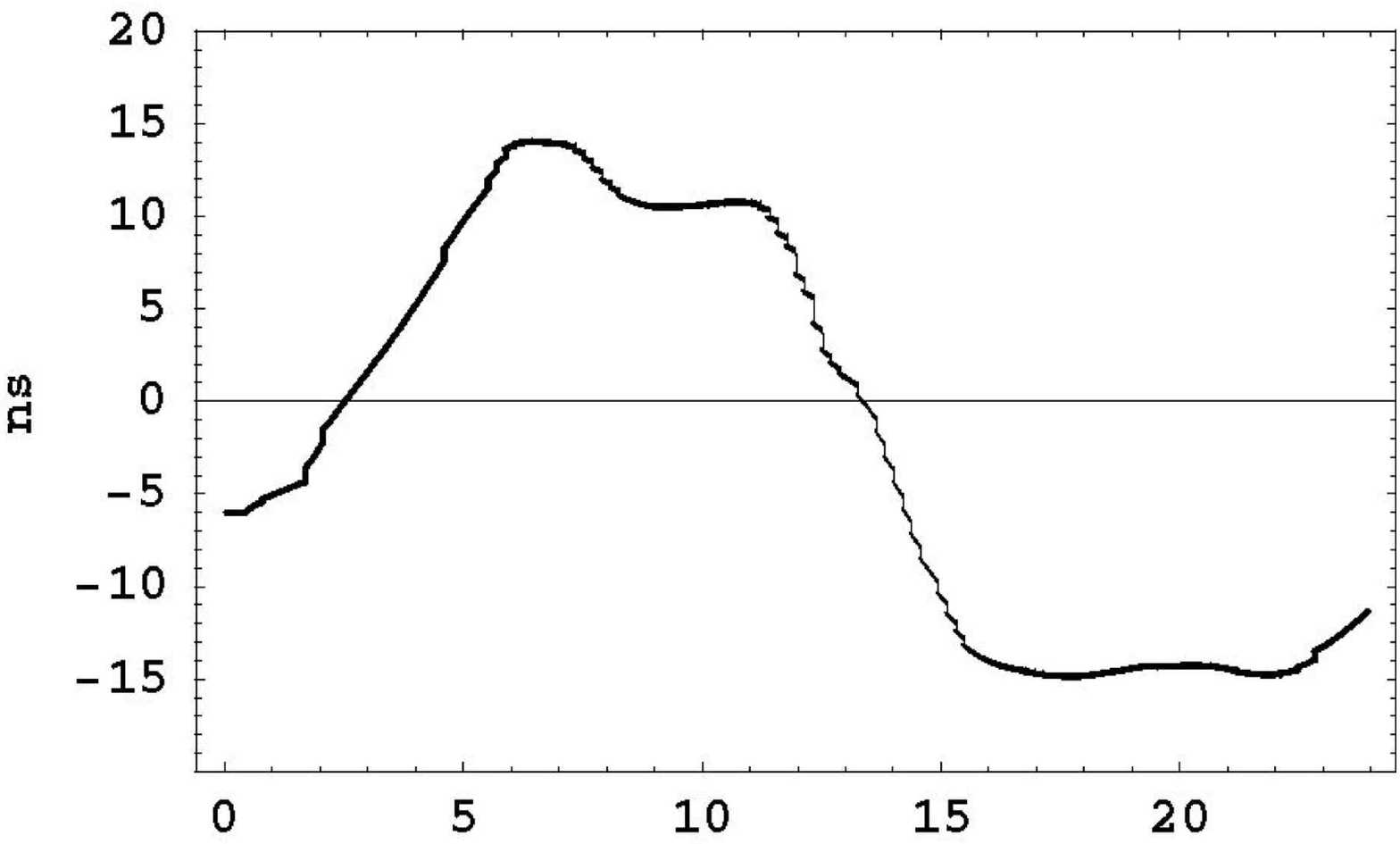}

\vspace{5mm}\hspace{4mm} \includegraphics[width=65mm,scale=0.83]{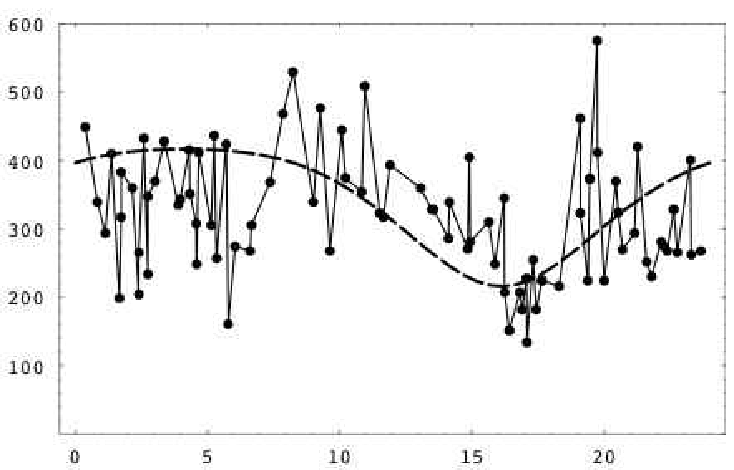}

\vspace{5mm} \hspace{-4mm} \includegraphics[width=75mm,scale=0.29]{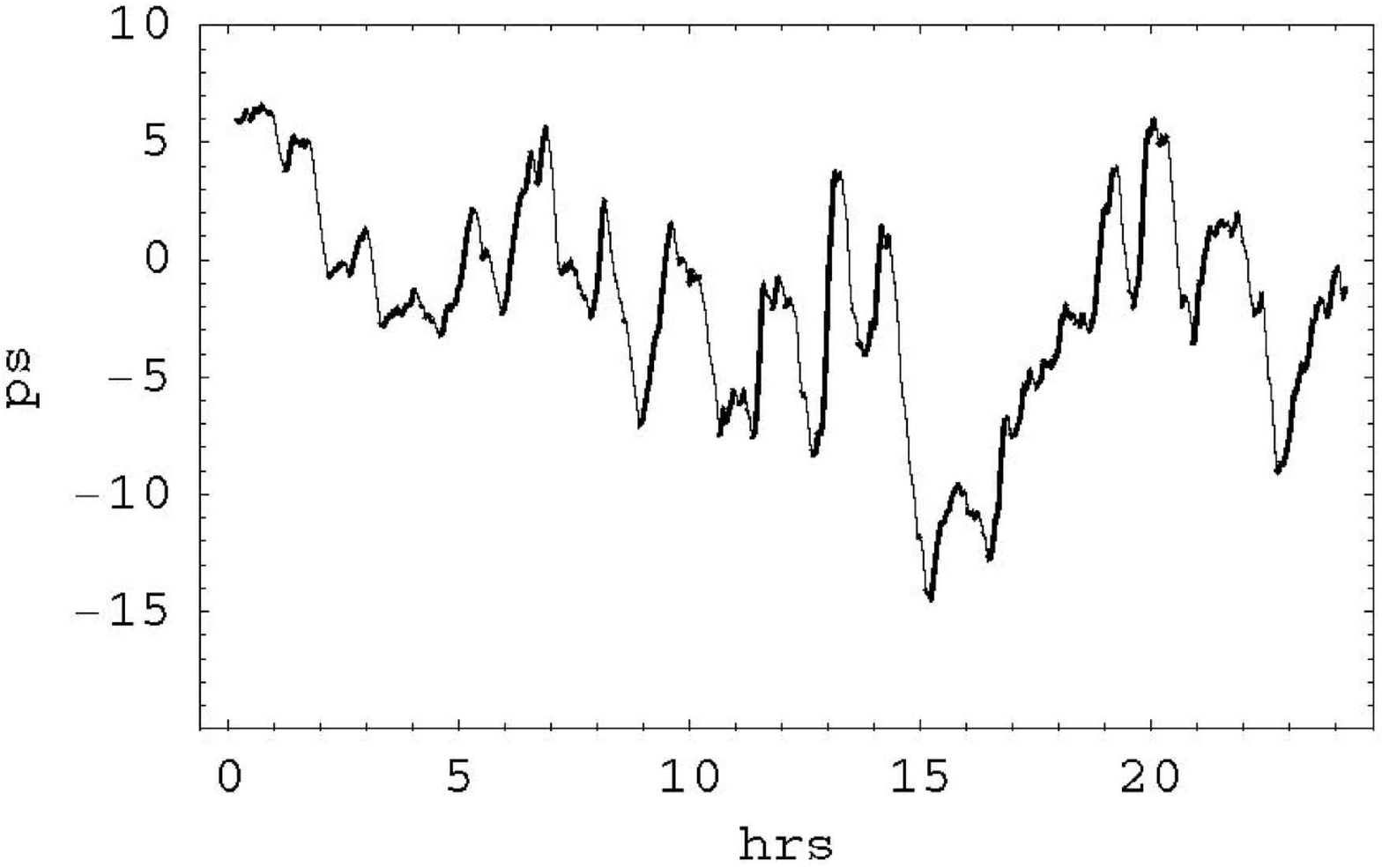}}
 \parbox{57mm}{\caption{\footnotesize{\it Top:} De Witte data, with sign reversed,  from the first sidereal day in 
Fig.\ref{fig:DeWitte}. This data gives a speed of approximately 430km/s. The  data appears to have been averaged over more than 1hr, but still shows wave effects. 
{\it Middle:} Absolute projected speeds $v_P$ in the Miller experiment  plotted against sidereal time in hours  for a composite day collected over a number of days in September 1925.   Maximum projected speed is 417\,km/s. The data shows considerable fluctuations.  The dashed curve shows the non-fluctuating  variation expected over one day as the Earth rotates, causing the projection onto  the plane of the interferometer of the  velocity of the average direction of the space flow to change.  If the data was plotted against solar time the form is shifted by many hours.  Note that the min/max occur at approximately 5\,hrs and 17\,hrs, as also seen by De Witte and  in the Cahill experiment.   {\it Bottom:}  Data from the Cahill experiment \cite{Coax}  for one sidereal day on approximately August 23, 2006.  We see similar variation with sidereal time, and also similar  wave structure.  This  data has been averaged over a running 1hr time interval to more closely match the time resolution of the Miller experiment. These fluctuations are believed to be real wave phenomena of the 3-space.  The new experiment gives a speed of 418\,km/s.  We see remarkable agreement between all three experiments.}\label{fig:SeptPlot}}
\end{figure}

The dominant effect in Fig.\ref{fig:DeWitte}  is caused by the rotation of the Earth, namely that the orientation of the coaxial cable with respect to the average direction of the flow past the Earth changes as the Earth rotates. This effect may be ap\-proximately unfolded from the data leaving the gravitational waves shown in Fig.\ref{fig:DeWitte}, \cite{Schrod,DeWitte}.  This is the first evidence that the velocity field describing the flow of space has a complex structure, and is indeed fractal. The fractal structure, i.\,e. that there is an intrinsic lack of scale to these speed fluctuations, is demonstrated by binning the absolute speeds   and counting the number of speeds  within each bin,  as discussed in \cite{Schrod, DeWitte}. The Miller data also shows evidence of turbulence of the same magnitude.  So far the data from four experiments, namely Miller, Torr and Kolen,  De Witte and Cahill, show turbu\-len\-ce in the flow of space past the Earth.  This is what can be called gravitational waves.  This can be understood by noting that fluctuations in the velocity field induce ripples in the mathematical construct known as spacetime, as in (\ref{eqn:GRE14}).  Such ripples in spacetime are known as gravitational waves.

\subsection{Cahill  Coaxial Cable Anisotropy Experiment}\label{subsect:cahill}

During 2006 Cahill \cite{Coax} performed another RF coaxial cable anisotropy experiment. 
This detector uses a novel timing scheme that overcomes the limitations associated with the two previous coaxial cable experiments.  The intention in such experiments is simply to measure the one-way travel time of RF waves propagating through the coaxial cable.  To that end one would apparently require two very accurate clocks at each end, and associated RF generation and detection electronics.
 However the major limitation is that even the best atomic clocks are not sufficiently accurate over even a day to make such \rule{-.5pt}{0pt}measurements \rule{-.5pt}{0pt}to \rule{-.5pt}{0pt}the \rule{-.5pt}{0pt}required \rule{-.5pt}{0pt}accuracy, \rule{-.5pt}{0pt}unless \rule{-.5pt}{0pt}the \rule{-.5pt}{0pt}cables are of order of a kilometre or so in length, and then tem\-pe\-ra\-ture control becomes a major problem. The issue is that the time variations are of the order of 25\,ps per 10 meters of cable. To measure that requires time measurements accurate to, say,  1\,ps.  But atomic clocks have accuracies over one day of around 100\,ps, implying that lengths of around 1 kilometre would be required, in order for the effect to well exceed timing errors. Even then the atomic clocks must be brought together every day to resynchronise them, or use De Witte's method of multiple atomic clocks.  The new experiment is based on the notion that optical fibers respond differently to coaxial cable with respect to  the speed of propagation of EM radiation.   Some results are shown in Fig.\ref{fig:SeptPlot}.

\section{Experimental and Observational Phenomena II}\label{sect:experimentalII}

\subsection{Gravitational Phenomena}\label{subsect:gravitational}

We have shown above that the dynamics of 3-space involves two constants: $G$ and $\alpha$. When generalising the Schr\"{o}dinger
and Dirac equations to take account of this 3-space we discovered  that we arrive at an explanation for the phenomenon of gravity  including the equivalence principle, as well as an explanation for the spacetime formalism. Here we explore various consequences of this new explanation for gravity particularly those effects which reveal the effects of the $\alpha$-dependent dynamics, in particular the bore hole anomaly which gives us the best estimate for the value of $\alpha$ from several bore hole experiments.  The dynamical  3-space also gives a completely new account of black holes; an account completely different from the putative black holes of GR. In particular these new black holes generate an acceleration $g$ that varies essentially as $1/r$, rather than as $1/r^2$ as in Newtonian gravity (NG) and GR.  This is a dramatic difference.  It explains immediately  the rotation of spiral galaxies, for which the rotation speed is essentially constant at the outer limits, whereas NG and GR   predict a $1/\sqrt{r}$ Keplerian form. It was this dramatic failure of NG and GR, and also in galactic clusters, that lead to the introduction of `dark matter' - to generate a greater gravitational acceleration.  The new theory of 3-space does not need this `dark matter'. The black hole phenomena  is complex, with minimal black holes induced by  matter, to primordial black holes that attract matter. In the former case, and where the matter, in the form of stars and so on, has an essentially  spherically symmetric distribution, it is possible to compute the effective mass of the induced minimal black holes. Observational data from these systems confirms the prediction. Other effects discussed are the gyroscope precession effect caused by the vorticity of the flow of 3-space past the earth. Finally we also discuss the cosmological Hubble expansion that arises from the 3-space dynamics. This gives an excellent parameter-free account of the redshift data from supernovae and gamma-ray bursts. GR requires `dark energy'  to fit that data, so here we see that the new 3-space dynamics does away with the need for `dark energy'.  Not discussed herein are anomalies in the Cavendish-like experiments to determine $G$ \cite{Cavendish},  the gravitational lensing effects predicted by the generalised Maxwell equations, and also a re-analysis of the precession of elliptical orbits, particularly that of Mercury, and various other gravitational effects, see \cite{Book}.

\subsection{Bore Hole Anomaly and the Fine Structure Constant}\label{subsect:bore}

\begin{figure}
\vspace{0mm}\parbox{85mm}{\includegraphics[width=84mm]{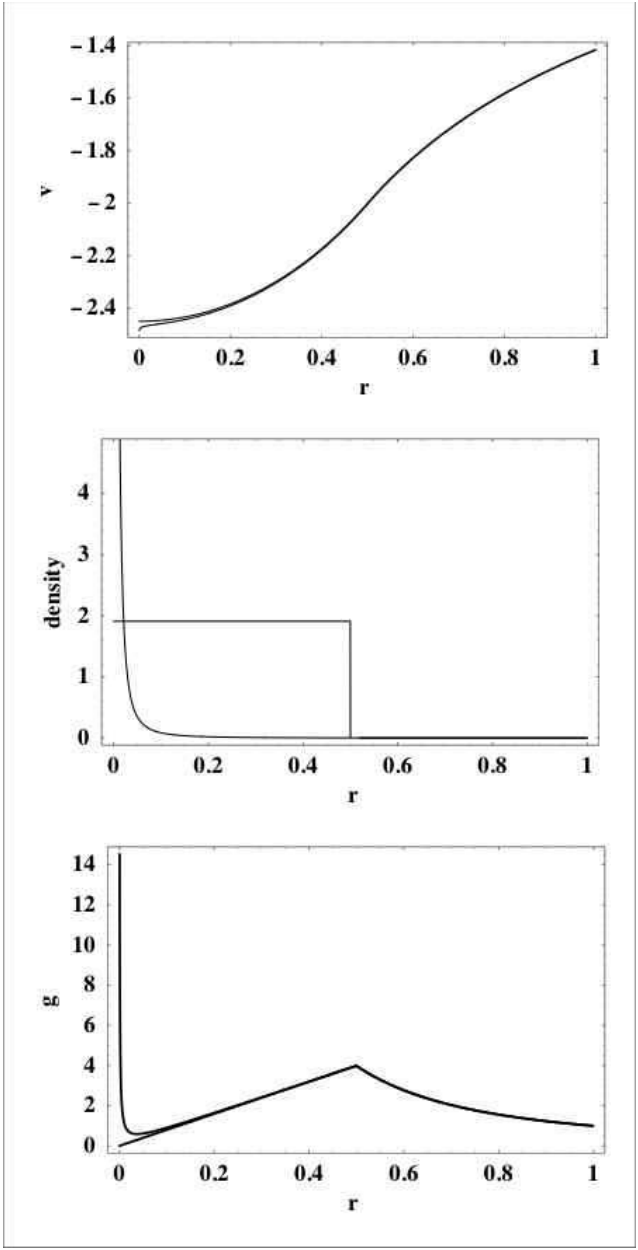}}\,
\parbox{47mm}{\caption{\footnotesize{  Upper plot shows speeds from numerical iterative  solution of (\ref{eqn:BoreE7}) for a solid sphere with
uniform density and radius $r=0.5$ for (i) upper curve the case $\alpha=0$ corresponding to Newtonian gravity, and (ii)
lower curve with $\alpha=1/137$. These solutions ony differ significantly near $r=0$. Middle  plot shows matter density
and `dark matter' density $\rho_{DM}$, from (\ref{eqn:BoreE8}), with arbitrary scales.  Lower plot shows the acceleration
from (\ref{eqn:HubE3}) for (i) the Newtonian in-flow  from the upper plot, and (ii) from the $\alpha=1/137$ case. The
difference is only significant near $r=0$. The accelerations begin to differ just inside the surface of the sphere at
$r=0.5$, according to (\ref{eqn:BoreE15}).  This difference is the origin of the bore hole $g$ anomaly, and permits the
determination of the value of $\alpha$ from observational data. This generic singular-$g$ behaviour, at $r=0$,
is seen in the earth, in globular clusters and in galaxies.}}\label{fig:earthplots}}
\end{figure}

 We now show that the Airy method \cite{Airy}  originally proposed for measuring $G$ actually  gives a technique for determining the value of $\alpha$ from earth based
bore hole gravity measurements. For a time-independent velocity field (\ref{eqn:HubE8}) may be written in the integral
form
\begin{equation}\label{eqn:BoreE6}
|{\bf v}({\bf r})|^2=2G\int d^3
r^\prime\frac{\rho({\bf r}^\prime)+\rho_{DM}({\bf r}^\prime)}{|{\bf r}-{\bf r}^\prime|}.
\end{equation}
When the matter density of the earth is assumed to be spherically symmetric, and that the velocity field is
now radial\footnote{This in-flow is additional to the observed velocity of the earth through 3-space.}  
(\ref{eqn:BoreE6}) becomes
\begin{equation}
v(r)^2=\frac{8\pi G}{r}\int_0^r s^2 \left[\rho(s)+\rho_{DM}(s)\right]ds +8\pi G\int_r^\infty s
\left[\rho(s)+\rho_{DM}(s)\right]ds, 
\label{eqn:BoreE7}\end{equation}
 where, with $v^\prime=dv(r)/dr$,
\begin{equation}
\rho_{DM}(r)= \frac{\alpha}{8\pi G}\left(\frac{v^2}{2r^2}+ \frac{vv^\prime}{r}\right).
\label{eqn:BoreE8}\end{equation}
Iterating (\ref{eqn:BoreE7}) once we find to 1st order in $\alpha$ that 
\begin{equation}
\rho_{DM}(r)=\frac{\alpha}{2r^2}\int_r^\infty s\rho(s)ds+O(\alpha^2),
\label{eqn:BoreE9}\end{equation}
so that in spherical systems the `dark matter' effect is concentrated near the centre, and we find that the total
`dark matter' is
\begin{equation}
M_{DM} \equiv 4\pi\int_0^\infty r^2\rho_{DM}(r)dr  =\frac{4\pi\alpha}{2}\int_0^\infty
r^2\rho(r)dr+O(\alpha^2) =\frac{\alpha}{2}M+O(\alpha^2) 
\label{eqn:BoreE10}\end{equation}
where $M$ is the total amount of (actual) matter. Hence to $O(\alpha)$   $M_{DM}/M=\alpha/2$ independently of the matter
density profile. This turns out to be  a very useful property as complete knowledge of the density profile is then
not required in order to analyse observational data. As seen in Fig.\ref{fig:earthplots} the singular behaviour of
both $v$ and $g$ means that there is a {\it black hole}\footnote{These are called {\it black holes} because there is
an event horizon, but in all other aspects differ from the {\it black holes} of General Relativity.} singularity at
$r=0$.

From (\ref{eqn:HubE3}), which is also the acceleration of matter \cite{Schrod}, the gravity acceleration\footnote{We
now  use the convention that $g(r)$ is positive if it is radially inward.} is found to be, to 1st order in
$\alpha$,  and using that $\rho(r)=0$ for $r>R$, where $R$ is the radius of the earth,
\begin{equation}
g(r)=\left\{ \begin{tabular}{ l} 
$\displaystyle{\frac{(1+\displaystyle{\frac{\alpha}{2}}) GM}{r^2}, \mbox{\ \ } r > R,}$  \\  
$\displaystyle{\frac{4\pi G}{r^2}\int_0^rs^2\rho(s)ds +\frac{2\pi\alpha G}{r^2}\int_0^r\left(\int_s^R s^\prime \rho(s^\prime)ds^\prime
\right) ds},\mbox{\  } r< R$.
\end{tabular}\right.   
\label{eqn:BoreE11}\end{equation}
This gives   Newton's `inverse square law' for $r > R$,  even when $\alpha \neq 0$, which explains why the
3-space self-interaction dynamics did not overtly manifest in the analysis of planetary orbits by Kepler and then
Newton. However inside the earth (\ref{eqn:BoreE11}) shows that  $g(r)$ differs from the Newtonian theory,
corresponding to $\alpha=0$,  as in Fig.\ref{fig:earthplots}, and it is this effect that allows the determination of
the value of  $\alpha$ from the Airy method.

Expanding  (\ref{eqn:BoreE11}) in $r$ about the surface, $r=R$, we obtain,  to 1st order in
$\alpha$ and  for an arbitrary density profile, but not  retaining any density gradients at the surface,
\begin{equation}
g(r)=\!\left\{ \begin{tabular}{ l} 
$\!\!\!\!\! \displaystyle{\frac{G_N M}{R^2}-\frac{2 G_N M}{R^3}(r-R),
\mbox{\ \ \ \  \ \ \ \ \  } r > R,}$ 
\\  \\ 
$\!\!\!\!\! \displaystyle{\frac{G_N M}{R^2}-\left(\frac{2 G_N M}{R^3}-4\pi(1-\frac{\alpha}{2})G_N\rho\right)\!(r-R),\mbox{\ \ \ } r<
R}$
\end{tabular}\right.   
\label{eqn:BoreE12}\end{equation}
where  $\rho$ is the matter density at the surface, $M$ is the total matter mass of the earth, and where we have
defined
\begin{equation}G_N\equiv(1+\frac{\alpha}{2})G.\label{eqn:BoreE13}\end{equation}
The corresponding Newtonian gravity expression is obtained by taking the limit $\alpha\rightarrow 0$,
\begin{equation}
g_N(r)=\!\left\{ \begin{tabular}{ l} 
$\!\!\!\!\!\displaystyle{\frac{G_N M}{R^2}-\frac{2 G_N M}{R^3}(r-R), \mbox{\ \ \ \  } r > R,}$  \\  \\
$\!\!\!\!\! \displaystyle{\frac{G_N M}{R^2}-\left(\frac{2 G_N M}{R^3}-4\pi G_N\rho\right)\!(r-R),\mbox{\ \  }r< R}$
\end{tabular}\right.   
\label{eqn:BoreE14}\end{equation}
Assuming Newtonian gravity (\ref{eqn:BoreE14})
then means that from  the measurement of difference between  the above-ground and  below-ground gravity gradients,
namely $4\pi G_N\rho$,  and also measurement of the matter density, permit the determination of  $G_N$.
This is the basis of the Airy method for determining $G_N$ \cite{Airy}.

When analysing the bore hole data it has been  found \cite{Ander89,Thomas90} that  the observed difference of the
gravity  gradients was inconsistent with $4\pi G_N\rho$ in  (\ref{eqn:BoreE14}), in that it was not given by the
laboratory value of $G_N$ and the measured matter density.  This is known as the bore hole
$g$ anomaly and which attracted much interest in the 1980's. 
The bore hole data papers  \cite{Ander89,Thomas90} report the discrepancy, i.e. the anomaly or the gravity residual as it
is called, between the Newtonian prediction and the measured below-earth gravity gradient. Taking the difference
between 
 (\ref{eqn:BoreE12})  and (\ref{eqn:BoreE14}), assuming the same unknown value of $G_N$ in both,  we obtain an expression for
the gravity residual
\begin{equation}
\Delta g(r)\equiv  g_N(r)-g(r)=\left\{ \begin{tabular}{ l} 
$\mbox{\ \ }0, \mbox{\ \ \ \ \ \ \ } r> R,$  \\   
$2\pi\alpha G_N\rho(r-R), \mbox{\  } r < R.$
\end{tabular}\right. 
\label{eqn:BoreE15}\end{equation}

When $\alpha\neq 0$ we have a two-parameter theory of gravity, and from (\ref{eqn:BoreE12}) we see that measurement
of the difference between the above ground and below ground gravity gradients is $4\pi(1-\frac{\alpha}{2}) G_N\rho$, and
this  is not sufficient to determine both $G_N$ and $\alpha$, given $\rho$, and so the Airy method is now understood
not to be a complete measurement by itself, i.e. we need to combine it with other measurements.  If we now use
laboratory Cavendish experiments to determine $G_N$, then from the bore hole  gravity residuals we can determine the
value of
$\alpha$, as already indicated in \cite{alpha,DM}.  These Cavendish
experiments can only determine $G_N$ up to corrections of order $\alpha/4$, simply because the analysis of the data from
these experiments assumed the validity of Newtonian gravity \cite{Book}. So the analysis of the bore hole residuals will give the
value of $\alpha$ up to $O(\alpha^2)$ corrections, which is consistent with the $O(\alpha)$ analysis reported above.

\begin{figure}
\vspace{3.2mm}\parbox{60mm}{\includegraphics[width=60mm]{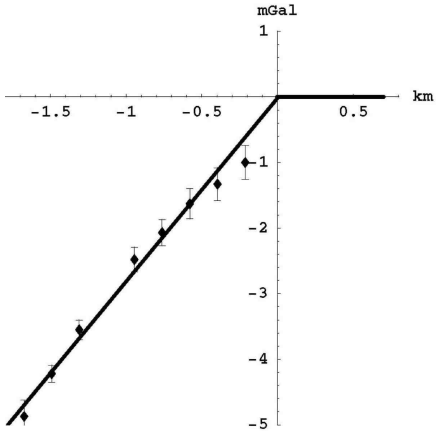}}\,
\parbox{70mm}{\caption{\footnotesize{  The data shows the gravity residuals for the Greenland Ice Shelf \cite{Ander89} Airy measurements of
the $g(r)$  profile,  defined as $\Delta g(r) = g_{Newton}-g_{observed}$, and measured in mGal (1mGal $ =10^{-3}$ cm/s$^2$)
and   plotted against depth in km. The bore hole effect is that Newtonian
gravity and the new theory differ only beneath the surface, provided that the measured above surface gravity gradient 
is used in  both theories.  This then gives the horizontal line above the surface. Using (\ref{eqn:BoreE15}) we obtain
$\alpha^{-1}=137.9 \pm  5$ from fitting the slope of the data, as shown. The non-linearity  in the data arises from
modelling corrections for the gravity effects of the   irregular sub ice-shelf rock  topography. \label{fig:Greenland}}}}
\end{figure}

\begin{figure}
\vspace{3.2mm}\parbox{65mm}{\includegraphics[width=65mm,scale=0.2]{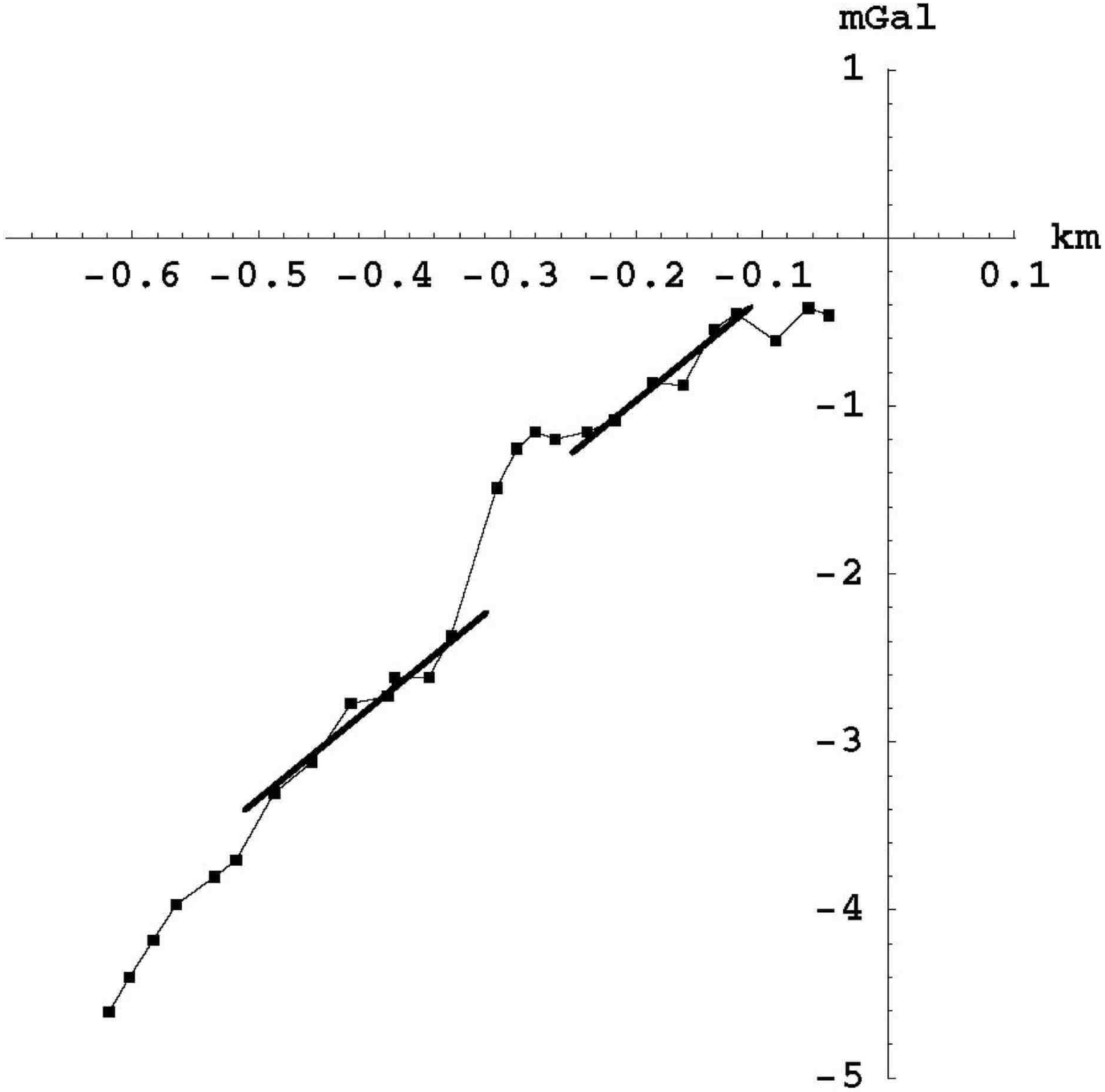}}
\parbox{70mm} {\includegraphics[width=65mm,scale=0.2]{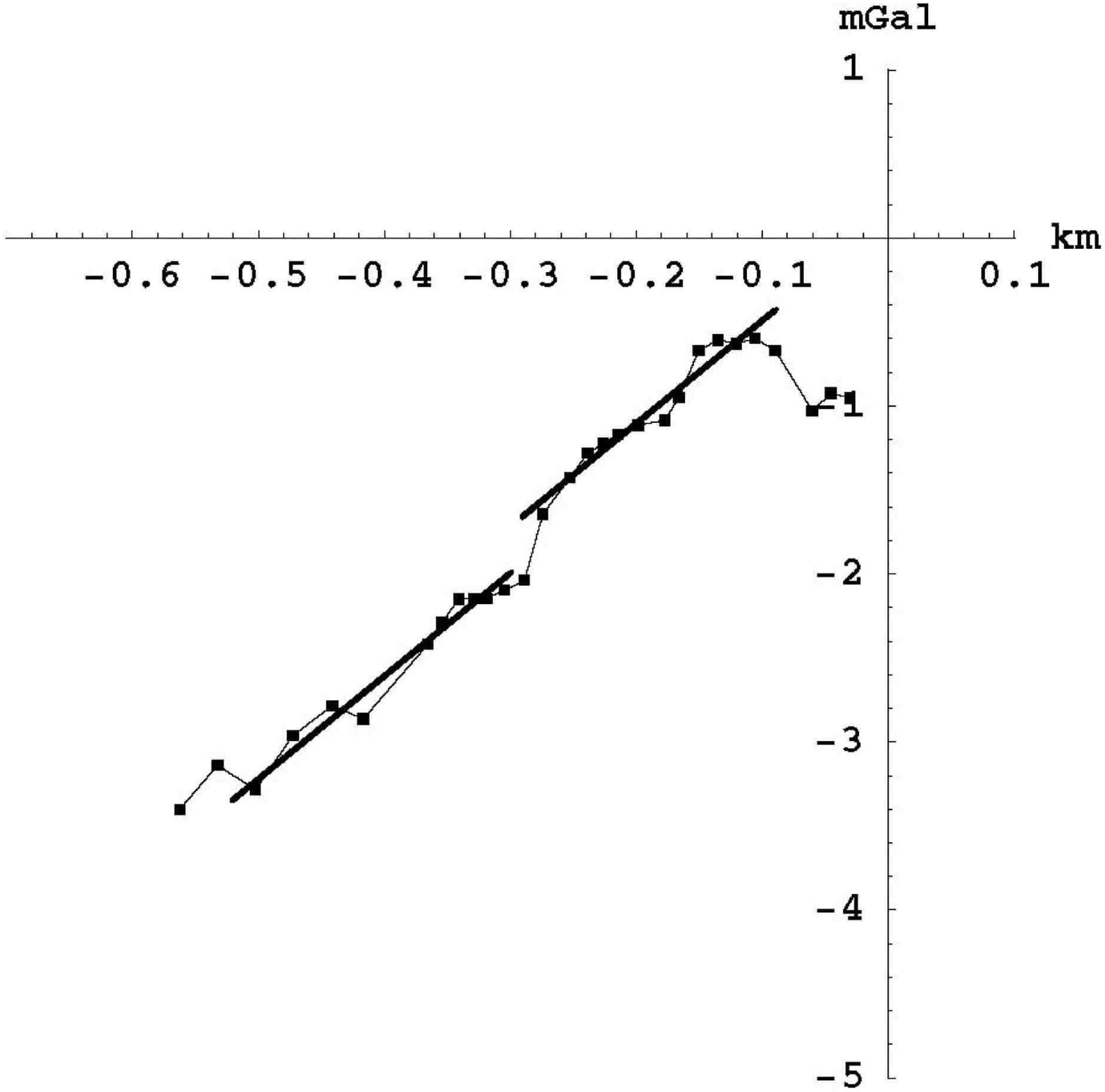}}
\caption{\footnotesize{  Gravity residuals from two of the Nevada bore hole experiments \cite{Thomas90} that give a best fit of $\alpha^{-1}=136.8\pm 3$ on using (\ref{eqn:BoreE15}). Some layering of the rock is evident.}}
\label{fig:Nevada}
\end{figure}

Gravity  residuals  from a bore hole
into the Greenland Ice Shelf  were determined  down to a depth of 1.5 km by Ander {\it et al.} \cite{Ander89} in
1989. The observations were made at the Dye 3 2033 m deep bore hole, which reached the basement rock. This bore hole is
60 km south of the Arctic Circle and 125 km inland from the Greenland east coast at an elevation of 2530 m. It was
believed that the ice  provided an opportunity to use the Airy method to determine $G_N$, but now it is
understood that in fact the bore hole residuals permit the determination of $\alpha$, given a laboratory value for
$G_N$. Various  steps were taken to remove unwanted effects, such as imperfect knowledge of the ice density and,
most dominantly, the terrain effects which arises from ignorance of the profile and density inhomogeneities of the
underlying rock.  The bore hole gravity meter  was calibrated by comparison with
 an absolute gravity meter. The ice density depends on pressure, temperature and air content, with the density rising to
its average value of $\rho=920$ kg/m$^3$ within some 200 m  of the surface, due to compression of the trapped air
bubbles. This surface gradient in the density has been modelled by the author, and is not large enough the affect the
results. The leading source  of uncertainty was from the gravitational effect of the bedrock
topography, and this was corrected for using Newtonian gravity.  The correction from this is actually the cause of the
non-linearity of the data points in Fig.\ref{fig:Greenland}.  A complete analysis would require that the effect of
this rock terrain be also computed using the new theory of gravity, but this was not done.    
Using $G_N=6.6742\times10^{-11}$ m$^3$s$^{-2}$kg$^{-1}$, which is the current CODATA value, we obtain from a least-squares fit of   the linear term in (\ref{eqn:BoreE15}) to the data
points in Fig.\ref{fig:Greenland} that $\alpha^{-1}=137.9\pm 5$, which equals the value of the fine structure
constant 
$\alpha^{-1}=137.036$ to within the errors, and for this reason we identify the  constant $\alpha$ in (\ref{eqn:BoreE15})
as being the fine structure constant. 
 The first  analysis \cite{alpha,DM} of the Greenland Ice Shelf data incorrectly  assumed that the ice density was
930 kg/m$^3$ which gave $\alpha^{-1}=139\pm 5$. However trapped air reduces the standard ice density to  the ice
shelf density of 920 kg/m$^3$, which brings the value of $\alpha$ immediately into better agreement with the
value of $\alpha=e^2/\hbar c$  known from quantum theory.

Thomas and Vogel \cite{Thomas90} performed another bore hole experiment at the Nevada Test Site in 1989 in which they
measured the gravity gradient as a function of depth, the local average matter density, and the above ground
gradient, also known as the free-air gradient.  Their intention was to test the extracted $G_{local}$ and compare
with other values of
$G_N$, but of course using the Newtonian theory. The Nevada bore holes, with typically 3 m diameter, were drilled as a
part of the U.S. Government tests of its nuclear weapons. The density of the rock is measured with a $\gamma-\gamma$
logging tool, which is essentially a $\gamma$-ray attenuation measurement, while in some holes the rock density was
measured with a coreing tool. The rock density was found to be  2000 kg/m$^3$, and is dry. This is the density used
in the analysis herein.   The topography for 1 to 2 km beneath the surface  is dominated by a series of overlapping
horizontal lava flows and alluvial layers. Gravity residuals from two of the bore holes are shown in  
Figs.\ref{fig:Nevada}. All gravity measurements were corrected for the
earth's tide, the terrain on the surface out to 168 km distance, and the evacuation of the holes. The gravity
residuals arise after allowing for, using Newtonian theory, the local lateral mass anomalies but assumed that the matter
beneath the holes occurs in homogeneous ellipsoidal layers.     We see in Fig.\ref{fig:Nevada} that the gravity
residuals are linear with depth, where the density is the average value of 2000 kg/m$^3$, but interspersed by layers
where the residuals show non-linear changes with depth. It is assumed here that these non-linear regions are caused
by variable density layers.  So in  analysing this data we have only used the linear regions, and a simultaneous
least-squares fit of the slope of (\ref{eqn:BoreE15}) to the slopes  of these four linear regions gives
$\alpha^{-1}=136.8\pm3$, which again is in extraordinary agreement with the value of $137.04$ from quantum
theory.   Here  we again used $G_N=6.6742\times10^{-11}$ m$^3$s$^{-2}$kg$^{-1}$, as for the
Greenland data analysis.  Zumberge {\it et al.} \cite{Zumberge1991} performed an extensive underwater Airy experiment, but failed to measure the above water $g$, so their results cannot be analysed in the above manner.

\subsection{Black Hole Masses and the Fine Structure Constant}\label{subsect:black}

\begin{figure}
\hspace{9mm}\includegraphics[scale=0.35]{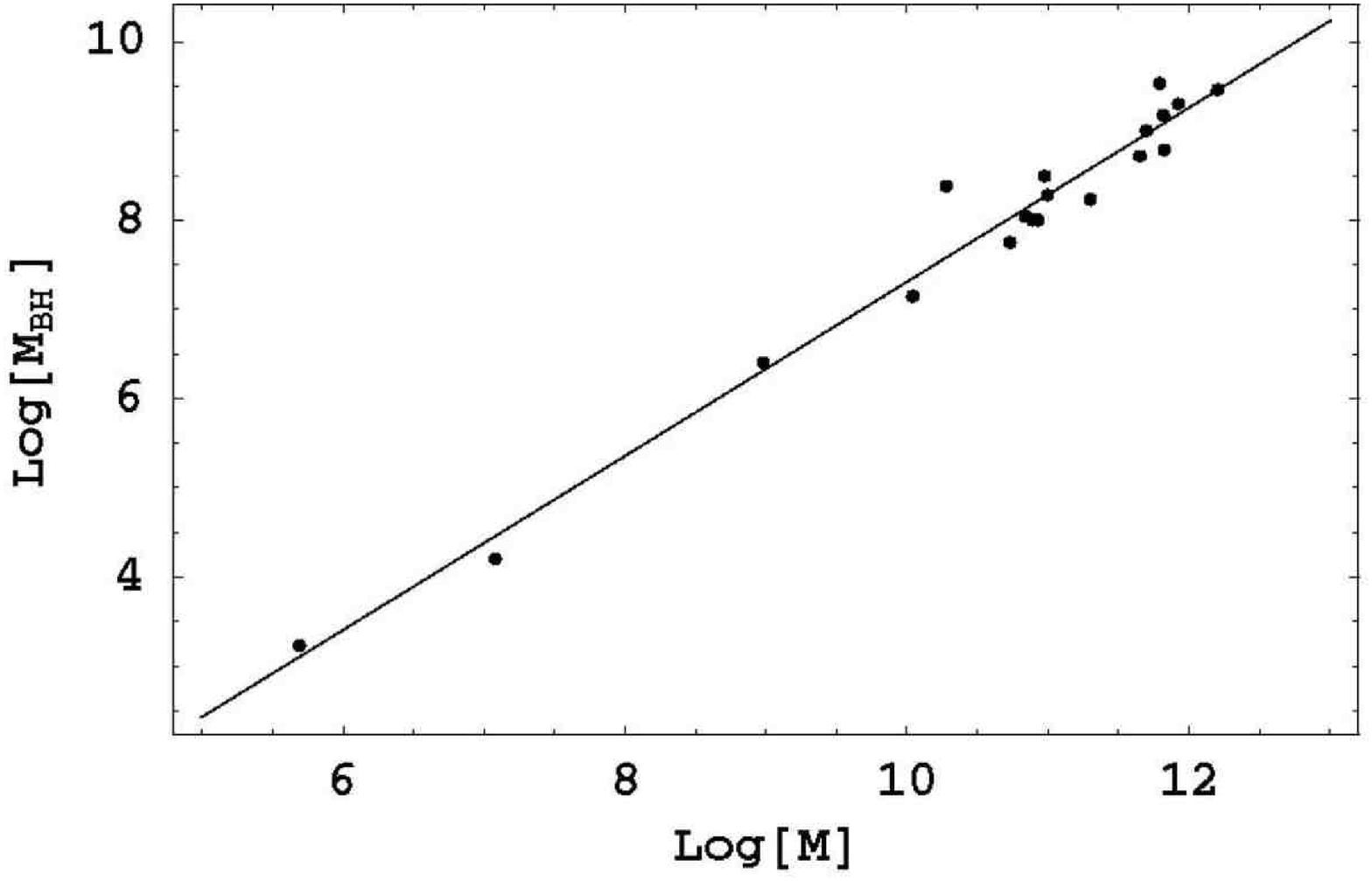}
\vspace{-4mm}
\caption{\footnotesize{The data shows $\mbox{Log}_{10}[M_{BH}]$ for the black hole masses $M_{BH}$  for
a variety of spherical matter systems with masses $M$, plotted against 
$\mbox{Log}_{10}[M]$, in solar masses $M_0$.  The straight line is the prediction from (\ref{eqn:bhmasses}) with $\alpha=1/137$. See \cite{newBH}  for references to the data. 
  \label{fig:blackholes}}}
\end{figure}

Equation (\ref{eqn:HubE1}) (with $\beta=-\alpha$) has `black hole' solutions.  The generic term `black hole' is used because they have a compact closed event horizon where the in-flow speed relative to the horizon equals the speed of light, but in other respects they differ from the putative black holes of General Relativity - in particular their gravitational acceleration is not inverse square law.  The evidence is that it is these new `black holes' from (\ref{eqn:HubE1}) that have been detected. There are two categories: (i) an in-flow singularity induced by the flow into a matter system, such as, herein, a spherical galaxy or globular cluster. These black holes are termed minimal black holes, as their effective mass is minimal, (ii) primordial naked black holes which then attract matter. These result in spiral galaxies, and the effective mass of the black hole is larger than required merely by the matter induced in-flow. These are therefore termed non-minimal black holes.   These explain the rapid formation of structure in the early universe, as the gravitational acceleration is approximately   $1/r$ rather than $1/r^2$. This is the feature that also explains the so-called `dark matter' effect in spiral galaxies.  We consider now  the minimal black holes. 

Equation  (\ref{eqn:HubE1})   has  exact analytic `black hole' solutions  where  $\rho=0$ (actually a one-parameter family - but we write in this form for comparison with the next section)
\begin{equation}
v(r) = K\left(\frac{1}{r}+\frac{1}{R_s}\left(\frac{R_s}{r}  \right)^{\displaystyle{\frac{\alpha}{2}}}  \right)^{1/2}
\label{eqn:Hubvexactb}\end{equation}
where the $1/r$ term can only arise if matter is present, and the 2nd term is the `black hole' effect.   The consequent `black hole' contribution to the total acceleration can be attributed to an effective mass $M_{DM}$, which we now also call $M_{BH}$.  To $O(\alpha)$ this effective mass is independent of the matter density profile, and is given by 
(\ref{eqn:BoreE10}),
\begin{equation}
M_{BH}=M_{DM} = 4\pi\int_0^\infty r^2\rho_{DM}(r)dr = \frac{\alpha}{2}M+O(\alpha^2)
\label{eqn:bhmasses}\end{equation}
This solution is applicable to the black holes at the centre of spherical star systems, where we identify $M_{DM}$ as $M_{BH}$.    So far black holes in 19  spherical star systems have been detected and together their masses are plotted in 
Fig.\ref{fig:blackholes} and compared with (\ref{eqn:bhmasses}) \cite{galaxies,newBH}.

This result applies to any spherically symmetric matter distribution. This means that the bore hole anomaly is indicative of an in-flow singularity at the centre of the earth.  This contributes some  0.4\% of the effective mass of the earth, as defined by Newtonian gravity.  However in star systems this minimal black hole effect is more apparent, and we label $M_{DM}$ as $M_{BH}$.  
  Essentially even in the non-relativistic regime the Newtonian theory of gravity, with its `universal' Inverse Square Law, is deeply flawed.

\subsection{Spiral Galaxies and the Rotation Anomaly}\label{subsect:spiral}

\begin{figure}[t]
\vspace{3.2mm}\parbox{65mm}{\includegraphics[width=65mm]{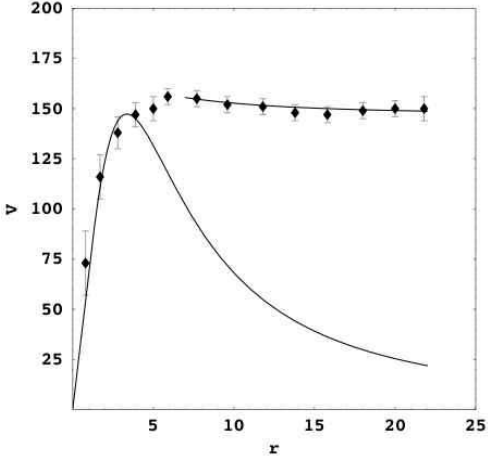}}\,
\parbox{68mm}{\caption{\footnotesize{  Data shows the non-Keplerian rotation-speed curve $v_O$ for the spiral galaxy NGC 3198 in km/s plotted
against radius in kpc/h. Lower curve is the rotation curve from the Newtonian theory  for an
exponential disk, which decreases asymptotically like $1/\sqrt{r}$. The upper curve shows the asymptotic form from
(\ref{eqn:Hubvorbital}), with the decrease determined by the small value of $\alpha$.  This asymptotic form is caused by
the primordial black holes at the centres of spiral galaxies, and which play a critical role in their formation. The
spiral structure is caused by the rapid in-fall towards these primordial black holes.}
\label{fig:NGC3198}}}\end{figure}

Equation (\ref{eqn:Hubvexactb}) gives also a direct explanation for the spiral galaxy rotation anomaly.    For a non-spherical system numerical solutions of (\ref{eqn:HubE1}) are required, but sufficiently far from the centre we find an exact non-perturbative two-parameter class of analytic solutions as in (\ref{eqn:Hubvexactb}).
There $K$ and $R_s$ are arbitrary constants in the $\rho=0$ region, but whose values are determined by matching to the solution in the matter region. Here $R_s$ characterises the length scale of the non-perturbative part of this expression,  and $K$ depends on $\alpha$, $G$ and details of the matter distribution.   From (\ref{eqn:HubE5})  and (\ref{eqn:Hubvexactb}) we obtain a replacement for  the Newtonian  `inverse square law',
\begin{equation}
g(r)=\frac{K^2}{2} \left( \frac{1}{r^2}+\frac{\alpha}{2rR_s}\left(\frac{R_s}{r}\right)
^{\displaystyle{\frac{\alpha}{2}}} 
\right),
\label{eqn:HubgNewl}\end{equation}
in the asymptotic limit.     The centripetal acceleration  relation for circular orbits 
$v_O(r)=\sqrt{rg(r)}$  gives  a `universal rotation-speed curve'
\begin{equation}
v_O(r)=\frac{K}{2} \left( \frac{1}{r}+\frac{\alpha}{2R_s}\left(\frac{R_s}{r}\right)
^{\displaystyle{\frac{\alpha}{2}}} 
\right)^{1/2}
\label{eqn:Hubvorbital}\end{equation}
 Because of the $\alpha$ dependent part this rotation-velocity curve  falls off extremely slowly with $r$, as is indeed observed for spiral galaxies. An example is shown in Fig.\ref{fig:NGC3198}. It was the inability of the  Newtonian  gravity and GR  to explain these observations that led to the  notion of `dark matter'.  So `dark matter' is not  a part of reality.

For the spatial flow in (\ref{eqn:Hubvexactb}) we may compute the effective `dark matter' density from (\ref{eqn:BoreE8})
 \begin{equation}
\rho_{DM}(r) = \frac{(1-\alpha)\alpha}{16\pi G}\frac{K^2}{R^3_s}\left(\frac{R_s}{r}  \right)^{2+\alpha/2} 
\label{eqn:DMDensity}\end{equation}
We see the standard $1/r^2$ behaviour usually attributed to `dark matter' in spiral galaxies. It should be noted  that the Newtonian component of  (\ref{eqn:Hubvexactb}) does not contribute, and that $\rho_{DM}({\bf r})$ is exactly zero in the limit $\alpha\rightarrow 0$.  So supermassive black holes and the spiral galaxy rotation anomaly are all $\alpha$-dynamics phenomena.

\subsection[Lense-Thirring Effect and the GPB Gyroscope Experiment]{Lense-Thirring Effect and the GPB Gyroscope \\ Experiment}\label{subsect:lense}

The Gravity Probe B (GP-B) satellite experiment was launched in April 2004. It has the
 capacity to measure the precession of four on-board gyroscopes to unprecedented
accuracy \cite{Schiff,PEV,Everitt,GPB}.  Such a precession is predicted by 
GR, with
 two components (i) a geodetic precession, and (ii) a `frame-dragging' precession
known as the Lense-Thirring effect.  The latter is particularly interesting effect
 induced by the rotation of the earth, and described in GR in terms of a
`gravitomagnetic' field.  According to GR this smaller effect will give
a precession of   0.042 arcsec per year for the GP-B gyroscopes.   Here we show that GR and the new theory make very
different predictions for the `frame-dragging' effect, and so the GP-B experiment will
be able to decisively test both theories. While predicting the same earth-rotation
induced precession, the new theory has an additional much larger `frame-dragging'
effect caused by the observed translational motion of the earth. As well the new
theory explains the `frame-dragging' effect in terms of  vorticity in a `substratum
flow'.  Herein the magnitude and signature of this new component of the  gyroscope
precession is predicted for comparison with data from GP-B when it becomes available.

\begin{figure}
\vspace{3.2mm}\parbox{60mm}{\includegraphics[width=60mm]{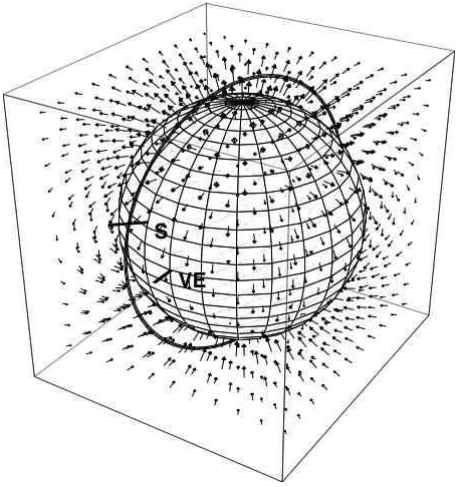}}\,
\parbox{70mm}{\caption{\footnotesize{  Shows the earth (N is up) and vorticity vector field component $\vec{\omega}$ induced by the rotation of the
earth, as in  (\ref{eqn:GPBrotation}). The polar orbit of the GP-B satellite is shown,    ${\bf S}$ is the gyroscope starting 
spin orientation, directed towards the guide  star IM Pegasi, RA = $22^h $ $53^\prime$ $ 2.26^{\prime\prime}$, Dec = $16^0$ $
50^\prime $ $28.2^{\prime\prime}$, and  ${\bf VE}$ is the vernal equinox.\label{fig:Rotation}}}}
\end{figure}

\begin{figure}
\vspace{3.2mm}\parbox{60mm}{\includegraphics[width=60mm]{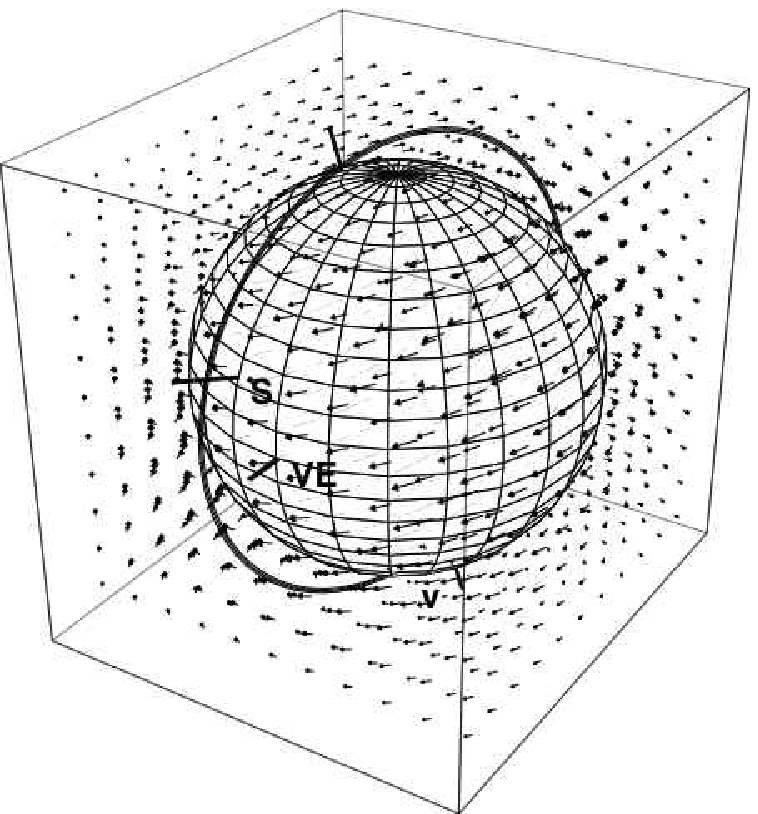}}\,
\parbox{70mm}{\caption{\footnotesize{  Shows the earth (N is up) and the much larger vorticity vector field component $\vec{\omega}$ induced by the
translation of the earth, as in  (\ref{eqn:GPBAMomega}). The polar orbit of the GP-B satellite is shown, and   ${\bf S}$ is the
gyroscope starting  spin orientation, directed towards the guide  star IM Pegasi,  RA = $22^h $ $53^\prime$ $ 2.26^{\prime\prime}$, Dec
= $16^0$ $ 50^\prime $ $28.2^{\prime\prime}$,   ${\bf VE}$ is the vernal equinox,
and ${\bf V}$ is the direction  $\mbox{RA} = 5.2^{h}$, $\mbox{Dec} = -67^0$ of the translational velocity ${\bf v}_c$.\label{fig:Absolute}}}}
\end{figure}

Here we consider one difference between the two theories, namely that associated with the vorticity
part of  (\ref{eqn:Vortomega}), leading to the `frame-dragging' or Lense-Thirring  effect. In GR the vorticity field
 is known as the `gravitomagnetic' field ${\bf B}=-c\:\vec{\omega}$. In  both GR and  the new
theory the vorticity is given by (\ref{eqn:VortCG4b}) but with a key difference: in GR ${\bf v}_R$ is {\it only} the
rotational velocity of the matter in the earth, whereas in the 3-space dynamics ${\bf v}_R$ is the vector sum of the  rotational velocity and the translational velocity of the
earth through the substratum.    

First consider the common but much smaller rotation induced `frame-dragging' or vorticity effect. Then
${\bf v}_R({\bf r})={\bf w}\times{\bf r}$ in (\ref{eqn:Vortomega}), where ${\bf w}$ is the angular
velocity of the earth, giving
\begin{equation}
\vec{\omega}({\bf r})=4\frac{G}{c^2}\frac{3({\bf r}.{\bf L}){\bf r}-r^2{\bf L}}{2 r^5},
\label{eqn:GPBrotation}\end{equation}
where ${\bf L}$ is the \index{angular momentum - earth} angular momentum of the earth, and ${\bf
r}$ is the distance from the centre. This component of the vorticity field is shown in
Fig.\ref{fig:Rotation}.  Vorticity may be detected by observing the precession of the GP-B
gyroscopes.  The vorticity term in 
 (\ref{eqn:GPBrotation}) leads to a torque on the angular momentum ${\bf S}$ of the gyroscope,
\begin{equation}
\vec{\tau}= \int d^3 r \rho({\bf r})\; {\bf r}\times(\vec{\omega}({\bf r}) \times{\bf v}_R({\bf r})),
\label{eqn:GPBtorque1}\end{equation}
where $\rho$ is its  density, and where
  ${\bf v}_R$ is used here to describe the rotation of the gyroscope.  Then $d{\bf S}=\vec{\tau}dt$ is the change in
${\bf S}$ over the time interval $dt$. In the above case 
${\bf v}_R({\bf r})={\bf s}\times{\bf r}$, where ${\bf s}$ is the angular velocity of the gyroscope.  
This gives
\begin{equation}
\vec{\tau}=\frac{1}{2}\vec{\omega}\times{\bf S}
\label{eqn:GPBtorque2}\end{equation}
and so $\vec{\omega}/2$ is the instantaneous angular velocity of precession of the gyroscope. This corresponds to
the well known fluid result that the vorticity vector is twice the angular velocity vector.   For GP-B the direction
of
${\bf S}$   has been chosen so that this precession is cumulative and, on averaging  over an orbit,
corresponds to some $7.7\times 10^{-6}$ arcsec per orbit, or 0.042 arcsec per year.  GP-B has been superbly
engineered so that measurements to a precision of 0.0005 arcsec are possible. 

\begin{figure}[t]
\hspace{15mm}\includegraphics[scale=1.3]{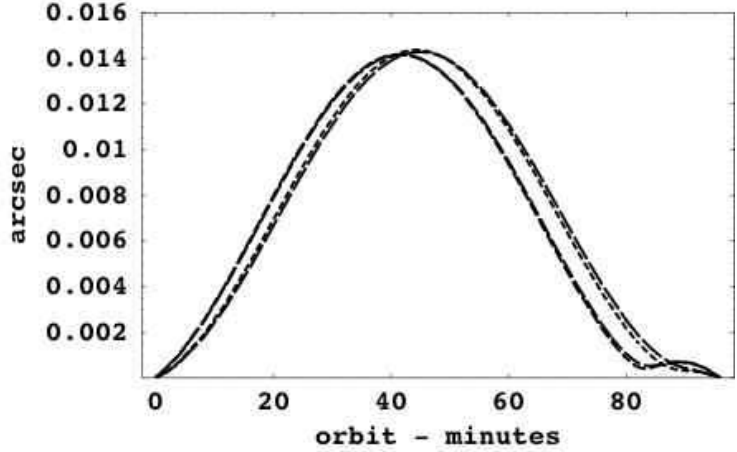}
\caption{\footnotesize{Predicted  variation of the precession angle  $\Delta \Theta=|\Delta {{\bf S}}(t)|/|{\bf S}(0)|$, in
arcsec, over one 97 minute GP-B orbit, from the vorticity induced by the translation of the earth, as given by
(\ref{eqn:GPBprecession}). The orbit time begins at location
${\bf S}$. Predictions are for the  months of April, August, September and February, labeled by  increasing dash
length.  The `glitches' near 80 minutes are caused by the angle effects in (\ref{eqn:GPBprecession}). These changes arise
from the effects of the changing orbital velocity of the earth about the sun.  The GP-B expected angle measurement
accuracy is 0.0005 arcsec.  }
\label{fig:Precession}}\end{figure}

However for the unique translation-induced precession if we  use $v_R \approx v_C = 430$ km/s in the
direction  $\mbox{RA} =5.2^{hr}$, $\mbox{Dec} =-67^0$, namely ignoring the effects of the orbital motion of the
earth, the observed flow past the earth towards the sun, and the flow into the earth, and effects of
the gravitational waves, then (\ref{eqn:Vortomega}) gives
\begin{equation}
\vec{\omega}({\bf r})=\frac{2GM}{c^2}\frac{{\bf v}_C\times{\bf r}}{r^3}.
\label{eqn:GPBAMomega}\end{equation}
This much larger component of the vorticity field is shown in Fig.\ref{fig:Absolute}.
The maximum magnitude of the speed of this precession  component is $\omega/2=gv_C/c^2=8 \times10^{-6}$arcsec/s, where here
$g$ is the gravitational acceleration at the altitude of the satellite.   This precession has a different signature: it  is
not cumulative, and is detectable by its variation over each single orbit, as its orbital average is zero, to first
approximation.   Fig.\ref{fig:Precession} shows   $\Delta \Theta=|\Delta {{\bf S}}(t)|/|{\bf S}(0)|$  over 
 one orbit, where, as in general,
\begin{equation}\Delta {{\bf S}}(t) =
\int_0^t dt^\prime \frac{1}{2}\vec{\omega}({\bf r}(t')) \times {\bf S}(t^\prime)
\approx \left(\int_0^t dt^\prime \frac{1}{2}\vec{\omega}({\bf r}(t'))\right) \times
{\bf S}(0).
\label{eqn:GPBprecession}\end{equation}  
Here $\Delta {{\bf S}}(t)$ is the integrated change in spin, and where
the approximation arises  because the change in
${\bf S}(t^\prime)$ on the RHS of (\ref{eqn:GPBprecession}) is negligible.   The plot in  Fig.\ref{fig:Precession}  shows
this effect to be some 30$\times$ larger than the expected GP-B errors, and so easily detectable, if it exists as
predicted herein.  This precession is about the instantaneous direction of the vorticity $\vec{\omega}({\bf r}((t))$
at the location of the satellite, and so is neither in the plane, as for the geodetic precession, nor perpendicular to
the plane of the orbit, as for the earth-rotation induced vorticity effect. 

Because the yearly orbital rotation  of the earth about the sun slightly effects 
${\bf v}_C$ \cite{AMGE} predictions for four months throughout the  year are shown  in
Fig.\ref{fig:Precession}. Such yearly effects were first seen in the Miller \cite{Miller}
experiment, see Fig.\ref{fig:MillerAz}.

\subsection{Cosmology: Expanding 3-Space and the Hubble Effect}\label{subsect:cosmology}

We now examine the predictions for the global expansion of the 3-space that follows from (\ref{eqn:HubE1}) (with $\beta=-\alpha$). We shall see that the solution gives an excellent parameter-free fit to the supernovae and gamma-ray bursts magnitude - redshift data \cite{Hubble}. This implies that there is no need to have a cosmological constant or `dark energy',  which are required by GR in order to fit this data. These also lead to the prediction that the universe expansion will accelerate in the future. This effect is also not required by the new 3-space dynamics.   So, like `dark matter', `dark energy' is an unnecessary and spurious notion.

\begin{figure}[t]
\hspace{12mm}{\includegraphics[width=100mm]{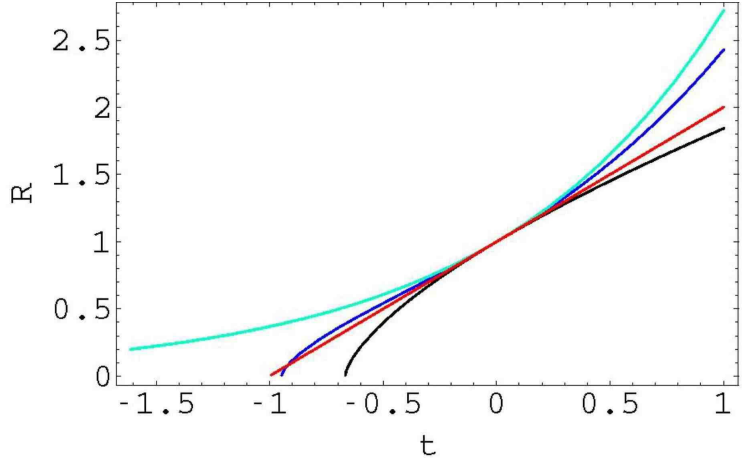}}\,
{\vspace{-3mm}\caption{\footnotesize{  Plot of the scale factor $R(t)$ vs $t$, with $t=0$ being `now' with $R(0)=1$, for the four cases discussed in the text, and corresponding to the plots in  Figs.\ref{fig:SN1} and \ref{fig:SN2}: (i) the upper  curve  is the `dark energy' only case, resulting in an exponential acceleration at all times, (ii) the bottom curve is the matter only prediction, (iii) the 2nd highest curve (to the right of $t=0$) is the best-fit  `dark energy' plus matter case showing a past deceleration and future exponential acceleration effect. The straight line plot   is the dynamical 3-space prediction showing a slightly  older universe compared to case (iii). We see that the best-fit `dark energy' - matter curve essentially converges on the dynamical 3-space result. All plots have the same slope at $t=0$, i.e. the same value of $H_0$. If the age of the universe is inferred to be some 14Gyrs for case (iii) then the age of the universe is changed to some 14.7Gyr for case (iv). 
 }\label{fig:Rtplot}}}
\end{figure}

Suppose that  we have a radially symmetric density $\rho(r,t)$ and that we look for a radially symmetric time-dependent flow ${\bf v}({\bf r},t) =v(r,t)\hat{\bf r}$ from (\ref{eqn:HubE1}) (with $\beta=-\alpha$).  Then $v(r,t)$ satisfies the equation,  with $v^\prime=\displaystyle{\frac{\partial v(r,t)}{\partial r}}$,
\begin{equation}
\frac{\partial}{\partial t}\left( \displaystyle{\frac{2v}{r}}+v^\prime\right)+vv^{\prime\prime}+2\frac{vv^{\prime}}{r}+(v^\prime)^2+\frac{\alpha}{4}\left(\frac{v^2}{r^2} +\frac{2vv^\prime}{r}\right)\nonumber \\ 
=- 4\pi G \rho(r,t)  \label{eqn:Hubradialflow}\end{equation}
Consider first the zero energy case $\rho=0$. Then we have a Hubble  solution $v(r,t)=H(t)r$, a centreless flow, determined by
\begin{equation}{\dot H}+\left(1+\frac{\alpha}{4}\right)H^2=0
\end{equation}
with ${\dot H}=\displaystyle{\frac{dH}{dt}}$.  We also introduce in the usual manner the scale factor $R(t)$ according to $H(t)=\displaystyle{\frac{1}{R}\frac{dR}{dt}}$. We then obtain the solution
\vspace{-2mm}
\begin{equation}
H(t)=\frac{1}{(1+\frac{\alpha}{4})t}=H_0\frac{t_0}{t}; \mbox{\ \  }  R(t)=R_0\left(\frac{t}{t_0} \right)^{4/(4+\alpha)}
\label{eqn:Hubspacexp}\end{equation}
where $H_0=H(t_0)$ and $R_0=R(t_0)$. 
We can write the  Hubble function $H(t)$ in terms of $R(t)$ via the inverse function $t(R)$, i.e. $H(t(R))$ and finally as $H(z)$, where the redshift observed now, $t_0$, relative to the wavelengths at time $t$, is  $z=R_0/R-1$. Then we obtain
\begin{eqnarray}
H(z)={H_0}(1+z)^{1+\alpha/4}
\label{eqn:HubH2a}\end{eqnarray}

\begin{figure}
\vspace{-20mm}
\hspace{0mm}\includegraphics[scale=0.4]{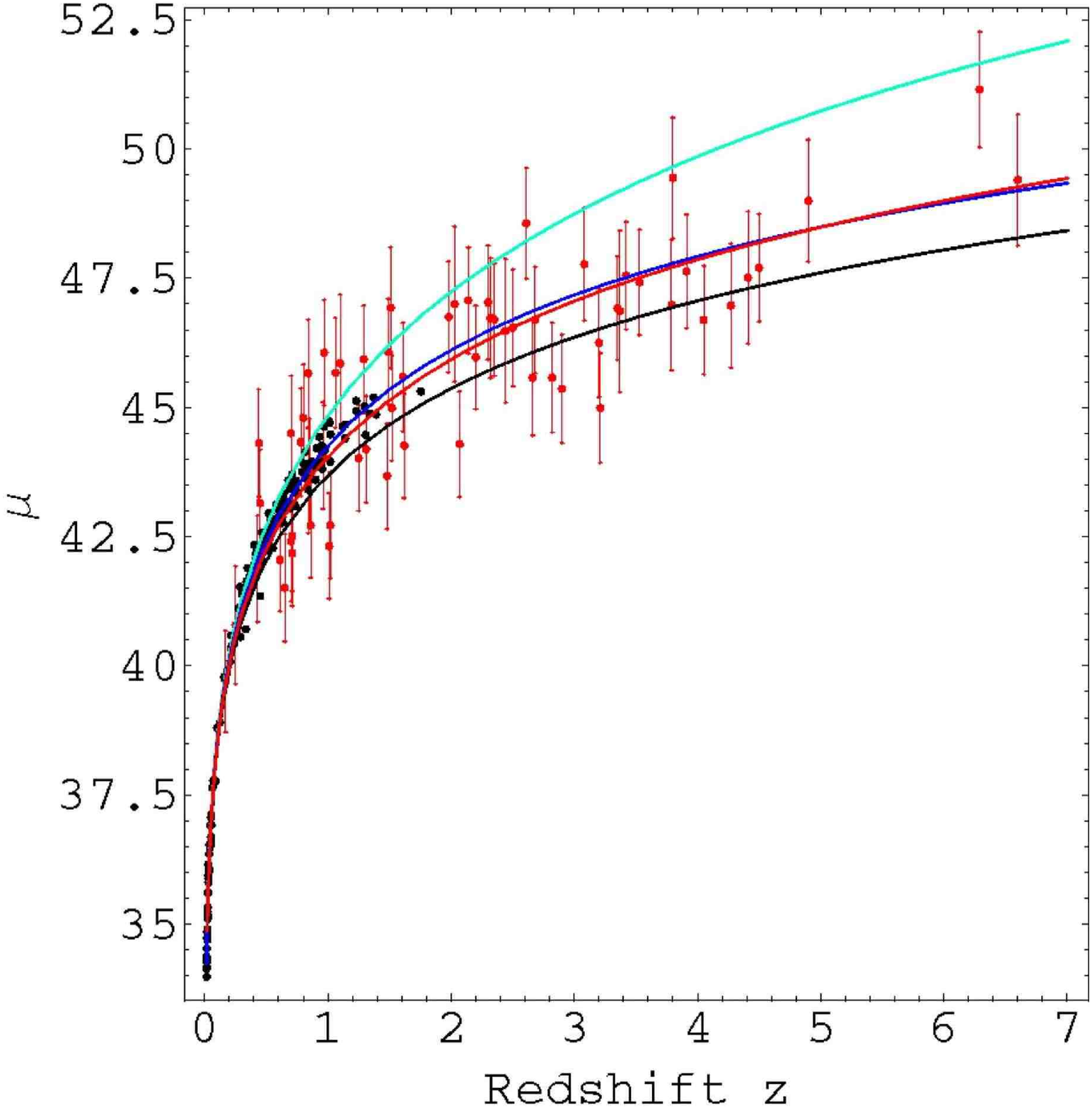}
\vspace{-10mm}\caption{\footnotesize{ Hubble diagram showing the combined supernovae data from Davis {\it et al.} \cite{Davis} using several data sets from   Riess {\it et al.} (2007)\cite{Riess} and Wood-Vassey {\it et al.}  (2007)\cite{WV} (dots without error bars for clarity - see Fig.\ref{fig:SN2} for error bars ) and the Gamma-Ray Bursts data (with error bars) from Schaefer \cite{GRB}.  Upper curve (green) is `dark energy' only $\Omega_\Lambda=1$, lowest curve (black) is matter only $\Omega_m=1$. Two middle curves show best fit of `dark energy'-matter (blue) and dynamical 3-space prediction (red), and are essentially indistinguishable.  However the theories make very different predictions for the future and for the age of the universe. We see that the best-fit `dark energy' - matter curve essentially converges on the dynamical 3-space prediction.}
\label{fig:SN1}}\end{figure}

\begin{figure}[t]
\vspace{3.0mm}\parbox{90mm}{\includegraphics[width=90mm]{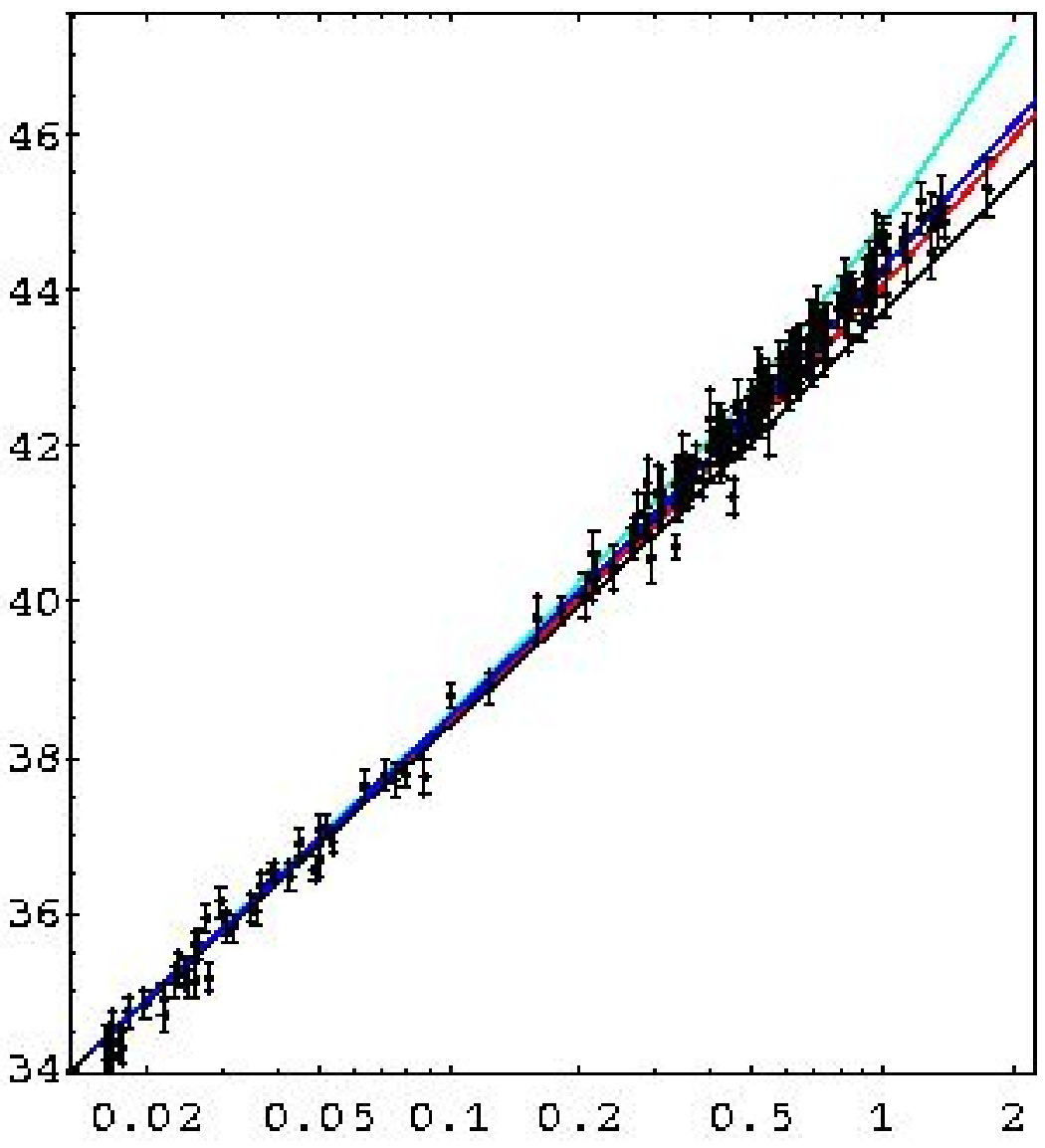}}\,
\parbox{40mm}{\caption{\footnotesize{  Hubble diagram as in Fig.\ref{fig:SN1} but plotted logarithmically  to reveal details for  $z<2$, and without GRB data.  Upper curve (green) is `dark energy' only $\Omega_\Lambda=1$. Next curve (blue) is  best fit of `dark energy'-matter. Lowest curve (black) is matter only $\Omega_m=1$. 2nd lowest curve (red) is  dynamical 3-space prediction. }\label{fig:SN2}}}
\end{figure}

We need to determine the distance travelled by the light from a supernova or gamma-ray burst (GRB) before detection, for this determines the apparent brightness. Using a choice of embedding-space coordinate system with $r=0$ at the location of a supernova or GRB the 
speed of light relative to this embedding space frame is $c+v(r(t),t)$, i.e $c$ wrt the space itself, as noted in Sect.\ref{sect:maxwell},  where $r(t)$ is the distance from the source. Then the distance travelled by the light at time $t$ after emission at time $t_1$ is determined implicitly by
\begin{equation}
r(t)=\int_{t_1}^t dt^\prime(c+v(r(t^\prime), t^\prime),
\label{eqn:Hubdistance1}\end{equation}
which has the solution, on using $v(r,t)=H(t)r$,
\begin{equation}
r(t)=c R(t)\int_{t_1}^t \frac{dt^\prime}{R(t^\prime)}
\label{eqn:Hubdistance2}\end{equation}
Expressed in terms of the observable redshift $z$ this gives
\begin{equation}
r(z)=c (1+z)\int_{0}^z \frac{dz^\prime}{H(z^\prime)}
\label{eqn:Hubdistance3}\end{equation}
 The effective dimensionless distance measure is given by
 \begin{equation}
d(z)=(1+z)\int_0^z \frac{H_0 dz^\prime}{H(z^\prime)}
\label{eqn:HubH1a}\end{equation}
 and the theory distance modulus is then defined by
\begin{equation}
\mu_{th}(z)=5\log_{10}(d(z))+m
\label{eqn:HubH1b}\end{equation}
Because all the selected supernova have the same absolute magnitude, the value of the constant  $m$  is determined
by fitting the low $z$ data. For the GRB data the magnitudes have been corrected  so that the data gives a best fit to the dark energy/matter plot in Fig.\ref{fig:SN1}.

Using the  Hubble expansion (\ref{eqn:HubH2a}) in (\ref{eqn:HubH1a}) and (\ref{eqn:HubH1b}) we obtain the  curve shown in Figs.\ref{fig:SN1} and \ref{fig:SN2}, yielding an excellent agreement with the supernovae and GRB data. Note that because $\alpha/4$ is so small it actually has negligible effect on these plots.  Hence the dynamical 3-space gives an immediate account of the universe expansion data, and does not require the introduction of  a cosmological constant or `dark energy', but which will be nevertheless discussed next.

When the energy density is not zero we need to take account of the dependence of $\rho(r,t)$ on the scale factor of the universe. In the usual manner we thus write
\begin{equation}
\rho(r,t)=\frac{\rho_{m}}{R(t)^3}+\frac{\rho_{r}}{R(t)^4}+\Lambda
\end{equation}
for matter, EM radiation and the cosmological constant or `dark energy' $\Lambda$, respectively, where the matter and radiation is approximated by a spatially uniform (i.e independent of $r$)  equivalent matter density. We argue here that $\Lambda$ - the dark energy density, like dark matter, is an unnecessary concept.
Then (\ref{eqn:Hubradialflow}) becomes for $R(t)$
\begin{eqnarray}
\frac{\ddot  R}{R}+\frac{\alpha}{4}\frac{{\dot R}^2}{R^2}
=-\frac{4\pi G}{3}\left(\frac{\rho_{m}}{R^3}+\frac{\rho_{r}}{R^4}+\Lambda \right)
\end{eqnarray}giving
\begin{equation}
{\dot R}^2=\frac{8\pi G}{3}\left(\frac{\rho_{m}}{R}+\frac{\rho_{r}}{R^2}+\Lambda R^2\right)-\frac{\alpha}{2}\int\frac{{\dot R}^2}{R}dR
\label{eqn:HubR2}\end{equation}
 In terms of ${\dot R}^2$ this has the solution
\begin{equation}{\dot R}^2\!=\!\frac{8\pi G}{3}\!\!\left(\!\frac{\rho_{m}}{(1-\frac{\alpha}{2})R}\!+\!\frac{\rho_{r}}{(1-\frac{\alpha}{4})R^2}\!+\!\frac{\Lambda R^2}{(1+\frac{\alpha}{4})}\!+\!b R^{-\alpha/2}\!\right)
\label{eqn:HubR3}\end{equation}
which is easily checked by substitution into (\ref{eqn:HubR2}), and where $b$ is an arbitrary integration constant. Finally we obtain from  (\ref{eqn:HubR3})
\begin{equation}
t(R)=\int^R_{R_0}\frac{dR}{\sqrt{\displaystyle{\frac{8\pi G}{3}}\left(\displaystyle{\frac{\rho_{m}}{R}+\frac{\rho_{r}}{R^2}}+\Lambda R^2+b R^{-\alpha/2}\right)}}
\label{eqn:HubR4}\end{equation} 
where now we have re-scaled parameters $\rho_m\rightarrow\rho_m/(1-\frac{\alpha}{2}), \rho_r\rightarrow\rho_r/(1-\frac{\alpha}{4})$ and $\Lambda\rightarrow\Lambda/(1+\frac{\alpha}{4})$. When $\rho_m=\rho_r=\Lambda=0$, (\ref{eqn:HubR4})
reproduces the expansion in (\ref{eqn:Hubspacexp}), and so the density terms in (\ref{eqn:HubR4}) give the modifications to  the dominant purely spatial expansion dynamics, which we have noted above  already gives an excellent account of the data.
From (\ref{eqn:HubR4})  we then obtain
\begin{equation}
H(z)^2={H_0}^2(\Omega_m(1+z)^3+\Omega_r(1+z)^4 +\Omega_\Lambda+\Omega_s(1+z)^{2+\alpha/2})
\label{eqn:HubH2}\end{equation}
with
\begin{equation}
\Omega_m+\Omega_r+\Omega_\Lambda+\Omega_s=1.
\end{equation}

Using the Hubble function  (\ref{eqn:HubH2}) in (\ref{eqn:HubH1a}) and (\ref{eqn:HubH1b}) we obtain  the plots in  Figs.\ref{fig:SN1} and \ref{fig:SN2} for four cases: (i) $\Omega_m=0,  \Omega_r=0, \Omega_\Lambda=1, \Omega_s=0$,  i.e a pure `dark energy' driven expansion, (ii)  $\Omega_m=1,  \Omega_r=0, \Omega_\Lambda=0, \Omega_s=0$  showing that a matter only expansion is not a good account of the data, (iii) from a least squares fit with $\Omega_s=0$ we find $\Omega_m=0.28,  \Omega_r=0, \Omega_\Lambda=0.68$  which led to the suggestion that `dark energy' effect was needed to fix the poor fit from (ii), and finally  (iv) $\Omega_m=0,   \Omega_r=0,  \Omega_\Lambda=0,  \Omega_s=1$, as noted above, that the spatial expansion dynamics alone gives a good account of the data. Of course the EM radiation term $\Omega_r$ is non-zero but small and determines the expansion during the baryogenesis initial phase, as does the spatial dynamics expansion term because of the $\alpha$ dependence. If the age of the universe is inferred to be some 14Gyrs for case (iii) then, as seen in Fig.\ref{fig:Rtplot}, the age of the universe is changed to some 14.7Gyr for case (iv). We see that the best-fit `dark energy' - matter curve essentially converges on the dynamical 3-space result.

The induced effective spacetime metric in (\ref{eqn:GRE14}) is for the Hubble expansion
\begin{equation}
ds^2=g_{\mu\nu}dx^\mu dx^\nu=dt^2-(d{\bf r}-H(t){\bf r})dt)^2/c^2 
\label{PGmetric}\end{equation}
 The occurrence of $c$ has nothing to do with the dynamics of the 3-space - it is related to the geodesics of relativistic quantum matter, as  shown in Sect.\ref{sect:general}. Changing  variables  ${\bf r}\rightarrow R(t){\bf r}$
we obtain
\begin{equation}
ds^2=g_{\mu\nu}dx^\mu dx^\nu=dt^2-R(t)^2d{\bf r}^2/c^2 
\label{Hubblemetric}\end{equation}
which is the usual   Friedman-Robertson-Walker (FRW) metric in the case of a flat spatial section. However when solving for $R(t)$ using the Hilbert-Einstein GR equations
the $\Omega_s$ term (with $\alpha\rightarrow 0$) is usually only present when the spatial curvature is 
non-zero.  So some problem appears to be present in the usual GR analysis of the FRW metric. However above we see that that term arises in fact even when the embedding space is flat.

\section{Conclusions}
We have briefly reviewed the extensive evidence for a dynamical 3-space, with the minimal dynamical equation now known and confirmed by numerous experimental and observational data.  This 3-space has been repeatedly detected since the Michelson-Morley experiment of 1887, and  they also detected `gravitational waves', which are just 3-space velocity fluctuations. As well the dynamical 3-space  has been indirectly detected by means of the dynamical equation explaining diverse phenomena.  We have shown that this equation has a Hubble expanding 3-space solution that in a parameter-free manner manifestly fits the recent supernovae and gamma-ray bursts redshift data. All of these successes imply  that `dark energy' and `dark matter' are unnecessary notions.  The Hubble solution leads to a uniformly expanding universe, and so  without acceleration: the claimed acceleration is merely a spurious artifact related to the  unnecessary `dark energy' notion.  This result gives an older age for the universe of some 14.7Gyr, and resolves as well various problems such as the fine turning problem, the horizon problem and other difficulties in the current modelling of the universe.  We have also shown why the spacetime formalism appeared to be so successful, despite having no ontological status. One key discovery has been that Newton's theory of gravity is flawed, except in the very special case of planets in orbit about a sun, which is of course the restricted manifestation of gravity that was available to Newton.  

At a deeper level the occurrence of $\alpha$ in (\ref{eqn:HubE1}) suggests that 3-space is actually a quantum system, and that (\ref{eqn:HubE1}) is merely a phenomenological description of that at the `classical' level. In which case the $\alpha$-dependent dynamics amounts to the detection of quantum space and quantum gravity effects, although clearly not of the form suggested by the quantisation of General Relativity. At a deeper level the information-theoretic {\it Process Physics} has given insights into the possible nature of reality as a limited self-referential system, in which quantum space and quantum matter are emergent phenomena, with both exhibiting non-local effects. In particular it implies that we have a `universal'  process time, as distinct from the current prevailing geometrical modelling of time.  These results all suggest that a radically different paradigm for reality is emerging, and in which we see a unification of quantum space and quantum matter, and with gravity an emergent phenomenon.

\vspace{5mm}
Special thanks to Tim Eastman,   Erich Weigold, Igor Bray and Lance McCarthy.

\end{document}